\documentclass[12pt,a4paper,twoside]{article} 
\usepackage{latexsym}
\usepackage{amsmath}
\usepackage{amssymb}
\usepackage{amsfonts}
\usepackage{ifthen}
\usepackage{mathrsfs}               
\usepackage{bbm}                    
\usepackage{bbold}                  
\usepackage{textcomp}
\usepackage{theorem}
\newcommand{\CC}{\mathbb{C}}

\newcommand{\NN}{\mathbb{N}}

\newcommand{\RR}{\mathbb{R}}

\newcommand{\ZZ}{\mathbb{Z}}
\newcommand{\supp}{\mathrm{supp}}

\newcommand{\dist}{\mathrm{dist}}
\newcommand{\Ran}{\mathrm{Ran}}

\newcommand{\Ker}{\mathrm{Ker}} 

\newcommand{\sgn}{\mathrm{sgn}}
\newcommand{\loc}{\mathrm{loc}}

\newcommand{\rot}{\mathrm{curl}}
\newcommand{\id}{\mathbbm{1}}
\newcommand{\klg}{\leqslant} 
\newcommand{\grg}{\geqslant}          
\newcommand{\ve}{\varepsilon}
\newcommand{\vp}{\varphi}

\newcommand{\vr}{\varrho}
\newcommand{\vt}{\vartheta}

\newcommand{\wt}[1]{\widetilde{#1}}
\newcommand{\SL}{\langle \,}    
\newcommand{\SR}{\, \rangle}    
\newcommand{\SPn}[2]{\langle \,#1\,|\,#2\, \rangle} 
\newcommand{\SPb}[2]{\big\langle \,#1\,\big|\,#2\, \big\rangle} 
\newcommand{\SPB}[2]{\Big\langle \,#1\,\Big|\,#2\, \Big\rangle}
\newcommand{\ol}[1]{\overline{#1}} 
\newcommand{\wh}[1]{\widehat{#1}}  
\newcommand{\mr}[1]{\mathring{#1}} 
\newcommand{\bigO}{\mathcal{O}}    

\newcommand{\DA}{D_{A}}             
\newcommand{\DAV}{D_{A,V}}
\newcommand{\DAVC}{D_{A,V_{\mathrm{C}}}}
\newcommand{\D}[1]{D_{#1}}
\newcommand{\Dt}{\check{D}_0}

\newcommand{\PO}{\Lambda^+_{0}}        
        
\newcommand{\PA}{\Lambda^+_{A}}
\newcommand{\PAV}{\Lambda^+_{A,V}}
\newcommand{\PAVC}{\Lambda^+_{A,V_{\mathrm{C}}}}
\newcommand{\PAVm}{\Lambda^-_{A,V}}
\newcommand{\PAVpm}{\Lambda^{\pm}_{A,V}}
\newcommand{\PAVj}{\Lambda^{+\,(j)}_{A,V}}
\newcommand{\PAVF}{\Lambda^F_{A,V}}

\newcommand{\PAVFN}{\Lambda^{F,N}_{A,V}}
\newcommand{\PP}{\Lambda^+}
\newcommand{\PPsharp}{\Lambda^{\sharp}}
\newcommand{\PPpm}{\Lambda^{\pm}}

\newcommand{\PAVN}{\Lambda^{+,N}_{A,V}}
\newcommand{\PAVNme}{\Lambda^{+,N-1}_{A,V}}
\newcommand{\Pp}[1]{\Lambda^+_{#1}}

\newcommand{\HH}{H}                  
\newcommand{\VE}{V_{\mathrm{E}}}     
\newcommand{\VC}{V_{\mathrm{C}}}
\newcommand{\VCt}{V_{\mathrm{C}}^{\mathrm{s}}}
\newcommand{\VCtt}{\check{V}_{\mathrm{C}}^{\mathrm{s}}}
\newcommand{\VCr}{V_{\mathrm{C}}^{\mathrm{r}}}
\newcommand{\VM}{V_{\mathrm{H}}}

\newcommand{\VME}{V_{\mathrm{HE}}}
\newcommand{\RA}[1]{R_A(#1)}       
\newcommand{\RAV}[1]{R_{A,V}(#1)}

\newcommand{\R}[2]{R_{#1}(#2)}
\newcommand{\dom}{\mathcal{D}}  
\newcommand{\formd}{\mathcal{Q}}
\newcommand{\core}{\mathscr{D}_N}
\newcommand{\coreNme}{\mathscr{D}_{N-1}}
\newcommand{\Th}{\sE^\cA}   
\newcommand{\ball}[2]{\mathcal{B}_{#1}(#2)} 
\newcommand{\Perm}{\mathfrak{S}}
\newcommand{\proof}{\noindent{\it Proof: }}
\newcommand{\qed}{\hfill$\blacksquare$}
\newcommand{\LO}{\mathscr{L}}      
\newcommand{\COI}{\mathscr{D}}   
\newcommand{\HR}{\mathscr{H}} 
\newcommand{\spec}{\mathrm{\sigma}}
\newcommand{\specac}{\mathrm{\sigma}_{\mathrm{ac}}}

\newcommand{\specess}{\mathrm{\sigma}_{\mathrm{ess}}}

\newcommand{\cA}{{\cal A}}\newcommand{\cN}{{\cal N}}

\newcommand{\cQ}{{\cal Q}}

\newcommand{\cM}{{\cal M}}\newcommand{\cY}{{\cal Y}}       
 
\newcommand{\sE}{\mathscr{E}}
\newcommand{\sR}{\mathscr{R}}
\newcommand{\sG}{\mathscr{G}}

\theoremstyle{change}
\newtheorem{theorem}{Theorem.}[section]
\newtheorem{lemma}[theorem]{Lemma.}
\newtheorem{proposition}[theorem]{Proposition.}
\newtheorem{corollary}[theorem]{Corollary.}

\newtheorem{example}[theorem]{Example.}
\newtheorem{remark}[theorem]{Remark.}
\newtheorem{hypothesis}{Hypothesis.}
\begin{document}

\begin{center}
{\Large{\bf
Spectral theory of no-pair Hamiltonians
}}
\end{center}

\vspace{0.6cm}

\begin{center}
{\sc {\large Oliver Matte and Edgardo Stockmeyer}  }\\

\smallskip

Mathematisches Institut der Universit\"at M\"unchen\\
Theresienstra{\ss}e 39, D-80333 M\"unchen, Germany.
\end{center}

\vspace{0.6cm}

\begin{abstract}
\noindent
We prove a HVZ theorem for a general class of no-pair Hamiltonians
describing an atom or positively charged ion
with several electrons in the presence of a
classical external magnetic field. Moreover, we show that there exist
infinitely many eigenvalues below the essential spectrum and that the
corresponding eigenfunctions are
exponentially localized. The novelty is that the electrostatic
and 
magnetic vector 
potentials as well as a non-local exchange potential
are included in the projection determining the model.  As a main
technical tool we derive various commutator estimates involving spectral
projections of Dirac operators with external fields.  
Our results apply to all nuclear charges $Z\klg 137$.

\smallskip

\end{abstract}

\section{Introduction}

The relativistic dynamics of a single electron moving in the potential
of a static nucleus, $\VC\klg0$, in the presence of an external
classical magnetic field $B=\rot\,A$ is
generated by the Dirac operator\footnote{
Energies are measured in units of $mc^2$,
$m$ denoting the rest mass of an electron and
$c$ the speed of light.
Length are measured in units of $\hslash/(mc)$,  
which is the Compton wave length divided by $2\pi$.
$\hslash$ is Planck's constant divided by $2\pi$.
In these units the square of the elementary charge
equals the fine structure constant, $e^2\approx1/137.037$.
}
\begin{equation}\label{intro0}
\DAVC\,:=\,\alpha\cdot(-i\nabla+A)\,+\,\VC\,.
\end{equation}
Here an electron is a state lying in the positive spectral subspace
of $\DAVC$. A ground state
of the one-electron atom modeled by $\DAVC$
can be characterized as
an energy minimizing bound state of the restriction of
$\DAVC$ to its positive spectral subspace, $\PAVC\,L^2(\RR^3,\CC^4)$, where
\begin{equation}\label{intro01}
\PAVC\,=\,\frac{1}{2}\,+\,\frac{1}{2}\,\sgn(\DAVC)\,.
\end{equation}
This is confirmed
by Dirac's interpretation of
the negative spectral subspace as a completely filled sea
of virtual electrons which, on account of Pauli's exclusion
principle, forces an additional electron
to attain a state of positive energy.
On the other hand it is well-known that there is no canonical
\`{a}-priori given atomic Hamiltonian
generating the relativistic time evolution
of $N>1$ interacting
electrons in the potential landscape of $\VC$
under the influence of $B$.
Guided by non-relativistic quantum mechanics
one might naively propose to start with
the formal expression
\begin{equation}
  \label{intro1}
  \,\sum_{j=1}^N \,\DAVC^{(j)}+\sum_{1\klg
j<k \klg N} W_{jk}\,,
\end{equation} 
where the superscript $(j)$ means that the operator below it
acts on the $j^{\mathrm{th}}$ electron and $W_{jk}\grg0$
is the interaction potential between the $j^{\mathrm{th}}$
and $k^{\mathrm{th}}$ electron.
It then turns out, however, that \eqref{intro1} suffers from
the phenomenon of continuum dissolution, that is,
the eigenvalue problem associated
to \eqref{intro1} has no normalizable solutions \cite{Su2}. 
A frequently used ansatz to find a reasonable and semi-bounded
Hamiltonian for an $N$-electron atomic system again
incorporates the concept of a Dirac sea.
Namely, one projects \eqref{intro1}
onto the $N$-fold antisymmetric tensor product
of a suitable one-electron subspace, i.e.,
one considers operators of the form
\begin{equation}\label{intro2}
 H_N \,:=\, \PAVN\, \,\Big(\sum_{j=1}^N \,\DAVC^{(j)}+\sum_{1\klg
j<k \klg N} W_{j\,k}\Big)\, \,\PAVN\,,
\end{equation}
where
\begin{equation}\label{intro2.1}
\PAVN\,:=\,\bigotimes_{j=1}^N \,\PAVj\,.
\end{equation}
Here $\PAV$ is defined as in \eqref{intro0}\&\eqref{intro01} but with $\VC$
replaced by a new potential $V$.
We emphasize that $\HH_N$ can formally
be derived from quantum electrodynamics 
by a procedure that neglects the creation and annihilation 
of electron-positron pairs \cite{Su1},
the latter being defined with respect to $\PAV$.  
Therefore, operators of the form \eqref{intro2} 
are often called no-pair Hamiltonians.
Obviously, the question arises how to choose the new potential $V$
determining the projection \eqref{intro2.1}
or, in other words, how to fix the Dirac sea for one electron
in the presence of the others in a physically efficient way.
Various possibilities are discussed 
from a physical and numerical point of view in \cite{Su1,Su2}.
The choice $V=0$ is refered to as the free picture, or
Brown-Ravenhall
model \cite{BrRa}. ($H_1$ is also sometimes called the
Bethe-Salpeter operator in this case \cite{BeSa}.) 
It has by now been studied in
many mathematical works
 \cite{BaMa,BaEv1,BaEv2,BaEv4,CaSi,EPS,GrTix,GrSi,HoeSi,
J-A1,J-A2,LSS,Mo1,Mo,MoVu,Tix1,Tix2,Tix3}. 
This is due to the fact that
the free projection, $\Pp{0,0}$, can be
calculated explicitely both in momentum and position space
\cite{BaMa,MoVu}.
The stability of the second kind in the free picture
with quantized electro-magnetic field (also included in the projection)
is treated in \cite{LL}.
The free picture is considered as one extreme case
in a family of intermediate pictures.
The opposite extreme case, called the Furry picture, 
is given by substituting the
(negative) Coulomb potential $\VC$ for $V$.
Other members of that family are obtained
by choosing $V$ to be equal to $\VC$ plus some
additional positive and in general non-local operator.
The additional non-local term may incorporate the interaction
with the remaining electrons and also take
vacuum polarization effects into account; compare
\cite{HLSS} where similar projections appear in a somewhat
different setting. 
We remark that the Furry and intermediate pictures give better
numerical results in comparison to the free picture
\cite{Su1,Su2}.

In this paper we do not aim to contribute to the subtle question
of optimizing the choice of $V$.
Rather we consider a general class of potentials 
which can be written as $V=\VC+\VM+\VE$, where 
$\VC$ may have several Coulomb-type singularities,
$\VM$
is a bounded potential function vanishing at infinity, and $\VE$
is a compact non-local operator that behaves nicely under
conjugation with exponential weights. Our goal then is
to establish some basic qualitative spectral properties
of $\HH_N$.  
First, we show that $H_N$ is well-defined 
on a natural dense subspace (which is not obvious)
and, thus, has a self-adjoint Friedrichs extension. 
We further locate its essential spectrum 
and, assuming that the number of electrons does not exceed
the sum of all nuclear charges, we 
prove that there exist infinitely many eigenvalues below the 
ionization threshold.
Moreover, we show that
the corresponding eigenfunctions are
exponentially localized. 
Our results apply to all nuclear charges $Z\klg137$.
The `easy part' of the HVZ theorem, i.e.,
the upper bound on the ionization threshold
holds for a certain class of
possibly unbounded magnetic fields. 
The `hard part', i.e.,
the lower bound on the ionization threshold
as well as our results on existence of eigenvectors
and their exponential localization are derived 
for bounded magnetic fields.

As an essential and novel technical input, necessary to obtain any of the 
results mentioned above, we derive various 
commutator estimates involving the spectral
projection $\PAV$, exponential weights, and cut-off functions.  
They describe the non-local properties of $\PAV$ in an $L^2$-sense
and might be of independent interest.

We remark that our commutator estimates would also allow to analyze
the Hamitonian $\HH_N$ in the free picture proceeding along the
discussion 
of the Furry and intermediate pictures presented here.
There would, however, appear a subtlety right in the beginning
of the analysis: Namely, for vanishing magnetic fields, it is known  
that the one-particle Brown-Ravenhall operator is stable
if and only if 
$Z\klg Z_c:=\frac{1}{e^2}\frac{2}{(2/\pi)+(\pi/2)}\approx124.2$ 
\cite{EPS}. 
For non-zero magnetic fields, such a critical value for the nuclear charge
is not known in general \cite{LSS}
and one had to study this question first or to
introduce the existence of a critical nuclear charge 
for $\HH_1$ as a hypothesis.
What is known is
that in the presence of an exterior $B$-field
one must include the vector potential in the projection for otherwise
the model is always unstable if $N>1$
\cite{GrTix}. (More precisely, the sum of the quadratic form and the
field energy is unbounded from below if 
the magnetic field is also
considered as a variable in the minimization.)

Finally, we comment on some closely related recent work.
In the free picture and for vanishing exterior magnetic field,
an HVZ theorem and the existence of infinitely many eigenvalues 
below the essential spectrum
have been proved in
\cite{MoVu}, for nuclear charges $Z\klg Z_c$. 
The case $N=2$ is also treated in \cite{J-A1}.
A more general HVZ theorem that applies to different particle
species and a wider class of interaction potentials and exterior fields
in the free picture has been established very recently
in \cite{Mo1}. Moreover, the reduction to irreducible representations
of the groups of rotation and reflection and 
permutations of identical particles is considered in \cite{Mo1}.
The exponential localization of eigenvectors
in the Brown-Ravenhall model is studied 
in \cite{Mo} improving and generalizing earlier
results from \cite{BaMa}.
In all these works the authors
employ explicite position space representations of $\Pp{0,0}$.
An HVZ theorem in the free picture with constant magnetic field
is established in
\cite{J-A2} again using explicite representations for the projection
based on Mehler's formula.
By employing
somewhat more abstract arguments we are able to study
commutator estimates for 
a wider class of projections in this paper.

\bigskip

\noindent
{\it The article is organized as follows.} 
In 
Section~\ref{sec-model} we introduce our mathematical model
precisely and state our main theorems.
Section~\ref{sec-commutators} is devoted to the study of some non-local
properties of $\PAV$
expressed in terms of various commutator estimates which form the basis
of the spectral analysis of $\HH_N$.
In Section~\ref{sec-exp-loc} we derive the exponential
localization estimate for $\HH_N$ and in
Section~\ref{ssec-lower-bound} infer a lower bound on the threshold
energy.
Section~\ref{sec-Weyl} deals with Weyl sequences 
and, finally, in Section~\ref{sec-eigenvalues} we
show that $\HH_N$ possesses infinitely many eigenvalues
below the threshold energy.
The main text is followed by an appendix containing
the proofs of some results of Section~\ref{sec-commutators}. 

\bigskip

\noindent
{\it Some frequently used notation.}
Open balls in $\RR^3$ with radius $R>0$ and center $z\in\RR^3$
are denoted by $\ball{R}{z}$. Spectral projections of a self-adjoint
operator, $T$, on some Hilbert space are denoted by $E_\lambda(T)$
and $E_I(T)$, if $\lambda\in\RR$ and $I$ is an interval.
$\dom(T)$ denotes the domain of the operator $T$ and
$\cQ(T)$ its form domain.
The characteristic function of a subset $\cM\subset\RR^n$
is denoted by $\id_\cM$.
$C,C',C'',\ldots$ denote constants whose values might change
from one estimate to another.


\section{The model and main results}
\label{sec-model}

In our choice of units
the free Dirac operator reads
$$
\D{0}\,:=\,-i\,\alpha\cdot\nabla+\beta
\,:=\,-i\sum_{j=1}^3\alpha_j\,\partial_{x_j}\:+\,\beta
\,.
$$
Here $\alpha=(\alpha_1,\alpha_2,\alpha_3)$
and $\beta=:\alpha_0$ are $4\times4$
hermitian matrices which satisfy the Clifford
algebra relations
\begin{equation}\label{Clifford}
\{\alpha_i\,,\,\alpha_j\}\,=\,2\,\delta_{ij}\,\id\,,
\qquad 0\klg i,j\klg3\,.
\end{equation}
In Dirac's representation, which we fix throughout the paper,
they are given as
$$
\alpha_j\,=\,\left(
\begin{array}{lr}
0&\sigma_j\\\sigma_j&0
\end{array}
\right),\quad j=1,2,3,\qquad
\beta\,=\,\left(
\begin{array}{lr}
\id&0\\0&-\id
\end{array}
\right)\,,
$$
where 
$\sigma_1,\sigma_2,\sigma_3$ 
are the 
standard 
Pauli matrices. $\D{0}$ is a self-adjoint
operator in the Hilbert space
$$
\HR\,:=\,L^2(\RR^3,\CC^4)
$$
with domain $H^1(\RR^3,\CC^4)$. Its spectrum is purely absolutely
continuous and given by the union of two halflines,
\begin{equation}\label{spec-D0}
\spec(\D{0})\,=\,
\specac(\D{0})\,=\,(-\infty,-1]\cup[1,\infty)\,.
\end{equation}
Next, we formulate our precise hypotheses on
the exterior electrostatic potential $\VC$
and on the potential $V$ determining the Dirac sea.
We think that,
at least with regards to the commutator estimates
in Section~\ref{sec-commutators}, it is interesting to keep
the conditions on $\VC$ and $V$ fairly general.

\begin{hypothesis}\label{hyp-Vogel}
There are a finite set $\cY\subset\RR^3$, $\#\cY<\infty$,
such that
$
\VC\in L_\loc^\infty(\RR^3\setminus\cY,\LO(\CC^4))
$ is almost everywhere hermitian 
and 
\begin{equation}\label{VCtozero}
\|\VC(x)\|\,\longrightarrow\,0\,,\qquad
|x|\to\infty\,.
\end{equation}
Moreover,
there exist
$\gamma\in(0,1)$
and $\ve>0$ such that the balls $\ball{\ve}{y}$,
$y\in\cY$, are mutually disjoint and, for $0<|x|<\ve$
and $y\in\cY$,
\begin{equation}\label{eq-Vogel1}
\|\VC(y+x)\|\,\klg\,\frac{\gamma}{|x|}\,.
\end{equation}
\end{hypothesis}

\begin{example}\label{ex-Coulomb}
{\em
The main example for a potential satisfying
Hypothesis~\ref{hyp-Vogel}
is certainly the Coulomb potential
generated by a finite number of static nuclei,
$$
\VC(x)\,=\,-\sum_{y\in\cY}\frac{e^2\,Z_y}{|x-y|}\:\id\,,
\qquad x\in\RR^3\setminus\cY\,.
$$
In this case
the restriction on the strength of the singularities of $\VC$
imposed in Hypothesis~\ref{hyp-Vogel} allows for all
nuclear charges, 
$Z_y\in\NN$, $Z_y\klg137$, for $y\in\cY$.
\hfill$\Box$
}
\end{example}

\begin{hypothesis}\label{hyp-V-komplett}
$V=\VC+\VM+\VE$, where $\VC$ fulfills Hypothesis~\ref{hyp-Vogel}
and 
$\VM\in L^\infty_\loc(\RR^3,\LO(\CC^4))$
is an almost everywhere hermitian matrix-valued function
dropping off to zero at infinity,
\begin{equation}\label{VMtozero}
\|\VM(x)\|\,\longrightarrow\,0\,,\qquad |x|\to\infty\,.
\end{equation}
$\VE$ is compact and has the following property:
There exist $m>0$ 
and some increasing function $c:[0,m)\to(0,\infty)$
such that,
for every 
$F\in C^1(\RR^3,\RR)$ with $|\nabla F|\klg a<m$,
\begin{eqnarray}
\forall\;\chi\in C^1(\RR^3,[0,1])\::\quad
\big\|\,\big[\,e^F\,\VE\,e^{-F}\,,\,\chi\,\big]\,\big\|&\klg&\label{bd-VE1}
c(a)\,\|\nabla\chi\|_\infty\,,
\\
\big\|\,[\VE\,,\,e^F]\,e^{-F}\,\big\|&\klg&\label{bd-VE2}
c(a)\,\|\nabla F\|_\infty\,,
\\
\lim_{R\to\infty}
\big\|\,
\id_{\RR^3\setminus\ball{R}{0}}\,e^F\,\VE\,e^{-F}\,
\id_{\RR^3\setminus\ball{R}{0}}
\,\big\|&=&\label{bd-VE3}
0
\,.
\end{eqnarray}
\end{hypothesis}

\begin{example}\label{ex-VH-VE}
{\em
(i) Possible choices for $\VM$ and $\VE$ satisfying the conditions of 
Hypothesis~\ref{hyp-V-komplett} are the Hartee and
non-local exchange potentials corresponding to a set of
exponentially localized orbitals
$\vp_1,\ldots,\vp_M\in\HR$, $|\vp_i(x)|\klg C\,e^{-m|x|}$, $1\klg i\klg M$,
for some $C\in(0,\infty)$. Their
Hartee potential is given as
$$
\VM(x)\,:=\,e^2\sum_{i=1}^M\Big(|\vp_i|^2*\frac{1}{|\cdot|}\Big)(x)\,,
\qquad x\in\RR^3\,.
$$
It incorporates the presence of $M$ electrons in a fixed state
into the Dirac sea by a
smeared out background density.
The exchange potential corresponding to $\vp_1,\ldots,\vp_M$ is
the integral operator with matrix-valued kernel
$$
\VE(x,y)\,:=\,e^2\sum_{i=1}^M\frac{\vp_i(x)\,\vp_i^*(y)}{|x-y|}\,.
$$
It is a correction to the Hatree potential accounting for
the Pauli principle. In the sense of quadratic forms it then holds
$\VC\klg V=\VC+\VM+\VE\klg0$, which justifies the notion
`intermediate picture'.

\smallskip

\noindent
(ii) More generally, we may set $\VM:=\vr*|\cdot|^{-1}$, for
some $0\klg \vr\in L^1\cap L^{5/3}(\RR^3)$.
In this case we find some $C\in(0,\infty)$ such that
$0\klg\VM\klg C/|\cdot|$. 
Moreover, standard theorems on integral operators show that every
kernel with values in the set of hermitian $(4\times4)$-matrices
satisfying
$$
\|\VE(x,y)\|\,\klg\,C'\,
\frac{e^{-m|x-y|}}{\SL x\SR^\rho|x-y|\SL y\SR^\rho}\,,
$$
for some $m,\rho,C'>0$,
yields a compact operator satisfying the conditions of
Hypothesis~\ref{hyp-V-komplett}.
\hfill$\Box$
}
\end{example}

\bigskip

As a first consequence of Hypothesis~\ref{hyp-V-komplett} we find,
for every locally bounded vector potential 
$A:\RR^3\to\RR^3$,
a distinguished
self-adjoint realization of the
Dirac operator
$$
\DAV\,=\,\alpha\cdot(-i\nabla+A)\,+\,\beta\,+\,V\,,
$$
whose essential spectrum is again contained
in $(-\infty,-1]\cup[1,\infty)$; see Lemma~\ref{le-BdMetc} below, 
where we recall some important
well-known facts on Dirac operators with singular potentials. 
Therefore, it makes sense to define the spectral projections,
\begin{equation}\label{def-PAV}
\PAV\,:=\,E_{[e_0,\infty)}(\DAV)\,,\quad
\PAVm\,:=\id-\PAV\,,
\end{equation} 
where
\begin{equation}\label{def-e0}
e_0\in\vr(\DAV)\cap(-1,1)\,.
\end{equation}
For later reference we introduce the parameter
\begin{equation}
\label{def-triangle0}
\triangle_0\,:=\,\min\big\{\,1-e_0\,,\,e_0+1\,\big\}\,.
\end{equation}
Many of our technical results
on $\DAV$, for instance, various commutator
estimates of Section~\ref{sec-commutators}
hold actually true under the mere assumption that the
components of $A$ are locally bounded.
Of course, if not
all eigenvalues
of $\DAV$ are larger than $-1$
and $e_0$ is chosen between
$-1$ and the lowest eigenvalue, the physical relevance of
the $N$-particle Hamiltonian $\HH_N$ 
becomes rather questionable.
We remark that such situations are not excluded by our hypotheses.
For instance, if $\VC$ is the Coulomb potential  
and the intensity of a constant exterior magnetic field is
increased, then the lowest eigenvalue of $\DAVC$ eventually
delves into the lower continuum \cite{DEL}. 
Nevertheless, our theorems hold for
any choice of $e_0$ as in \eqref{def-e0}.

In order to define the atomic no-pair
Hamiltonian precisely we first set
$$
\HR_N\,:=\,\bigotimes_{i=1}^N\HR\,,\quad
\HR_N^+\,:=\,\PAVN\,\HR_N\,,\quad N\in\NN\,,\quad
\quad \HR^+\,:=\,\HR_1^+\,,
$$
where $\PAVN$ is given by \eqref{intro2.1} and \eqref{def-PAV}.
We let $W:\RR^3\times\RR^3\to[0,\infty]$ denote the
interaction potential between two electrons. 

\begin{hypothesis}\label{hyp-W}
There is some $\wt{\gamma}>0$ such that,
for all $x,y\in\RR^3$, $x\not=y$,
\begin{equation}\label{eq-hyp-W}
0\,\klg\,W(x,y)\,=\,W(y,x)\,\klg\,\wt{\gamma}\,|x-y|^{-1}\,.
\end{equation}
\end{hypothesis}

When we consider $N$ electrons located at $x_1,\ldots,x_N\in\RR^3$
we denote their common position variable by $X=(x_1,\ldots,x_N)$.
Furthermore, we often write $W_{jk}$ for the 
maximal multiplication operator
in $\HR_N$ induced by the function 
$(\RR^{3})^N\ni X\mapsto W(x_j,x_k)$.
For $N\in\NN$,
we introduce a symmetric, semi-bounded operator
acting in $\HR_N^+$
by
\begin{eqnarray}\label{def-core}
\dom(\mr{\HH}_N)&:=&\PAVN\,\core\,,\quad
\core\,:=\,\bigotimes_{i=1}^N\COI\,,\quad 
\COI\,:=\,C_0^\infty(\RR^3,\CC^4)\,,
\\
\mr{\HH}_N\,\Phi&:=&\label{def-Hcirc}
\PAVN\,\Big(\sum_{i=1}^N\D{A,\VC}^{(i)}+\!\!\!\sum_{1\klg i<j\klg N}\!
W_{ij}\,\Big)\PAVN\,\Phi\,,\quad \Phi\in\dom(\mr{\HH}_N)\,.
\end{eqnarray}

\begin{proposition}\label{prop-Friedrichs}
Assume that $V$ fulfills Hypothesis~\ref{hyp-V-komplett},
$W$ fulfills Hypothesis~\ref{hyp-W}, and that 
$A\in L^\infty_\loc(\RR^3,\RR^3)$
satisfies $\|e^{-\tau_0|x|}\,A\|_\infty<\infty$,
for some $0\klg\tau_0<\min\{\triangle_0,m\}$.
Then the operator $\mr{\HH}_N$ given by \eqref{def-core}
and \eqref{def-Hcirc} is well-defined, symmetric, and
semi-bounded from below. 
\end{proposition}

\proof
The only claim that is not obvious is that $W_{ij}\,\PAVN\,\phi$
is again square-integrable,
for every $\phi\in\core$.
This follows, however, from Corollary~\ref{cor-core}.
\qed

\bigskip

We denote the Friedrichs extension of $\mr{\HH}_N$
by $\HH_N$.
Note that
we do not require the elements in the domain of $\HH_N$
to be anti-symmetric
since in our proofs it is convenient to consider
$\HH_N$ as an operator acting in the full tensor product.
Of course, in the end we shall be interested in the restriction
of $\HH_N$ to the anti-symmetric (fermionic) subspace of
$\HR_N^+$. We denote the anti-symmetrization operator on $\HR_N$
by $\cA_N$,
\begin{equation}\label{def-AN}
(\cA_N\,\Phi)(X)\,=\,\frac{1}{N!}\sum_{\sigma\in\Perm_N}
\sgn(\sigma)\,\Phi(x_{\sigma(1)},\dots,x_{\sigma(N)})\,,
\qquad \Phi\in\HR_N\,,
\end{equation}
where $\Perm_N$ is the group of permutations of $\{1,\dots,N\}$,
and define the no-pair Hamiltonian by
$$
\HH_N^\cA\,:=\,\HH_N\!\!\upharpoonright_{\cA_N\HR_N^+}\,.
$$
Our first main result is the following theorem, where
\begin{equation}\label{def-EA}
\Th_N\,:=\,\inf\spec(\HH_N^\cA)\,,\quad N\in\NN\,,
\qquad \Th_0\,:=\,0\,.
\end{equation}

\begin{theorem}[Exponential localization]\label{main-thm-exp}
Assume that $V$ and $W$ fulfill
Hypotheses~\ref{hyp-V-komplett}
and~\ref{hyp-W}, respectively.
If $A\in C^1(\RR^3,\RR^3)$ 
and $B=\rot A$ is bounded and if $I\subset\RR$ is an interval
with $\sup I<\Th_{N-1}+1$, then there exists $b\in(0,\infty)$
such that $\Ran\big(E_I(\HH_N^\cA)\big)\subset\dom(e^{b|X|})$ and
$$
\big\|\,e^{b|X|}\,E_I(\HH_N^\cA)\,\big\|\,<\,\infty\,.
$$
\end{theorem}

\proof
This theorem is proved in Section~\ref{sec-exp-loc}.
\qed

\begin{remark}\label{rem-exp}
{\em
In the case $N=1$ the assertion of Theorem~\ref{main-thm-exp}
still holds true under the assumptions of 
Proposition~\ref{prop-Friedrichs}. This follows from the proof
of Theorem~\ref{main-thm-exp}. In fact, for $N=1$, we do not have
to control error terms involving the interaction $W$ which is
the only reason why $B$ is assumed to be bounded in 
Theorem~\ref{main-thm-exp}. If also $V=\VC$, then we obtain an
exponential localization estimate for an eigenvector, $\phi_E$, with
eigenvalue $E\in(-1,1)$ of the
Dirac operator $\DAVC$. The estimate on the decay rate which could
be extracted from our proof is, however, suboptimal due to error
terms coming from the projections;
see also \cite{BeGe} for decay estimates for Dirac operators.
\hfill$\Box$
}
\end{remark}

\bigskip

Next, we introduce a hypothesis which is used to prove the
`easy part' of the HVZ theorem below.

\begin{hypothesis}\label{hyp-Weyl}
(i) For every $\lambda\grg1$,
there exist radii, $1\klg R_1<R_2<\ldots$, $R_n\nearrow\infty$,
and $\psi_1,\psi_2,\ldots\in\COI$
such that 
\begin{equation}\label{Weyl-DA}
\|\psi_n\|=1\,,\quad
\supp(\psi_n)\subset\RR^3\setminus\ball{R_n}{0}\,,
\quad\lim_{n\to\infty}
\big\|\,(\DA-\lambda)\,\psi_n\,\big\|=0\,.
\end{equation}
(ii) $A\in C^1(\RR^3,\RR^3)$ and $B=\rot A$ has the following
property:
There are $b_1\in(0,\infty)$
and $0\klg\tau<\min\{\triangle_0,m\}$ ($m$ and $\triangle_0$
are the parameters
appearing in Hypothesis~\ref{hyp-V-komplett} and \eqref{def-triangle0})
such that, for all $x,y\in\RR^3$,
\begin{equation}\label{var-bd-B}
\big|\,B(x)-B(y)\,\big|\,\klg\,b_1\,e^{\tau|x-y|}\,.
\end{equation}
\end{hypothesis}

\begin{example}[\cite{HNW}]\label{ex1-HNW}
{\em
We recall a result from \cite{HNW} which provides a large class
of examples where
Hypothesis~\ref{hyp-Weyl}(i) is fulfilled:
Suppose that $A\in C^\infty(\RR^3,\RR^3)$, $B=\rot A$,
and set, for $x\in\RR^3$ and $\nu\in\NN$,
\begin{equation*}\label{def-epsnu}
\epsilon_0(x)\,:=\,|B(x)|\,,\qquad
\epsilon_\nu(x)\,:=\,
\frac{\sum_{|\alpha|=\nu} |\partial^\alpha B(x)|}{
1+\sum_{|\alpha|<\nu} |\partial^\alpha B(x)|}\,.
\end{equation*}
Suppose further that
there exist $\nu\in\NN_0$, $z_1,z_2,\ldots\in\RR^3$, and 
$\rho_1,\rho_2,\ldots>0$ such that $\rho_n\nearrow\infty$,
the balls $\ball{\rho_n}{z_n}$, $n\in\NN$, are mutually
disjoint and
\begin{equation*}\label{hyp-Weyl-eq2}
\sup\big\{\,\epsilon_\nu(x)\,\big|\:x\in\ball{\rho_n}{z_n}\,\big\}
\,\longrightarrow\,0\,,\qquad n\to\infty\,.
\end{equation*}
Then there is a Weyl sequence, $\psi_1,\psi_2,\ldots$, that satisfies
the conditions of Hypothesis~\ref{hyp-Weyl}(i).
\hfill$\Box$
}
\end{example}

\bigskip

The fact that the additional assumption of Part~(ii)
of the next theorem yields a lower bound on the
ionization threshold is an observation made in
\cite{Gr} for Schr\"odinger operators.

\begin{theorem}[HVZ]\label{main-thm-HVZ}
Assume that $V$ and $W$ fulfill
Hypotheses~\ref{hyp-V-komplett}
and~\ref{hyp-W}, respectively. Then the following assertions
hold true:
\begin{enumerate}
\item[(i)] If Hypothesis~\ref{hyp-Weyl} is fulfilled also, then
$\specess(\HH_N^\cA)\,\supset\,[\,\Th_{N-1}+1\,,\,\infty)$.
\item[(ii)]  Assume additionally
that, for every
interval $I\subset\RR$ with $\sup I<\Th_{N-1}+1$,
there is some $g\in C(\RR,(0,\infty))$
such that $g(r)\to\infty$, as $r\to\infty$, and
$g(|X|)\,E_I(\HH_N^\cA)\in\LO(\cA_N\HR_N)$.
Then 
$\specess(\HH_N^\cA)\,\subset\,[\,\Th_{N-1}+1\,,\,\infty)$.
In particular, this inclusion is valid
if $A\in C^1(\RR^3,\RR^3)$
and $B=\rot A$ is bounded.
\end{enumerate} 
\end{theorem}

\noindent{\it Proof:}
(i) follows directly from Lemma~\ref{le-Weyl-HN} and
(ii) follows from Theorem~\ref{thm-EI-compact}
and Theorem~\ref{main-thm-exp}.
\qed

\bigskip

To show the existence of infinitely many eigenvalues below the
bottom of the essential spectrum of $\HH_N^\cA$
we certainly need a condition
on the relationship between $\VC$, $W$, and the magnetic field.
To formulate it we set, for $\delta,R>0$,
\begin{eqnarray}
S_\delta(R)&:=&\label{def-Sdelta}\big\{\,
x\in\RR^3\,:\;(1-\delta)\,R\klg|x|\klg(1+\delta)\,R\,\big\}
\\
v_\star(\delta,R)&:=&\label{def-vstar}
\sup_{x\in S_\delta(R)}\;\sup_{|v|=1}\SPb{v}{\VC(x)\,v}_{\CC^4}\,.
\end{eqnarray}

\begin{hypothesis}\label{hyp-ex}
(i) $V$ fulfills Hypothesis~\ref{hyp-V-komplett}.

\smallskip

\noindent
(ii) $A\in C^1(\RR^3,\RR^3)$ and $B=\rot A$ is bounded.

\smallskip

\noindent
(iii) 
There exist radii $1\klg R_1<R_2<\ldots$, $R_n\nearrow\infty$,
some constant $\delta\in(0,1/N)$, 
and a sequence of spinors,
$\psi_1,\psi_2,\ldots\in\COI$, with
vanishing lower spinor components,
$\psi_n=(\psi_{n,1},\psi_{n,2},0,0)^\top$, $n\in\NN$,
such that
\begin{equation}\label{supp-psin-ex}
\|\psi_n\|=1\,,\quad
\supp(\psi_n)\,\subset\,\big\{\,R_n<|x|<(1+\delta/2)\,R_n\,\big\}\,,
\qquad 2R_n\,\klg\,R_{n+1}\,,
\end{equation}
for all $n\in\NN$, and 
\begin{equation}\label{Weyl-DA-ex}
\big\|\,\big(\DA-1\big)\,\psi_n\,\big\|\,=\,\bigO(1/R_n)\,,
\qquad n\to\infty\,.
\end{equation}
(iv) $W$ fulfills Hypothesis~\ref{hyp-W} and,
for every $\delta\in(0,\frac{1}{N})$, we find some $\ve\in(0,1)$ 
such that
$$
\limsup_{n\to\infty}R_n\,\Big(v_\star\big(\delta,R_n\big)
\,+\,
(N-1)\!\sup_{|x-y|\grg(1-\ve)R_n}W(x,y)
\Big)\,<\,0\,.
$$
\end{hypothesis}

\begin{example}{\em
$V=\VC+\VM+\VE$ and $W$ fulfill
Hypothesis~\ref{hyp-ex}(i) and~(iv), if $\VC$ is given as in 
Example~\ref{ex-Coulomb} with $\sum_{y\in\cY}Z_y\grg N$,
$\VM$ and $\VE$ are given as in Example~\ref{ex-VH-VE}(i) or~(ii),
and $W$ is the Coulomb repulsion,
$W(x,y)=e^2/|x-y|$.

Hypothesis~\ref{hyp-ex}(iii) is fulfilled under the following strengthened
version of the condition given in \cite{HNW}:
Suppose again that
$A\in C^\infty(\RR^3,\RR^3)$,
$B=\rot A$, and let $\ball{\rho_n}{z_n}$ denote the balls appearing
in Example~\ref{ex1-HNW}. Suppose additionally that there
is some $C\in(0,\infty)$ such that
$\rho_n<|z_n|\klg C\,\rho_n$, for all $n\in\NN$,
and that either
$$
\sup\big\{\,|B(x)|\,:\;x\in\ball{\rho_n}{z_n}\,\big\}\,\klg\,
C/|z_n|^2\,,\qquad n\in\NN\,,
$$
or
\begin{equation*}\label{hyp-HNW-ex}
\forall\;n\in\NN\::\;\;|B(z_n)|\grg1/C\quad
\textrm{and}\quad
\sup\big\{\,\epsilon_\nu(x)\,\big|\:x\in\ball{\rho_n}{z_n}\,\big\}
\,=\,o(\rho_n^{-\nu})\,.
\end{equation*}
Then we find a Weyl sequence $\psi_1,\psi_2,\ldots$
satisfying the conditions in Hypothesis~\ref{hyp-ex}(iii).
This follows by inspecting and adapting all relevant proofs in \cite{HNW}.
We leave this procedure to the reader since it is straightforward
but a little bit lengthy.
\hfill$\Box$
}
\end{example}

\begin{theorem}[Existence of bound states]\label{main-thm-ev}
Assume that $V$, $W$, and $A$ fulfill Hypothesis~\ref{hyp-ex}.
Then $\HH_N^\cA$ has infinitely
many eigenvalues below the infimum of its essential spectrum,
$\inf\specess(\HH_N^\cA)=\Th_{N-1}+1$.
\end{theorem}

\noindent{\it Proof:} This theorem 
is proved in Section~\ref{sec-eigenvalues}.
\qed


\section{Spectral projections of the Dirac operator}
\label{sec-commutators}

In this section we study spectral projections of Dirac operators
with singular potentials in magnetic fields.
We start by recalling some basic well-known 
facts about Dirac operators in Subsection~\ref{ssec-prop-DAV}.
A crucial role is played by the resolvent identity stated
in that subsection which applies to Coulomb singularities
corresponding to nuclear charges up to $Z\klg137$.
In Subsection~\ref{ssec-conj} we derive some norm estimates
on resolvents of Dirac operators which are conjugated
with exponential weight functions. We 
verify that
the decay rate is not smaller than the distance of the 
real part of the
spectral parameter to the essential spectrum.
The simple Neumann-type argument we employ to prove this
for non-vanishing electrostatic potentials
might be new.
In
Subsection~\ref{ssec-commutators}
we derive the main technical tools of this paper, namely,
various commutator estimates involving spectral projections
of singular Dirac operators. Finally,
in Subsection~\ref{ssec-PA-PAV} we study the difference
of projections with and without electrostatic potentials.


\subsection{Basic properties of Dirac operators with singular
potentials in magnetic fields}\label{ssec-prop-DAV}

In the next lemma we collect various well-known results
on Dirac operators which play an important role
in the whole paper.
To this end we let 
$H_c^s:=H_c^{s}(\RR^3,\CC^4)$ denote all elements of
$H^s:=H^s(\RR^3,\CC^4)$, $s\in\RR$, having compact support.
Moreover, we denote the canonical extension of $\D{0}$ 
to an element of $\LO(H^{1/2},H^{-1/2})$ by $\Dt$.
It shall sometimes be convenient to consider the singular part
of $\VC$,
\begin{equation}\label{def-wtVC}
\VCt(x)\,:=\,\sum_{y\in\cY}\vr(x-y)\,\VC(x)\,,
\qquad x\in\RR^3\,,
\end{equation}
where $\vr\in C_0^\infty(\RR^3,[0,1])$ equals $1$ on $\ball{\ve/2}{0}$
and $0$ outside $\ball{\ve}{0}$.
Here $\ve$ is the parameter appearing in Hypothesis~\ref{hyp-Vogel}.
We let $\VCt(x)=S(x)\,|\VCt|(x)$
denote the polar decomposition of $\VCt(x)$.
By Hardy's inequality we know that $\VCt$ 
is a bounded operator
from $H^1(\RR^3,\CC^4)$ to $L^2(\RR^3,\CC^4)$. 
By duality and interpolation
it possesses a unique extension 
$\VCtt\in\LO(H^{1/2},H^{-1/2})$.
Given some $A\in L_\loc^\infty(\RR^3,\RR^3)$ we set
$
A^{\mathrm{s}}:=(1-\vt)\,A
$,
where $\vt\in C_0^\infty(\RR^3,[0,1])$ is equal to $1$
on some ball containing $\supp(\VCt)$.
We let $\alpha\cdot A^{\mathrm{s}}(x)=U(x)\,|\alpha\cdot A^{\mathrm{s}}(x)|$
denote
the polar decomposition of $\alpha\cdot A^{\mathrm{s}}(x)$
and note that the operator sum
$\Dt+\alpha\cdot A^{\mathrm{s}}+\VCtt$ 
is well-defined as an
element of $\LO(H^{1/2}_c,H^{-1/2}_c)$, for every
$A\in L_\loc^\infty(\RR^3,\RR^3)$.

\begin{lemma}[\cite{BdMP,Ne1,Ne2,RTdA}]\label{le-BdMetc-0}
Assume that $A\in L^\infty_\loc(\RR^3,\RR^3)$ and $\VC$
fulfills
Hypothesis~\ref{hyp-Vogel}.
Then there is unique self-adjoint operator,
$\D{A^{\mathrm{s}},\VCt}$, such that:
\begin{enumerate}
\item[(i)] 
$\dom\big(\D{A^{\mathrm{s}},\VCt}\big)\subset 
H_\loc^{1/2}(\RR^3,\CC^4)$.
\item[(ii)] For all 
$\psi\in H_c^{1/2}(\RR^3,\CC^4)$ and 
$\phi\in\dom\big(\D{A^{\mathrm{s}},\VCt}\big)$,
\begin{eqnarray*}
\lefteqn{
\SPn{\psi}{\D{A^{\mathrm{s}},\VCt}\,\phi}
\,=\,
\SPb{|\D{0}|^{1/2}\,\psi}{\sgn(\D{0})\,|\D{0}|^{1/2}\,\phi}
}
\\
& &+\,
\SPb{|\alpha\cdot A^{\mathrm{s}}|^{1/2}\,\psi}{U\,
|\alpha\cdot A^{\mathrm{s}}|^{1/2}\,\phi}
\,+\,\SPb{|\VCt|^{1/2}\,\psi}{S\,
|\VCt|^{1/2}\,\phi}
\,.
\end{eqnarray*}
\end{enumerate}
\end{lemma}

\noindent{\it Notes:}
In \cite[Proposition~4.3]{RTdA} it is observed that
the claim follows from \cite[Theorem~1.3]{BdMP}
and \cite{Ne1,Ne2}.
\qed

\bigskip

Consequently, we may define a self-adjoint operator,
\begin{equation}\label{def-DAV}
\DAV\,:=\,\D{A^{\mathrm{s}},\VCt}\,+\,\alpha\cdot(A-A^{\mathrm{s}})
\,+\,(\VC-\VCt)
\,+\,\VM\,+\,\VE
\end{equation}
on the domain 
$\dom(\DAV)
=\dom(\D{A^{\mathrm{s}},\VCt})$.
Notice that in \eqref{def-DAV}
we only add bounded operators to $\D{A^{\mathrm{s}},\VCt}$.
We state some of its properties in the following lemma
where
\begin{eqnarray}\label{lisa3}
\RAV{z}&:=&(\DAV-z)^{-1}\,,\qquad z\in\vr(\DAV)
\,.
\end{eqnarray}

\begin{lemma}[\cite{BdMP,Ne1,Ne2,RTdA}]\label{le-BdMetc}
Assume that $A\in L^\infty_\loc(\RR^3,\RR^3)$ and that $V$
fulfills
Hypothesis~\ref{hyp-V-komplett}.
Then the following assertions hold true:

\smallskip

\noindent(a)
$\id_{\ball{R}{0}}\,(\DAV-i)^{-1}$ 
is compact, for every $R>0$.

\smallskip

\noindent(b)
$\specess(\DAV)=\specess(\DA)$,
$\spec(\DA)\subset
(-\infty,-1]\cup[1,\infty)$.

\smallskip

\noindent(c)
$\DAV$ is essentially self-adjoint on 
\begin{equation}\label{def-De}
\dom_{e}\,
:=\,\big\{\,\phi\in H^{1/2}_c(\RR^3,\CC^4)\,:\:
\Dt\,\phi+\alpha\cdot A\,\phi+
\VCtt\,\phi\in L^2(\RR^3,\CC^4)
\,\big\}
\end{equation}
and, for $\phi\in\dom_e$, $\DAV\,\phi$ is given as a sum of four
vectors in $H^{-1/2}$, 
$$
\DAV\,\phi\,=\,
\Dt\,\phi\,+\,\alpha\cdot A\,\phi\,+\,\VCtt
\,\phi\,+\,(V-\VCt)\,\phi\,.
$$
Moreover, $\dom_{e}=\dom(\DAV)\cap\mathscr{E}'$, where $\mathscr{E}'$
denotes the dual space of $C^\infty(\RR^3,\CC^4)$.

\smallskip

\noindent
(d) For $\chi\in C_0^\infty(\RR^3)$
and $\phi\in\dom(\DAV)$, we have
$\chi\phi\in\dom_e\subset\dom(\DAV)$
and
$$
\big[\,\DAV\,,\,\chi\,\big]\,\phi\,=\,-i(\alpha\cdot\nabla\chi)\,\phi
\,+\,\big[\,\VE\,,\,\chi\,\big]\,\phi\,.
$$
In particular, for $z\in\vr(\DAV)$,
\begin{eqnarray}
\big[\,\chi\,,\,\RAV{z}\,\big]&=&\RAV{z}\,
\big[\,\DAV\,,\,\chi\,\big]\,
\RAV{z}\label{commutatorR}
\\
&=&\nonumber
\RAV{z}\,\big(-i(\alpha\cdot\nabla\chi)\,\phi
\,+\,\big[\,\VE\,,\,\chi\,\big]\big)\,\RAV{z}\,.
\end{eqnarray}
(e) If $A$ is bounded, then 
$\dom(\D{A,V})\subset H^{1/2}(\RR^3,\CC^4)$.
\end{lemma}

\noindent{\it Notes: }
Since $\VE$ is compact it is clear that all assertions hold
true as soon as they hold
for $\VE=0$, which we assume in the following.
To prove (a) we write
$$
\id_{\ball{R}{0}}\,(\DAV-i)^{-1}
\,=\,
\big(\id_{\ball{R}{0}}\,|\D{0}|^{-1/2}\big)\big(|\D{0}|^{1/2}\,\chi
\,(\DAV-i)^{-1}\big)\,,
$$
where $\chi\in C_0^\infty(\RR^3,[0,1])$ equals $1$ in a neighbourhood
of $\ball{R}{0}$. Then we use that 
$\id_{\ball{R}{0}}\,|\D{0}|^{-1/2}$ is compact
and that $|\D{0}|^{1/2}\,\chi\,(\DAV-i)^{-1}$ 
is bounded by Lemma~\ref{le-BdMetc-0} and the closed graph theorem.
By standard arguments we obtain
the identity $\specess(\DAV)=\specess(\DA)$
from (a) since $V$ drops off
to zero at infinity; see, e.g., \cite[\textsection4.3.4]{Th}.
The inclusion $\spec(\DA)\subset(-\infty,-1]\cup[1,\infty)$
follows from supersymmetry arguments; see, e.g., \cite[\textsection5.6]{Th}.
The assertions in~(c) follow from \cite[\textsection2]{BdMP},
(d) follows from \cite[Lemma~G]{BdMP}, and (e) from \cite{Ne1}.
\qed

\bigskip

Next, we recall the useful
resolvent identity \eqref{jim1adjoint} (see, e.g., \cite{GeMa,Xia})
which is used very often in the sequel.
The vector potential $\wt{A}$ in \eqref{jim1adjoint} 
could for instance be the gradient of some gauge potential
or just be equal to zero.
We recall another well-known resolvent identity \cite{Ne1}
in the beginning of Appendix~\ref{app-proofs}.

\begin{lemma}\label{le-jimbo}
Assume that $V$ fulfills Hypothesis~\ref{hyp-V-komplett},
and that $A,\wt{A}\in L_\loc^\infty(\RR^3,\RR^3)$.
Let $\wt{V}$ be either $\VCt$ (given by
\eqref{def-wtVC}) or $0$, let  
$z\in\vr(\D{\wt{A},\wt{V}})\cap\vr(\DAV)$ and 
$\chi\in C^\infty(\RR^3,\RR)$ 
be constant outside some ball in $\RR^3$,
and assume that $(\VC-\wt{V})\,\chi $ and $\alpha\cdot(A-\wt{A})\,\chi$
are bounded.
Then
\begin{eqnarray}
\chi\,\RAV{z}&=&\chi\,\R{\wt{A},\wt{V}}{z}
\nonumber\,+\,\R{\wt{A},\wt{V}}{z}
\,i\alpha\cdot(\nabla\chi)\,\big(\R{\wt{A},\wt{V}}{z}-\RAV{z}\big)
\\
& &\label{jim1adjoint}
\,-\,\R{\wt{A},\wt{V}}{z}\,
\chi\,\big(V-\wt{V}+\alpha\cdot(A-\wt{A})\big)\,\RAV{z}\,.
\end{eqnarray}
\end{lemma}

\proof
Let $\phi\in\wt{\dom}_e:=\{\psi\in H_c^{1/2}|\,\Dt\psi+\alpha\cdot\wt{A}+
\wt{V}\psi\in L^2\}$.
Since $\chi$ can be written as $\chi=c+\vt$, for some
$c\in\RR$ and $\vt\in C_0^\infty(\RR^3,\RR)$, Lemma~\ref{le-BdMetc}(c)\&(d)
imply
that $\chi\,\phi\in\wt{\dom}_e$.
By the definition of $\dom_e$ in \eqref{def-De}
and the assumptions on $\chi$
it further follows that $\chi\,\phi\in\dom_e\subset\dom(\DAV)$
and 
$$
\D{\wt{A},\wt{V}}\,\chi\,\phi\,=\,\DAV\,\chi\,\phi\,+\,
\big\{-V+\wt{V}-\alpha\cdot(A-\wt{A})\big\}\,\chi\,\phi\,.
$$
Therefore, we obtain
\begin{eqnarray*}
\lefteqn{
\big(\R{\wt{A},\wt{V}}{z}-\RAV{z}\big)\,\chi\,(\D{\wt{A},\wt{V}}-z)\,\phi
}
\\
&=&
\big(\R{\wt{A},\wt{V}}{z}-\RAV{z}\big)\,
\big(\,(\D{\wt{A},\wt{V}}-z)\,\chi
\,+\,i\alpha\cdot(\nabla\chi)\,\big)\,\phi
\\
&=&
\chi\,\phi\,-\,\RAV{z}\,\big(\DAV-z-V+\wt{V}-\alpha\cdot(A-\wt{A})\big)\,
\chi\,\phi
\\
& &
\,+\,\big(\R{\wt{A},\wt{V}}{z}-\RAV{z}\big)\,i\alpha\cdot(\nabla\chi)\,\phi
\\
&=&
\RAV{z}\,\big(V-\wt{V}+\alpha\cdot(A-\wt{A})\big)\,\chi\,\R{\wt{A},\wt{V}}{z}\,
(\D{\wt{A},\wt{V}}-z)\phi
\\
& &
\,+\,
\big(\R{\wt{A},\wt{V}}{z}
-\RAV{z}\big)\,i\alpha\cdot(\nabla\chi)\,\R{\wt{A},\wt{V}}{z}\,
(\D{\wt{A},\wt{V}}-z)\phi
\,.
\end{eqnarray*}
As $\D{\wt{A},\wt{V}}$ is essentially self-adjoint
on $\wt{\dom}_e$, 
we know that 
$(\D{\wt{A},\wt{V}}-z)\,\wt{\dom}_e$
is dense, which together with the calculation above implies 
\begin{eqnarray}
\big(\R{\wt{A},\wt{V}}{z}-\RAV{z}\big)\,\chi&=&
\big(\R{\wt{A},\wt{V}}{z}-\RAV{z}\big)\,
i\alpha\cdot(\nabla\chi)\,\R{\wt{A},\wt{V}}{z}\label{jim1}
\\
& &\nonumber
\,+\,\RAV{z}\,\big(V-\wt{V}+\alpha\cdot(A-\wt{A})\big)
\,\chi\,\R{\wt{A},\wt{V}}{z}\,.
\end{eqnarray}
Taking the adjoint of \eqref{jim1} (with $z$ replaced by $\ol{z}$)
we obtain \eqref{jim1adjoint}.
\qed


\subsection{Conjugation of $\RAV{z}$ with exponential weights}
\label{ssec-conj}

As a preparation for the localization estimates
for the spectral projections we shall now study the
conjugation of $\RAV{z}$ with exponential weight
functions $e^F$ acting as multiplication operators
on $\HR$. 
To this end we recall that $e_0\in(-1,1)$ is an element
of the resolvent set of $\DAV$ and set
\begin{eqnarray}
\label{def-delta0}
\delta_0&:=&\inf\big\{\,|e_0-\lambda|\,:\:\lambda\in\sigma
(\DAV)\,\big\}\,>\,0\,,
\\
\Gamma&:=&e_0\,+\,i\RR\,.\label{def-Gamma}
\end{eqnarray}
Notice that the decay rate in the following lemma is 
determined only
by the decay rate $m$ appearing in Hypothesis~\ref{hyp-V-komplett} 
and the number $\triangle_0$ defined in \eqref{def-triangle0}.
In the next proof and
henceforth we shall often use the abbreviations
\begin{equation}\label{def-DA-RA}
\DA\,:=\,\D{A,0}\,,
\qquad\RA{z}\,:=\,(\DA-z)^{-1}\,,\qquad z\in\vr(\DA)\,.
\end{equation}

\begin{lemma}\label{le-conjugation-RAV}
Assume that $A\in L^\infty_\loc(\RR^3,\RR^3)$
and that $V$ fulfills Hypothesis~\ref{hyp-V-komplett}.
Let $0<a<\min\{\triangle_0,m\}$. Then there
is some $C_{a}\in(0,\infty)$ such that, for all
$F\in C^\infty(\RR^3,\RR)$ with $F(0)=0$,
$F\grg0$ or $F\klg0$, 
$
\|\nabla F\|_\infty\klg a
$,
and all $z=e_0+i\eta\in\Gamma$, 
\begin{equation}\label{harry}
\big\|\,e^{F}\,\RAV{z}\,e^{-F}\,\big\|\,\klg\,
\frac{C_a}{\sqrt{1+\eta^2}}\,.
\end{equation}
\end{lemma}

\proof
First, we assume that $F$ is 
constant outside some ball in $\RR^3$.
Then it suffices to treat the case $F\grg0$, since otherwise
we could consider the adjoint of $e^{F}\,\RAV{z}\,e^{-F}$. 
To begin with we recall 
the identity
\begin{equation}\label{lisa9}
\big\|\,\alpha\cdot v\,\big\|_{\LO(\CC^4)}\,=\,|v|\,,\qquad v\in\RR^3\,,
\end{equation}
which follows for instance from the $C^*$-equality and the 
Clifford algebra relations \eqref{Clifford}.
In particular, 
$$
\|\alpha\cdot\nabla F\|\,\|\RA{z}\|
\klg\, \frac{a}{\sqrt{\triangle_0^2+\eta^2}}\:<\,1\,,
\qquad z=e_0+i\eta\in\Gamma\,.
$$
Since $F$ is smooth and constant outside some
compact set Lemma~\ref{le-BdMetc}(d) permits to get, for $z\in\Gamma$,
\begin{equation}\label{RA-conjugated}
e^F\,\RA{z}\,e^{-F}
\,=\,
\RA{z}\sum_{\ell=0}^\infty\{-i(\alpha\cdot\nabla F)\,\RA{z}\}^\ell
\,,
\end{equation}
whence
\begin{equation}\label{marah1}
\big\|\,
e^F\,\RA{z}\,e^{-F}\,\big\|
\,\klg\,\frac{1}{1-a/\triangle_0}
\,\frac{1}{\sqrt{\triangle_0^2+\eta^2}}\,,
\qquad z=e_0+i\eta\in\Gamma
\,.
\end{equation}
Next, we pick some $R>\max\{|y|:\,y\in\cY\}$
and
$\chi\in C^\infty(\RR^3,[0,1])$
such that $\chi\equiv0$ on $\ball{R}{0}$,
$\chi\equiv1$ on $\RR^3\setminus\ball{R+2}{0}$,
and $\|\nabla\chi\|_\infty\klg1$.
We set $\ol{\chi}:=1-\chi$.
Furthermore, we let $\wt{\chi}$ denote the characteristic
function of $\RR^3\setminus\ball{R}{0}$.
We choose $R$ so large 
(depending on $a$, but not on $F$; recall \eqref{bd-VE3})
that 
\begin{equation}\label{marah2}
\sup_{|x|\grg R}\|\VC(x)\|\,+\,\sup_{|x|\grg R}\|\VM(x)\|\,+\,
\big\|\,\wt{\chi}\,e^F\,\VE\,e^{-F}
\,\wt{\chi}\,\big\|
\,\klg\,\frac{\triangle_0}{2}\:\Big(1-\frac{a}{\triangle_0}\Big)\,.
\end{equation}
Conjugating \eqref{jim1adjoint} with exponential weights 
and rearranging the terms we find, for $z\in\Gamma$,
\begin{eqnarray*}
\lefteqn{
\big\{\id+e^F\,\RA{z}\,e^{-F}\,(\wt{\chi}\,\VC\,+\,\wt{\chi}\,\VM
\,+\,\chi\,e^F\,\VE\,e^{-F}\,\wt{\chi})\big\}
\,\chi\,e^F\,\RAV{z}\,e^{-F}
}
\\
&=&
\chi\,e^F\,\RA{z}\,e^{-F}\,-\,
\big(e^F\,\RA{z}\,e^{-F}\big)
\,\big(e^Fi\alpha\cdot\nabla\chi\big)
\,\big(\RAV{z}-\RA{z}\big)\,e^{-F}
\\
& &
\;-\,\big(e^F\,\RA{z}\,e^{-F}\big)\,\big(\chi\,e^F\,\VE\,e^{-F}\,
\ol{\chi}\big)\,(\id_{\ball{R+2}{0}}\,e^F)
\,\RAV{z}\,e^{-F}
\,.
\end{eqnarray*}
Here the operator $\{\cdots\}$ on the left side
can be inverted by means of a Neumann series
and $\|\{\cdots\}^{-1}\|\klg2$ by \eqref{marah1}\&\eqref{marah2}.
Furthermore, we observe that, by the choice of $\chi$,
the assumption on $F$,
and \eqref{lisa9},
$$
\big\|\,e^Fi\alpha\cdot\nabla\chi\,\big\|
\,\klg\,e^{a(R+2)},\quad
\|\RA{z}\|\,\klg\,1\,,\quad\|\RAV{z}\|\,\klg\,
\frac{1}{\delta_0}\,,\quad
\|e^{-F}\|\,\klg\,1\,.
$$
Using these remarks together with \eqref{bd-VE2}
and \eqref{marah1} we obtain
\begin{equation*}
\big\|\,\chi\,e^F\,\RAV{z}\,e^{-F}\,\big\|
\,\klg\,
\frac{C_a'\,e^{R+2}}{\sqrt{\triangle_0^2+\eta^2}}
\,.
\end{equation*}
This estimate implies the assertion 
if $F$ is constant outside some ball
since, certainly,
$\|\ol{\chi}\,e^F\|\klg e^{a(R+2)}$ and, for $z=e_0+i\eta\in\Gamma$,
\begin{equation}\label{marah4}
\|\RAV{z}\|\,\klg\,\frac{1}{\sqrt{\delta_0^2+\eta^2}}\,.
\end{equation}
Let us now assume that $F\grg0$ is 
not necessarily bounded.
Let $F_1,F_2,\dots\in C^\infty(\RR^3,[0,\infty))$ be  
constant outside some ball and such that $F_n=F$ on $\ball{n}{0}$
and $F_n\to F$.
Then $e^{-F_n}\RAV{z}\,e^{F_n}\,\phi\to e^{-F}\,\RAV{z}\,e^F\phi$
by the dominated convergence theorem,
for every $\phi\in\COI$. 
Since $e^{-F_n}\RAV{z}\,e^{F_n}$ obeys the estimate \eqref{harry}
uniformly in $n$, we see that the densely defined operator
$e^{-F}\,\RAV{z}\,e^F\!\!\upharpoonright_{\COI}$
is bounded and satisfies \eqref{harry}, too.
But this is the case if and only if its adjoint,
$e^F\,\RAV{z}\,e^{-F}=\big(e^{-F}\,\RAV{z}\,e^F\big)^*$,
is an element of $\LO(\HR)$ and satisfies
\eqref{harry} as well.
\qed

\bigskip

In the applications of the previous lemma
the following observation is very useful.

\begin{lemma}\label{le-Cauchy}
Assume that $A\in L^\infty_\loc(\RR^3,\RR^3)$
and that $V$ fulfills Hypothesis~\ref{hyp-V-komplett}.
Let $0<a<\min\{\triangle_0,m\}$. Then there
is some $C_{a}'\in(0,\infty)$ such that, for all
$F\in C^\infty(\RR^3,\RR)$ with $F(0)=0$,
$F\grg0$ or $F\klg0$,
$\|\nabla F\|_\infty\klg a$, which are constant outside
some ball in $\RR^3$, and for all 
$\phi\in\HR$,
\begin{equation}\label{Cauchy1}
\int_\Gamma\big\|\,|\DAV|^{1/2}
e^{F}\,\RAV{z}\,e^{-F}\,\phi\,\big\|^2\,|dz|\,\klg\,
C_a'\,\|\phi\|^2\,,
\end{equation}
and, for $\phi\in\dom\big(|\DAV|^{1/2}\big)$,
\begin{equation}\label{Cauchy2}
\int_\Gamma\big\|\,
e^{F}\,\RAV{z}\,e^{-F}\,|\DAV|^{1/2}
\,\phi\,\big\|^2\,|dz|\,\klg\,
C_a'\,\|\phi\|^2\,.
\end{equation}
\end{lemma}

\proof
For later reference we additionally pick
some $\chi\in C^\infty(\RR^3,\RR)$
which is constant outside some large ball
and infer from Lemma~\ref{le-BdMetc}(e) 
that, for $z\in\Gamma$, 
\begin{eqnarray}\label{lisa6}
\big[\,\RAV{z}\,,\,\chi\,e^{F}\,\big]&=&
\RAV{z}\,\big\{\,i\alpha\cdot(\nabla\chi+\chi\,\nabla F)
\\
& &\nonumber
\quad\,+\,[\,\chi\,e^{F}\,,\,\VE\,]\,e^{-F}\,\big\}\,e^F\,\RAV{z}\,.
\end{eqnarray}
The special case $\chi\equiv1$ implies
\begin{eqnarray}\label{harry1}
e^F\,\RAV{z}\,e^{-F}&=&\RAV{z}
\,-\,\RAV{z}\,\big\{\,i\alpha\cdot\nabla F
\\
& &\nonumber\,+\,[\,e^{F}\,,\,\VE\,]\,e^{-F}\,\big\}
\,e^F\,\RAV{z}\,e^{-F}\,.
\end{eqnarray}
Taking the adjoint and replacing $F$ by $-F$ 
and $\ol{z}$ by $z$ we also get
\begin{eqnarray}\label{harry2}
e^F\,\RAV{z}\,e^{-F}&=&\RAV{z}
\,-\,e^F\,\RAV{z}\,e^{-F}\,\big\{\,i\alpha\cdot\nabla F
\\
& &\nonumber\,+\,e^{F}\,[\VE\,,\,e^{-F}]\,\big\}\,\RAV{z}\,.
\end{eqnarray}
Now, let $T$ be a self-adjoint operator on some Hilbert space
$\mathscr{K}$
such that $(-\delta_0,\delta_0)\subset\vr(T)$.
Then, for $\phi\in\mathscr{K}$,
\begin{equation}
\label{lisa8}
\int_\RR\big\|\,|T|^{1/2}\,(T-i\eta)^{-1}\,\phi\big\|^2\,d\eta
\,=\,
\int_\RR\int_\RR\frac{|\lambda|}{\lambda^2+\eta^2}\,d\eta\,
d\|E_\lambda(T)\,\phi\|^2
\,=\,\pi\,\|\phi\|^2\,,
\end{equation}
and it is elementary to check that, for $\eta\in\RR$,
\begin{equation}\label{ralf3}
\big\|\,|T|^{1/2}\,(T-i\eta)^{-1}\,\big\|\,=\,
\frac{\delta_0^{1/2}\,\id_{(-\delta_0,\delta_0)}(\eta)}{
\sqrt{\delta_0^2+\eta^2}}\:+\:
\frac{\id_{(-\delta_0,\delta_0)^c}(\eta)}{\sqrt{2|\eta|}}
\end{equation}
Using \eqref{lisa8} and \eqref{ralf3} with $T=\DAV-e_0$
and taking \eqref{bd-VE3}, \eqref{harry}, \eqref{lisa9}, and \eqref{marah4}
into account we readily derive the asserted estimate \eqref{Cauchy1}
from \eqref{harry1}.
The second estimate \eqref{Cauchy2}
it obtained analogously by means of
\eqref{harry2}.
\qed


\subsection{Commutators}
\label{ssec-commutators}

In this subsection we derive the crucial technical
prerequisits for the spectral analysis of $\HH_N$,
namely various commutator estimates involving the projection
$\PAV$, cut-off functions, and exponential weights
$e^F$. 
Our standard assumptions on the cut-off and weight
functions are
\begin{equation}\label{hyp-chi}
\chi\in C^\infty(\RR^3,[0,1])\;\textrm{is constant outside some ball.}
\end{equation}
and
\begin{equation}\label{hyp-F}
\left.\begin{split}
&F\in C^\infty(\RR^3,\RR)\,,\quad
F\grg0\;\;\textrm{or}\;\;F\klg0\,,\quad 
F(0)\,=\,0\,,\quad|\nabla F|\,\klg\,a\,,
\\
&F\;\,\textrm{is constant outside some ball.}
\end{split}\right\}
\end{equation}
To shorten the presentation we generalize our estimates
to unbounded $F$ only if this is explicitely used in this article.

\begin{proposition}\label{prop-lisa}
Assume that $A\in L^\infty_\loc(\RR^3,\RR^3)$ and that
$V$ fulfills Hypothesis~\ref{hyp-V-komplett}
and
let $0<a_0<\min\{\triangle_0,m\}$. Then there is some constant
$C_{a_0}\in(0,\infty)$ such that, for all $a\in[0,a_0]$
and
$\chi,F$ satisfying \eqref{hyp-chi} and \eqref{hyp-F},
\begin{equation}
\big\|\,|\DAV|^{1/2}\,[\PAV\,,\,\chi\,e^{F}]\,e^{-F}\,|\DAV|^{1/2}
\,\big\|\label{lisa2}
\,=\,C_{a_0}\big(\|\nabla\chi\|_\infty+a
\big)\,.
\end{equation}
\end{proposition}

\proof
We shall employ the identity
\begin{equation}\label{lisa4}
\big[\PAV\,,\,\chi\,e^{F}\big]\,=\,\frac{ 1}{2}\,
\big[\sgn\big(\DAV-e_0\big)\,,\,\chi\,e^{F}\,\big]
\end{equation}
and the representation of the sign function as a Cauchy principal value
(see, e.g., \cite[Page~359]{Kato}),
\begin{equation}\label{lisa5}
\sgn\big(\DAV-e_0\big)\,\psi\,=\,
\int_{\Gamma} \RAV{z}\,\psi\,\frac{dz}{\pi}
\,:=\,\lim_{R\to\infty}\int_{-R}^R\R{A,V}{e_0+i\eta}\,\psi
\,\frac{d\eta}{\pi}
\,,
\end{equation}
for $\psi\in\HR$,
where $\Gamma$ is defined in \eqref{def-Gamma}.
Taking also \eqref{bd-VE1}, \eqref{bd-VE2}, and
\eqref{lisa6} in to account
we obtain
\begin{eqnarray}
\lefteqn{
\big|\SPb{|\DAV|^{1/2}\,\phi}{[\PAV\,,\,\chi\,e^{F}]\,e^{-F}\,
|\DAV|^{1/2}\psi}\big|\nonumber
}
\\
&\klg&\nonumber
\int_\Gamma\big\|\,|\DAV|^{1/2}\,\RAV{\bar{z}}\,\phi\,\big\|\label{lisa7a}
\,\big\|\,i\alpha\cdot(\nabla\chi+\chi\,\nabla F)\,+\,
[\chi\,e^{F}\,,\,\VE]\,e^{-F}
\,\big\|
\\\nonumber
& &\quad\cdot\;\big\|\,
e^{F}\,\RAV{z}\,e^{-F}\,|\DAV|^{1/2}
\,\psi\,\big\|
\,\:\frac{|dz|}{2\pi}
\\
&\klg&\nonumber
C_{a_0}'\,
\big(\|\nabla\chi\|_\infty+\|\chi\nabla F\|_\infty+a\big)
\Big(\int_\Gamma\big\|\,|\DAV|^{1/2}\,\RAV{z}\,\phi\,\big\|^2
\,\frac{|dz|}{2\pi}\Big)^{1/2}
\\
& &\quad\cdot
\Big(\int_\Gamma\big\|\,e^{F}\,\RAV{z}\,e^{-F}\,|\DAV|^{1/2}
\,\psi\,\big\|^2
\,\frac{|dz|}{2\pi}\Big)^{1/2}
\,,
\end{eqnarray}
for $\phi,\psi\in\dom(|\DA|^{1/2})\supset\Ran(\RA{z})$.
By virtue of \eqref{Cauchy1}\&\eqref{Cauchy2}
we first infer that
$$
[\PAV\,,\,\chi\,e^{F}]\,e^{-F}\,|\DAV|^{1/2}\,
\psi\in\dom\big((|\DAV|^{1/2})^*\big)\,=\,
\dom(|\DAV|^{1/2})\,.
$$
We conclude by recalling that 
an operator $T:\dom(T)\to\mathscr{K}$ 
on some Hilbert space 
$\mathscr{K}$ is bounded if and only if 
\begin{equation}\label{op-norm}
\sup
\Big\{\,\big|\SPb{\phi}{T\,\psi}\big|
\::\;\phi\in X,\,\psi\in\dom(T)\,,\;\|\phi\|=\|\psi\|=1\,\Big\}
\end{equation}
is finite, in which case it is equal to the
norm of $T$. Here $X\subset\mathscr{K}$ is a subspace
with $\ol{X}\supset\Ran(T)$.
\qed

\bigskip

Given some suitable weight function, $F$, we abbreviate
\begin{equation}\label{def-PAVF}
\PAVF\,:=\,e^F\,\PAV\,e^{-F}\,.
\end{equation}

\begin{corollary}\label{cor-eFPAVe-F}
Assume that $A\in L^\infty_\loc(\RR^3,\RR^3)$
and that $V$ fulfills Hypothesis~\ref{hyp-V-komplett}.
Let $0<a<\min\{\triangle_0,m\}$. Then there is some
$C(a)\in(0,\infty)$ such that, for all 
$F\in C^\infty(\RR^3,\RR)$ 
satisfying $F(0)=0$, $F\grg0$ or $F\klg0$, 
and $\|\nabla F\|_\infty\klg a$,
we have $\PAVF\in\LO(\HR)$ and
$
\big\|\,\PAVF\,\big\|\klg C(a)
$.
\end{corollary}

\proof
First, we assume that $F$ satisfies \eqref{hyp-F}. In this case
the claim follows from Proposition~\ref{prop-lisa} because
$[e^F\,,\,\PAV]\,e^{-F}=\PAVF-\PAV$.
If $F$ is unbounded, then we apply an approximation
argument similar to the one at the end of the proof of
Lemma~\ref{le-conjugation-RAV}.
\qed

\begin{corollary}\label{cor-ruth}
Assumme that $A\in L^\infty_\loc(\RR^3,\RR^3)$ and that
$V$ fulfills Hypothesis~\ref{hyp-V-komplett}
and
let $0<a_0<\min\{\triangle_0,m\}$. Then
there is some constant
$C\in(0,\infty)$ such that, for all $a\in[0,a_0]$,
$\chi,F$ satisfying \eqref{hyp-chi},\eqref{hyp-F}, and 
$\|\nabla\chi\|_\infty\klg1$,
$L\in\LO(\HR)$, and $\vp\in\HR$,
\begin{align}
\Big|\SPb{\vp}{\,\PAVF\,\chi\,L\,\chi\,\PAVF\,\vp}
-&
\SPb{\vp}{\chi\,\PAV\,L\,\PAV\,\chi\,\vp}\Big|\nonumber
\\
&\label{ruth1L}\klg\,
\big(a+\|\nabla\chi\|_\infty\big)\,C\,\|L\|\,\|\vp\|^2\,.
\end{align}
Moreover,
for all $\vp\in\dom(\DAV)$,
\begin{eqnarray}
\lefteqn{
\Big|\SPb{\vp}{\,\PAVF\,\chi\,\DAVC\,\chi\,\PAVF\,\vp}
-\label{ruth1}
\SPb{\vp}{\chi\,\PAV\,\DAVC\,\PAV\,\chi\,\vp}\Big|
}
\\
&\klg&\nonumber
\big(a+\|\nabla\chi\|_\infty\big)\,\inf_{0<\ve\klg1}\Big\{
\ve\,\SPb{\vp}{\chi\,\PAV\,\DAVC\,\PAV\,\chi\,\vp}
+C\,\ve^{-1}\,\|\vp\|^2\Big\}.
\end{eqnarray}
If $\VC=V=0$, then
\eqref{ruth1} still holds true, 
if $\chi\,\DA\,\chi$ is replaced
by $\chi\,|\DA|\,\chi$ on the left side.
\end{corollary}

\proof
The proof of
\eqref{ruth1L} is a rather obvious application of
Proposition~\ref{prop-lisa} and in fact a simpler
analogue of the derivation of \eqref{ruth1} below.
Once \eqref{ruth1L} is verified, it suffices to prove \eqref{ruth1}
with $\DAVC$ replaced by $\DAV$ since $\VM$ and $\VE$ are bounded.
Without loss of generality we may further assume that $\DAV$
is positive on the range of $\PAV$. For otherwise we could add
a suitable constant to $\DAV$. To prove \eqref{ruth1} we first recall
that
$\PAV$ maps the domain of $\DAV$ into itself and
by Lemma~\ref{le-BdMetc}(d) we know that multiplication with
$\chi$ or $e^{\pm F}$ leaves $\dom(\DAV)$ invariant, too.
We thus have the following identity on $\dom(\DAV)$,
\begin{eqnarray*}
\lefteqn{
\PAVF\,\chi\,\DAV\,\chi\,\PAVF
\,-\,\chi\,\PAV\,\DAV\,\chi
}
\\
&=&
e^F\,\big[\PAV\,,\,\chi\,e^{-F}\big]\,\DAV\,\PAV\,\chi
\,
+\,
\chi\,\PAV\,\DAV\,\big[\chi\,e^F\,,\,\PAV\big]\,e^{-F}
\\
& &\;+\,
e^F\,\big[\PAV\,,\,\chi\,e^{-F}\big]\,\DAV\,
\big[\chi\,e^F\,,\,\PAV\big]\,e^{-F}.
\end{eqnarray*}
It follows that the absolute value on the left side of \eqref{ruth1}
is less than or equal to
\begin{equation*}
\begin{split}
\big\|\,|\DAV|^{1/2}\,\PAV\,\chi\,&\vp\,\big\|\,\Big\{
\sum_{\sharp=\pm}\big\|\,|\DAV|^{1/2}\,\big[\PAV\,,\,\chi\,e^{\sharp\,F}
\big]\,e^{-\sharp F}\,\big\|
\Big\}\,\|\vp\|
\\
&+\,
\prod_{\sharp=\pm}\big\|\,|\DAV|^{1/2}\,\big[\PAV\,,\,\chi\,e^{\sharp\,F}
\big]\,e^{-\sharp F}\,\big\|\,\|\vp\|^2\,,
\end{split}
\end{equation*}
which together with Proposition~\ref{prop-lisa} implies \eqref{ruth1}.
The last statement of the lemma is valid since the argument above
works equally well with $|\DAV|$ in place of $\DAV$ because
$\PAV\,|\DAV|=\PAV\,\DAV$.
\qed

\begin{lemma}\label{cor-H1/2}
Assume that $A\in L^\infty_\loc(\RR^3,\RR^3)$
satisfies $\|A\,e^{-\tau_0|x|}\|_\infty<\infty$,
for some $0\klg\tau_0<\min\{m,\triangle_0\}$,
and that $V$ fulfills Hypothesis~\ref{hyp-V-komplett}.
Then
$\PAV\,\phi\in H^{1/2}(\RR^3,\CC^4)$, for every $\phi\in\COI$.
\end{lemma}

\proof
Let $\phi\in\COI$.
We pick some 
$\chi\in C_0^\infty(\RR^3,[0,1])$ with $\chi\equiv1$
on $\supp(\phi)$.
Furthermore, we pick
$\zeta\in C_0^\infty(\RR^3,[0,1])$
such that $\zeta\equiv1$ on $\supp(\chi)\cup\ball{R}{0}$,
where $R>\max\{|y|:\,y\in\cY\}$. 
We set
$\ol{\zeta}:=1-\zeta$.
Since $\dom(\DAV)\subset H^{1/2}_\loc(\RR^3,\CC^4)$
and the spectral projection $\PAV$ maps the domain
of $\DAV$ into itself
it follows that 
$
\zeta\,\PAV\,\phi
\in H^{1/2}(\RR^3,\CC^4)
$.
Furthermore, we pick a (smooth, locally finite) 
partition of unity on $\RR^3$,
$\{J_\nu\}_{\nu\in\NN}$, $\sum_{\nu=1}^\infty J_\nu=1$,
such that $\sum_{\nu=1}^\infty|\nabla J_\nu|\klg C$,
for some constant $C\in(0,\infty)$.
Setting $\zeta_\nu:=J_\nu\,\ol{\zeta}$, $\nu\in\NN$,
$\wt{\phi}:=\PAV\,\big(\DAV-i\big)\,\phi$,
and
using \eqref{jim1adjoint},
we obtain
\begin{eqnarray}
\ol{\zeta}\,\PAV\,\phi&=&\sum_{\nu=1}^\infty \zeta_\nu
\,\RAV{i}\,\PAV\,\nonumber
\big(\DAV-i\big)\,\phi
\\
&=&
\ol{\zeta}\,\R{0}{i}\,
\wt{\phi}
\label{sophia1}
-\sum_{\nu=1}^\infty\R{0}{i}\,i\alpha\cdot(\nabla\zeta_\nu)
\big(\RAV{i}-\R{0}{i}\big)
\,\wt{\phi}
\\ \label{sophia2}
& &\;\;\quad-\sum_{\nu=1}^\infty\R{0}{i}\,\zeta_\nu\,(V+\alpha\cdot A)
\,\RAV{i}\,\wt{\phi}\,.
\end{eqnarray}
Here the sum in \eqref{sophia1} commutes with the first resolvent
and the strong limit
$\sum_{\nu=1}^\infty i\alpha\cdot(\nabla\zeta_\nu)$
defines a bounded operator on $L^2(\RR^3,\CC^4)$.
To treat \eqref{sophia2} we first use that 
$\sum_{\nu=1}^\infty\zeta_\nu\,V=\ol{\zeta}\,V$ is bounded.
Next, we pick some $F \in C^\infty(\RR^3,[0,\infty))$ 
vanishing on
some ball containing $0$ and
$\supp(\phi)$ and satisfying $F(x)=a\,|x|-a'$, for $x$ outside some
sufficiently large ball with $\tau_0<a<\min\{m,\triangle_0\}$, $a'>0$.
Then we write
\begin{eqnarray*}
\lefteqn{
(\alpha\cdot A)\,\RAV{i}\,\wt{\phi}
\,=\,
\big(e^{-F}\,\alpha\cdot A\big)\,\big(e^{F}\,\RAV{i}\,e^{-F}\big)\times
}
\\
& &\quad\times\big(e^{F}\,\PAV\,e^{-F}\big)\,\big(
\D{A,\VC+\VM}+i\alpha\cdot\nabla F+e^F\,\VE\, e^{-F}\,\big)\,\phi\,.
\end{eqnarray*}
Using \eqref{bd-VE2}, Lemma~\ref{le-conjugation-RAV}, and 
Corollary~\ref{cor-eFPAVe-F} we see that 
$(\alpha\cdot A)\,\RAV{i}\,\wt{\phi}$ 
is an element of $L^2(\RR^3,\CC^4)$.
These remarks imply that 
$\ol{\zeta}\,\PAV\,\phi$ belongs to $\Ran\big(\R{0}{i}\big)
+\Ran\big(\ol{\zeta}\R{0}{i}\big)=H^1(\RR^3,\CC^4)$.
\qed

\bigskip

We may now conclude that $\HH_N$ is well-defined on the dense
domain $\core$ defined in \eqref{def-core}.

\begin{corollary}\label{cor-core}
Assume that $A\in L^\infty_\loc(\RR^3,\RR^3)$ 
satisfies $\|A\,e^{-\tau_0|x|}\|_\infty<\infty$,
for some $0\klg\tau_0<\min\{m,\triangle_0\}$,
and that
$V$ fulfills Hypothesis~\ref{hyp-V-komplett}.
Then,
for $\Psi\in\core$ and $1\klg i<j\klg N$,
$$
\int_{\RR^{3N}}\frac{1}{|x_i-x_j|^2}\:
\big|\PAVN\,\Psi(X)\big|^2\,dX\,<\,\infty\,.
$$
\end{corollary}

\proof
Let $\phi,\psi\in \COI$.
Thanks to Lemma~\ref{cor-H1/2} we know that both
$\PAV\,\phi$ and $\PAV\,\psi$ belong to $H^{1/2}(\RR^3,\CC^4)$
and, hence, to $L^3(\RR^3,\CC^4)$
by the Sobolev inequality for $|\frac{1}{i}\nabla|$.
An application of the Hardy-Littlewood-Sobolev
inequality thus
yields
$$
\int_{\RR^6}\frac{1}{|x-y|^2}
\,\big|\PAV\,\phi(x)\big|^2\,\big|\PAV\,\psi(y)\big|^2
\:dx\,dy\,<\,\infty\,.
$$
This estimate clearly implies the full assertion.
\qed

\bigskip

\bigskip

In our applications it is important to control commutators
that are multiplied with square-roots
of the electron-electron interactions
$W(x_i,x_j)$. 
In order to formulate an appropriate estimate we set
\begin{equation}\label{def-Wy}
W_y(x)\,:=\,W(x,y)\,=\,W(y,x)\,,\qquad x,y\in\RR^3\,,
\end{equation}
in what follows.
The proof of the next proposition
looks somewhat lengthy and is hence postponed to 
Appendix~\ref{app-proofs}.
This is due to the fact that the
singularity of $W_y$
may be located anywhere and that we 
allow for unbounded magnetic fields. 
We remark that, even in the case $V=0$,
a diamagnetic inequality is not very
useful in this context since, for unbounded magnetic fields,
one cannot compare $|-i\nabla+A|$ with 
$|\DA|$. We tackle this problem by a procedure that involves
a partition of unity, local gauge transformations, and exponential decay
estimates which control the correlation between different
regions in position space. 
As a result we obtain a commutator estimate which can be chosen to
depend
only on the local magnitude of $|B|$ 
either at the singularity $y$ or on the support of the 
involved cut-off
function.
For any function $\chi$ on $\RR^3$ we use the notation
\begin{equation}\label{def-infty-norm-chi}
\|B\,\|_{\infty,\chi}\,:=\,\sup\big\{|B(x)|:\,x\in\supp(\chi)\big\}\,.
\end{equation}

\begin{proposition}\label{prop-W1/2}
Assume that $A\in C^1(\RR^3,\RR^3)$ and
$B=\rot A$ satisfies \eqref{var-bd-B} and that $V$
fulfills Hypothesis~\ref{hyp-V-komplett}.
Let $0\klg a_0<\min\{m,\triangle_0\}$
and $\cN\subset\RR^3$ be a neighbourhood of the set of singularities,
$\cY$, of $\VC$.
Then there is some constant, $C_{a_0,\cN}\in(0,\infty)$,
such that, for all $a\in[0,a_0]$, all
$\chi,F$ satisfying \eqref{hyp-chi},\eqref{hyp-F}
which are constant on 
$\cN$, and all $y\in\RR^3$,
\begin{eqnarray}
\lefteqn{\nonumber
\big\|\,W_y^{1/2}\,\big[\,\PAV\,,\,e^F\,\chi\,\big]\,e^{-F}\,\big\|
}
\\
&\klg&\label{gustav99}
C_{a_0,\cN}\,\Big(1+\min\big\{
|B(y)|\,,\,\|B\,\|_{\infty,\chi}\big\}\Big)\,
\big(a+\|\nabla\chi\|_\infty\big)\,.
\end{eqnarray}
If $\VE=0$, then $\|B\,\|_{\infty,\chi}$ 
can be replaced
by $\|B\,\|_{\infty,\chi\nabla F+\nabla\chi}$ in \eqref{gustav99}.
\end{proposition}

\begin{corollary}\label{cor-ruth-W}
Assume that $A\in C^1(\RR^3,\RR^3)$ and $B=\rot A$
is bounded and that $V$ fulfills Hypothesis~\ref{hyp-V-komplett}.
Then we find, for every $\ve>0$, some constant $C_{a_0,\ve}\in(0,\infty)$
such that, for all $F$ satisfying \eqref{hyp-F},
$\vp\in\core$, and $1\klg i\klg N$,
\begin{eqnarray*}
\lefteqn{
\Big|\SPb{\vp}{\id\otimes e^F\,\PAVN\,W_{iN}\,\PAVN\,\id\otimes e^{-F}\,\vp}
\,-\,
\SPb{\vp}{\PAVN\,W_{iN}\,\PAVN\,\vp}\Big|
}
\\
&\klg&
a\,\big\{\,\ve\,\SPb{\vp}{\PAVN\,W_{iN}\,\PAVN\,\vp}\,+\,C_{a_0,\ve}
\,\|\vp\|^2\big\}\,,
\end{eqnarray*}
where $e^F$ acts only on the last variable.
\end{corollary}

\proof
This corollary is proved 
by means of Proposition~\ref{prop-W1/2}
in the same way as Corollary~\ref{cor-ruth}.
We also recall that $\PAVN\,\core\subset\dom(W_{ij})$.
\qed

\bigskip

The technique used in the proof
of Proposition~\ref{prop-W1/2}
also yields the following
result whose proof can be found in Appendix~\ref{app-proofs}, too:

\begin{lemma}\label{le-josephine}
Assume that $A\in C^1(\RR^3,\RR^3)$ and
$B=\rot A$ satisfies \eqref{var-bd-B} and that $V$
fulfills Hypothesis~\ref{hyp-V-komplett}.
Then there is some constant $C\in(0,\infty)$
such that, for all $\psi\in \dom(\DAV)$,  
\begin{equation}\label{petra}
\big\|\,W_y^{1/2}\,\PAV\,\psi\,\big\|\,\klg\,C\,
\Big(1+\min\big\{|B(y)|\,,\,\|B\,\|_{\infty,\psi}\big\}\Big)
\big\|\,\big(\DAV-i\big)\,\psi\,\big\|\,.
\end{equation}
\end{lemma}


\subsection{Differences of projections}
\label{ssec-PA-PAV}

In our applications
it is eventually necessary to have some control on
the difference between $\PAV$ and
$$
\PA\,:=\,\Pp{A,0}\,.
$$

\begin{lemma}\label{le-PA-PAV}
Assume that $A\in L_\loc^\infty(\RR^3,\RR^3)$
and that $V$ fulfills Hypothesis~\ref{hyp-V-komplett}.
Then there is some $C\in(0,\infty)$
such that, for all
$\zeta\in C^\infty(\RR^3,[0,1])$ which are constant outside some ball
such that $\zeta\,\VC$ is bounded,
$$
\big\|\,|\DA|^{1/2}
\zeta\,\big(\PA-\PAV\big)\,\big\|\,\klg\, C
\,\big(\|\zeta\,V\|\,+\,\|\nabla\zeta\|_\infty\big)\,.
$$
In particular, $\zeta\,\PAV\,\vp\in\dom(|\DA|^{1/2})$, for
every $\vp\in\dom(\DA)$.
\end{lemma}

\proof
Due to \eqref{lisa5} the norm in the statement (if it exists)
is bounded from above by
$$
\sup_{{
\phi\in\dom(|\DA|^{1/2}),\,\psi\in\HR
\atop
\|\phi\|=\|\psi\|=1
}}
\int_\Gamma\Big|\SPB{|\DA|^{1/2}\,
\phi}{\zeta\,\big(\R{A}{z}-\R{A,V}{z}\big)\,\psi}
\Big|\,\frac{|dz|}{\pi}
\,.
$$
We next use \eqref{jim1adjoint}, \eqref{marah4}, and \eqref{Cauchy2}
to conclude that the asserted bound holds true.
\qed

\bigskip

We note the following trivial consequence of the previous lemma:
Namely, we pick some $\theta\in C_0^{\infty}(\RR^3,[0,1])$
with $\theta\equiv1$ on $\ball{1}{0}$ and
$\theta\equiv0$ outside $\ball{2}{0}$, and set
$\theta_R(x):=\theta(x/R)$, for $R\grg1$, $x\in\RR^3$.
By virtue of Hypothesis~\ref{hyp-V-komplett}
and Lemma~\ref{le-PA-PAV} we then have, for every $\zeta$
as in the statement of Lemma~\ref{le-PA-PAV},
\begin{equation}\label{conv-PA-PAV}
\big\|\,|\DA|^{1/2}
(1-\theta_R)\,\zeta\,\big(\PA-\PAV\big)\,\big\|
\,\klg\,C
\,\Big(\|(1-\theta_R)\,V\|\,+\,\frac{\|\nabla\theta\|_\infty}{R}\Big)
\,\longrightarrow\,0,
\end{equation}
as $R$ tends to infinity.

\begin{corollary}\label{cor-PA-PAV}
Assume that $A\in L_\loc^\infty(\RR^3,\RR^3)$ and that
$V$ fulfills Hypothesis~\ref{hyp-V-komplett}. 
Then there is some $C\in(0,\infty)$
such that, for every $\zeta\in C^\infty(\RR^3,\RR)$,
which is constant outide some ball and such that $\zeta\,\VCt=0$, 
and every $\vp\in\COI$,
\begin{eqnarray}
\lefteqn{\label{ottmar1}
\Big|\SPb{\vp}{\PAV\,\zeta\,\DAVC\,\zeta\,\PAV\,\vp}
\,-\,\SPb{\vp}{\PA\,\zeta\,\DA\,\zeta\,\PA\,\vp}
\Big|
}
\\
&\klg&\nonumber
\big(\|\zeta\,V\|+\|\nabla\zeta\|_\infty\big)
\inf_{0<\ve\klg1}
\Big\{\ve\,\SPb{\vp}{\PA\,\zeta\,|\DA|\,\zeta\,\PA\,\vp}
\,+\,\frac{C}{\ve}\,\|\vp\|^2\Big\}\,,
\end{eqnarray} 
and
\begin{eqnarray}
\lefteqn{\label{ottmar2}
\Big|\SPb{\vp}{\zeta\,\PAV\,\DAVC\,\PAV\,\zeta\,\vp}
\,-\,\SPb{\vp}{\zeta\,\PA\,\DA\,\PA\,\zeta\,\vp}
\Big|
}
\\
&\klg&\nonumber
\big(\|\zeta\,V\|+\|\nabla\zeta\|_\infty\big)\inf_{0<\ve\klg1}
\Big\{\ve\,\SPb{\vp}{\zeta\,\PA\,\DA\,\PA\,\zeta\,\vp}
\,+\,\frac{C}{\ve}\,\|\vp\|^2\Big\}\,.
\end{eqnarray} 
The last estimate still holds true (with a new constant $C$),
if $\DA$ and $\DAVC$ are replaced by $\DA-1$ and $\DAVC-1$,
respectively.
\end{corollary}

\proof
Let $\vp\in\COI$
and let $\theta_R$ be the cut-off function constructed in the
paragraph preceeding \eqref{conv-PA-PAV}.
On account of Lemma~\ref{cor-H1/2} we know
that $\PAV\,\vp\in H^{1/2}$ and, hence,
we infer from Lemma~\ref{le-BdMetc}(c) that 
$\theta_R\,\zeta\,\PAV\,\vp \in H_c^{1/2}$ belongs to the domain of $\DA$.
Applying also the formula appearing in (ii) of Lemma~\ref{le-BdMetc-0}
and using $\zeta\,\VCt=0$
we obtain
\begin{eqnarray*}
\SPb{\vp}{\PAV\,\zeta\,\D{A,\VCt}\,\zeta\,\PAV\,\vp}
&=&
\lim_{R\to\infty}
\SPb{\theta_R\,\zeta\,\PAV\,\vp}{\D{A,\VCt}\,\zeta\,\PAV\,\vp}
\\
&=&
\lim_{R\to\infty}
\SPb{\DA\,\theta_R\,\zeta\,\PAV\,\vp}{\zeta\,\PAV\,\vp}\,.
\end{eqnarray*}
Writing $\delta\PP:=\PAV-\PA$ 
we further get
\begin{eqnarray*}
\lefteqn{
\SPb{\DA\,\theta_R\,\zeta\,\PAV\,\vp}{\zeta\,\PAV\,\vp}
}
\\
&=&
\SPb{\DA\,\theta_R\,\zeta\,\PA\,\vp}{\zeta\,\PA\,\vp}
\,+\,
\SPb{\DA\,\theta_R\,\zeta\,\PA\,\vp}{\zeta\,\delta\PP\,\vp}
\\
& &
\;+\;
\SPb{\theta_R\,\zeta\,\delta\PP\,\vp}{\DA\,\zeta\,\PA\,\vp}
\,+\,
\SPb{\DA\,\theta_R\,\zeta\,\delta\PP\,\vp}{\zeta\,\delta\PP\,\vp}\,.
\end{eqnarray*}
By virtue of Lemma~\ref{le-PA-PAV} we know that
$\zeta\,\delta\PP\,\vp\in\dom(|\DA|^{1/2})$ and it is easy to see
that $\DA\,\theta_R\,\zeta\,\PA\,\vp\to\DA\,\zeta\,\PA\,\vp$,
as $R\to\infty$.
Using also \eqref{conv-PA-PAV} we arrive at
\begin{eqnarray*}
\lefteqn{
\Big|
\SPb{\vp}{\PAV\,\zeta\,\D{A,\VCt}\,\zeta\,\PAV\,\vp}
\,-\,\SPb{\DA\,\zeta\,\PA\,\vp}{\zeta\,\PA\,\vp}\Big|
}
\\
&\klg&
2\,\big\|\,|\DA|^{1/2}\,\zeta\,\PA\,\vp\,\big\|
\,\big\|\,|\DA|^{1/2}\,\zeta\,\delta\PP\,\vp\,\big\|
\,+\,
\big\|\,|\DA|^{1/2}\,\zeta\,\delta\PP\,\vp\,\big\|^2\,.
\end{eqnarray*}
Therefore, we obtain \eqref{ottmar1} by applying
Lemma~\ref{le-PA-PAV} once again.
\eqref{ottmar2} follows from a straightforward
combination of \eqref{ottmar1} and Corollary~\ref{cor-ruth}.
The last statement of Corollary~\ref{cor-PA-PAV} 
follows from \eqref{ottmar2} and Lemma~\ref{le-PA-PAV}.
\qed


\section{Exponential localization}
\label{sec-exp-loc}

In this section we prove
Theorem~\ref{main-thm-exp}.
To this end we adapt an argument from \cite{BFS1}
and some useful improvements of the latter from \cite{Gr}
to our non-local situation. 
In the proof below we present the general strategy of the argument.
In doing so we refer to three technical lemmata
whose proofs are postponed to the end of this section.
Throughout this section
we always assume that the assumptions
of Theorem~\ref{main-thm-exp} are fulfilled.

\bigskip

\noindent
{\it Proof of Theorem~\ref{main-thm-exp}: }
Since $\HH_N$ is
bounded from below we may suppose that $\inf I>-\infty$. 
By assumption we have $\sup I<\Th_{N-1}+1$. 
Moreover, we consider $\HH_N$ as an operator on 
the unprojected $N$-particle space $\HR_N$.
In this case we have to keep in mind that
$0$ becomes an infinitely degenerated eigenvalue of $\HH_N$. 
Our goal is to show that there are $b,C\in(0,\infty)$
such that $\int_{\RR^{3N}}e^{2b|X|}|\Phi(X)|^2dX\klg C$, for
all normalized $\Phi\in\Ran(E_I(\HH_N))$ such that $\Phi=\cA_N\,\Phi$
and $\Phi=\PAVN\,\Phi$.
Borrowing an idea from \cite{Mo} we simplify the problem
by using the bounds
$$
e^{2b|X|}\,\klg\, \max_{j= 1, \dots, N}
e^{2b\sqrt{N}|x_j|}\,\klg\,
 \sum_{j=1}^Ne^{2b\sqrt{N}|x_j|}\,,
\qquad X=(x_1,\ldots,x_N)\in(\RR^3)^N\,,
$$
and the anti-symmetry of $\Phi=\cA_N\,\Phi$.
(We are not aiming to derive good estimates on the decay rate here.)
Indeed,
it suffices to show that there exist $a,C'\in(0,\infty)$ such that
\begin{equation}\label{exp-est-Phi}
\int_{\RR^{3N}}e^{2a|x_N|}\,\big|\Phi(X)\big|^2
\,dX\,\klg\, C\,,
\end{equation}
for all $\Phi=\cA_N\,\Phi=\PAVN\,\Phi\in E_I(\HH_N)$, $\|\Phi\|=1$.
Then 
Theorem~\ref{main-thm-exp} holds true with $b=a/\sqrt{N}$.
Furthermore, it suffices to show that
\eqref{exp-est-Phi} holds true with $a|x_N|$ replaced by $F(x_N)$,
for every (bounded) $F:\RR^3\to\RR$ satisfying \eqref{hyp-F}.
This is in fact an obvious consequence of the monotone convergence
theorem applied to the integrals 
$\int_{\RR^{3N}}e^{2F_n(x_N)}\big|\Phi(X)\big|^2\,dX$ with
$\Phi$ as above, where
$F_1,F_2,\ldots$ is a suitable increasing sequence
of functions satisfying \eqref{hyp-F} and converging to $a|x_N|$.
Therefore, it suffices to find some $a>0$ such that
\begin{equation}\label{main-est-PP}
\big\|\,(\cA_{N-1}\otimes e^{F})\,E_I(\HH_N)\,(\cA_{N-1}\otimes\id)\,
\PAVN\,\big\|\,<\,\infty\,,
\end{equation}
for every $F$ satisfying \eqref{hyp-F},
where $\cA_{N-1}$ denotes anti-symmetrization of the first $N-1$
variables and $e^F$ 
acts only on the
$N^{\mathrm{th}}$ electron variable.

We start by introducing a comparison operator. To this end we  
pick some $\chi\in C^\infty(\RR^3,[0,1])$ such that
$\chi\equiv1$ outside $\ball{2}{0}$ and $\chi\equiv0$
on $\ball{1}{0}$ and set $\chi_R:=\chi(\,\cdot\,/R)$
and $\ol{\chi}_R:=1-\chi_R$, for $R\grg1$. 
Furthermore, we define orthogonal projections
\begin{eqnarray*}
P_{N-1}&:=&\cA_{N-1}\,\PAVNme\
\,,\qquad\, P_{N-1}^\bot\,=\,\id_{\HR_{N-1}}\,-\,P_{N-1}\,,
\\
Q_N&:=&(\cA_{N-1}\otimes\id)\,\PAVN\,,\quad Q_N^\bot\,=\,
\id_{\HR_N}\,-\,Q_N\,.
\end{eqnarray*}
Then the comparison operator is defined, \`a-priori 
on the
domain $\core\subset\HR_N$, by
\begin{eqnarray}\nonumber
\wt{\HH}_{N}&:=&Q_N\,\HH_N\,Q_N
\,+\,
\HH_{N-1}^\cA\otimes\PAVm\,+\,\Th_{N-1}
\,P_{N-1}^\bot\otimes\id
\\
& & \nonumber
+\,P_{N-1}\otimes\,\PAV\,(1-\Th_1)\,\ol{\chi}_R\,\PAV\,+\,Q_N^\bot
\\
&=&
\HH_{N-1}^\cA\otimes\id +\Th_{N-1}\,
P_{N-1}^\bot\otimes\id
\label{ana100a}
\\
& &
+\; P_{N-1}\otimes 
\PAV\,\big\{\,
\DAVC\,+\,(1-\Th_1)\,\ol{\chi}_R\,\big\}\,\PAV\label{ana100b}
\,+\,Q_N^\bot
\\
& &
+\,\sum_{i=1}^{N-1}Q_N\,
W_{iN}\,Q_N\,.\label{ana100c}
\end{eqnarray}
We denote the Friedrichs extension of $\wt{\HH}_{N}$
again by the same symbol. Notice
that on $\core$ we have $\HH_{N-1}^\cA\otimes\id +\Th_{N-1}\,
P_{N-1}^\bot\otimes\id\grg \Th_{N-1}\,\id_{\HR_N}$. Furthermore,
Lemma~\ref{piola} below this implies that
\begin{equation*}
P_{N-1}\otimes \PAV\,\big\{\,
\DAVC\,+\,(1-\Th_1)\,\ol{\chi}_R\,\big\}\,\PAV\,+\,Q_N^\bot
\,\grg\, \id\,-\,o(1)\,P_{N-1}\otimes\id\,,
\end{equation*}
as $R$ tends to infinity.
We now pick some $\ve>0$ with
$\sup I<\Th_{N-1}+1-\ve$. Then the above remarks imply
\begin{equation}\label{ina1}
\SPn{\psi}{\wt{\HH}_{N}\,\psi}\,\grg\,(\Th_{N-1}+1-\ve/2)\,\|\psi\|^2\,,
\qquad \psi\in\core\,,
\end{equation}
for all sufficiently large $R\grg1$.
Next, we define  
$$
\HH_N^\prime
\,:=\,Q_N\,\HH_N\,Q_N+\HH_{N-1}^\cA\otimes\PAVm\,.
$$ 
Then 
$\wt{\HH}_N$ and $\HH_N^\prime$ have the same domain 
since they differ by a
bounded operator on their common form core $\core$. 
We further pick some $\wt{\chi}_I\in C_0^\infty(\RR,[0,1])$,
such that $\wt{\chi}_I\equiv1$ on $I$ and
$
\supp(\wt{\chi}_I)\,\subset\,(-\infty, \Th_{N-1}+1-\ve)
$.
Then $\wt{\chi}_I(\wt{\HH}_N)=0$ by \eqref{ina1} 
and we observe that
\begin{equation}\label{ina1a}
Q_N\,E_I(\HH_N)\,Q_N\,=\,Q_N\,E_I(\HH_N')\,Q_N
\,=\,\big(\wt{\chi}_I(\HH_N')\,-\,\wt{\chi}_I(\wt{\HH}_{N})\big)\,Q_N\,.
\end{equation}
We preserve the symbol $\wt{\chi}_I$ to denote
an almost
analytic extension of $\wt{\chi}_I$
(see, e.g., \cite{DiSj}) to a smooth, compactly supported
function on the complex plane
satisfying
\begin{equation}
\label{frieda}
\begin{split}
&\supp(\wt{\chi}_I)\,\subset\,(-\infty, \Th_{N-1}+1-\ve)\,
+\,i(-\delta, \delta)\,,
\\
& \partial_{\ol{z}}\wt{\chi}_I(z)\,=\,\bigO_N\big(|\Im z|^N\big)\,,
\quad N\in\NN\,,
\end{split}
\end{equation}
where $\partial_{\ol{z}}=\frac{1}{2}(\partial_{\Re z}+i\partial_{\Im z})$.
Here we may choose $\delta>0$ as small as we please.
We shall apply the Helffer-Sj\"ostrand formula (see, e.g., \cite{DiSj}),
\begin{equation*}\label{HS-for}
\wt{\chi}_I(T)\,=\,\int_\CC\,(z-T)^{-1}\,d\wt{\chi}_I(z),\qquad
d\wt{\chi}_I(z)\,:=\,
\frac{i}{2\pi}\,\partial_{\ol{z}}\wt{\chi}_I(z)\,dz\wedge d\ol{z},
\end{equation*}
which holds for every self-adjoint operator $T$ on some Hilbert space.
By means of  (\ref{ina1a}) we then find the representation
\begin{equation}\label{ina2}
Q_N\,E_I(\HH_N)\,Q_N
\,=\,\int_\CC\big[(\HH_N^\prime-z)^{-1}-(\wt{\HH}_N-z)^{-1}\big]\,
d\wt{\chi}_I(z)\,Q_N\,.
\end{equation}
For some $F$ as
in \eqref{hyp-F} (which acts only on the last variable
in what follows) we abbreviate
$$
\PAVFN\,:=\,e^F\,\PAVN\, e^{-F}\,.
$$ 
Then \eqref{ina2} and 
the second
resolvent identity together with the trivial identities
$Q_N^\bot\,Q_N=0=(P_{N-1}^\bot\otimes\id)\,Q_N$
yield
\begin{eqnarray}
\lefteqn{
\big\|\,(\cA_{N-1}\otimes e^F)\,E_I(\HH_N)\,
(\cA_{N-1}\otimes\id)\,\PAVN\,\big\|\nonumber
}
\\
&\klg&\nonumber
\int_\CC
\Big\|
e^F\,(\wt{\HH}_N-z)^{-1}\,P_{N-1}\otimes\big\{\PAV\,
(1-\Th_1)\,\overline{\chi}_R\,\PAV\big\}\times
\\
& &\qquad\qquad\nonumber
\times\,Q_N\,(\HH_N^\prime-z)^{-1}\,Q_N\,
\Big\|\,|d\wt{\chi}_I(z)|
\\\nonumber
&\klg& (1-\Th_1)\int_\CC
\big\|\,e^F(\wt{\HH}_N-z)^{-1}e^{-F}\,\big\|\,\big\|\,\PAVFN\,\big\|
\,\big\|\,e^F\,\ol{\chi}_R\,\big\|\,\frac{|d\wt{\chi}_I(z)|}{|\Im z|}
\\
&\klg&
C_{a,R}\int_\CC \label{mona3}
\big\|\,e^F(\wt{\HH}_N-z)^{-1}e^{-F}\,\big\|\,
\frac{|d\wt{\chi}_I(z)|}{|\Im z|}.
\end{eqnarray}
In the last step we apply
Proposition \ref{prop-lisa} and
$\|\,e^F\,\ol{\chi}_R\,\|\klg e^{2aR}$. 
By \eqref{frieda} $|d\wt{\chi}_I(z)|/|\Im z|$ is a finite measure.
To conclude the proof of Theorem~\ref{main-thm-exp}
it thus remains to show that the norm of
$e^F(\wt{\HH}_N-z)^{-1}e^{-F}$ is uniformly  bounded in 
all $z\in\supp(\wt{\chi}_I)\setminus\RR$ 
and $F$ satisfying \eqref{hyp-F}. This is done in the 
rest of this proof.

Since $F$ satisfies \eqref{hyp-F} we know that
$\id_{N-1}\otimes e^F$ is an isomorphism on $\HR_N$.
Therefore, the densely defined operators 
$e^F\,\wt{\HH}_N\,e^{-F}$ and $\wt{\HH}_N$
have the same resolvent set and 
\begin{equation}\label{def-RF}
\sR_F(z)\,:=\,
e^F\,(\wt{\HH}_N-z)^{-1}\,e^{-F}=(e^F\wt{\HH}_Ne^{-F}-z)^{-1},
\qquad z\in\vr(\wt{\HH}_N)
\,.
\end{equation}
In particular, 
$e^F\wt{\HH}_N\,e^{-F}$ is closed because its resovent set is not
empty. 
Using the identity ${\sR_F(z)^{*}}^{-1}={\sR_F(z)^{-1}}^*$
we readily verify that 
$(e^F\,\wt{\HH}_N\,e^{-F})^*=e^{-F}\,\wt{H}_N\,e^F$.
Since
$e^{\pm F}$ maps $\core$ into itself we further have
\begin{equation}\label{mona4}
\core\subset\dom(e^{\pm F}\wt{\HH}_Ne^{\mp F})=e^{\pm F}
\dom(\wt{\HH}_N)\subset
e^{\pm F}\formd(\wt{\HH}_N)\,.
\end{equation} 
The following two lemmata, whose proofs
are postponed 
to the end of this section, show that
$e^F\wt{\HH}_Ne^{-F}$ is a small
form perturbation of $\wt{\HH}_N$. We define $T:\core\to\HR_N$ by
\begin{equation}
  \label{mona5}
T\,\vp\,:=\,e^F\,\wt{\HH}_N\,e^{-F}\,\varphi\,-\,\wt{\HH}_N\,\varphi\,,
\qquad \vp\in \core\,.
\end{equation}

\begin{lemma}\label{identity}
Assume that $F:\RR^3\to\RR$ satisfies \eqref{hyp-F}.
Then we have, as $a>0$ tends to zero,
\begin{equation}
  \label{mona6}
|\SPn{\varphi}{T\,\varphi}|\,\klg\, a\,\SPn{\varphi}{\wt{\HH}_N\,\varphi}+
\bigO(a)\,\SPn{\varphi}{\varphi}\,,
\qquad \vp\in \core\,.
\end{equation}
\end{lemma}

\begin{lemma}\label{conservation-formdomain}
There exist constants $c_1,c_2\in(0,\infty)$ 
such that, for all $F:\RR^3\to\RR$ satisfying \eqref{hyp-F}
and all $\varphi \in \core$, 
\begin{equation}
  \label{mona7}
\big|\SPb{e^{\pm F}\,\varphi}{\wt{\HH}_N\,e^{\pm F}\,\varphi}\big|\,\klg\,
c_1\,\|e^{\pm F}\|^2\,
\SPn{\varphi}{\wt{\HH}_N\,\varphi}+c_2\,\|e^{\pm F}\|^2\,\|\vp\|^2\,.
\end{equation}
In particular, $e^{\pm F}\formd(\wt{\HH}_N)\subset\formd(\wt{\HH}_N)$.
\end{lemma}

If $a<1/2$ then
Lemma \ref{identity} implies that 
$\big(e^F\,\wt{\HH}_N\,e^{-F}\big)\!\!\upharpoonright_{\core}$ 
has a distinguished, sectorial, closed extension, $\wt{\HH}_{N}^F$, 
that is
the only closed extension having the
properties $\dom(\wt{\HH}_{N}^F)\subset \formd(\wt{\HH}_{N})$,
$\dom(\wt{\HH}_{N}^{F*})\subset \formd(\wt{\HH}_{N})$,
and $i\eta\in\vr(\wt{\HH}_{N}^F)$, for all $\eta\in\RR$ with
sufficiently large absolute value; see
\cite{Kato}. 
Thanks to \eqref{def-RF}, \eqref{mona4}, 
and  Lemma
\ref{conservation-formdomain},
we know
that
$e^F\wt{\HH}_Ne^{-F}$ {\em is} a closed extension enjoying
these properties, whence 
$$
\wt{\HH}_{N}^F\,=\,e^F\,\wt{\HH}_N\,e^{-F}\,.
$$
We are now prepared to derive a uniform bound
on the norm under the integral sign in \eqref{mona3}. 
For $z\in \supp(\wt{\chi}_I)$ and 
$\varphi\in \core$, we obtain
\begin{equation}
  \label{mona8}
\begin{split}
\Re\,\SPb{\varphi}{(\wt{\HH}_{N}^F-z)\,\varphi}\,&=\,
\SPb{\varphi}{(\wt{\HH}_{N}-\Re z)\,\varphi}\,+\,
\Re\SPn{\varphi}{T\,\varphi}
\\
&\grg\,(1-a)\,\SPB{\varphi}{\Big(\wt{\HH}_N-\frac{\Re z}{1-a}
\Big)\,\varphi} \,-\,\bigO(a)\,\|\,\varphi\,\|^2\,.
\end{split}
\end{equation}
By \eqref{ina1} and \eqref{frieda} we thus find 
$a\in(0,1/2)$ and $R\in[1,\infty)$ 
such that, for all $z\in \supp(\wt{\chi}_I)$ and $\vp\in\core$, 
$$
\Re\,\SPb{\varphi}{(\wt{\HH}_{N}^F-z)\,\varphi}\,
\grg \, \frac{\ve}{4}\,\|\vp\|^2\,.
$$
This inequality implies that,
for $z\in\supp(\wt{\chi}_I)$, the numerical range of $\wt{\HH}_{N}^F-z$
is contained in the half space $\{\zeta\in\CC:\,\Re \zeta\grg\ve/4\}$
\cite[Theorem~VI.1.18 and Corollary~VI.2.3]{Kato}.
Moreover, by \eqref{def-RF} the deficiency of $\wt{\HH}_{N}^F-z$ is zero,
for all $z\in\CC\setminus\RR$, and we may hence estimate the norm
of $(\wt{\HH}_{N}^F-z)^{-1}$ by the inverse distance of $z$
to the numerical range of $\wt{\HH}_{N}^F$ \cite[Theorem~V.3.2]{Kato}.
We thus arrive at 
$$
\big\|\,(\wt{\HH}_{N}^F-z)^{-1}\,\big\|\,\klg\,\frac{4}{\ve}\,,
\qquad z\in \supp(\wt{\chi}_I)\,,
$$
which together with \eqref{mona3} proves Theorem~\ref{main-thm-exp}.
\qed

\bigskip

\begin{lemma}\label{piola}
For every sufficiently large $R\grg1$, there is some $c_R\in(0,\infty)$
such that $c_R\to0$, as $R\to\infty$, and, for all
$\varphi\in \COI$,
\begin{equation}
  \label{eq:ana9}
\SPb{\varphi}{\PAV\,\big[\DAVC+(1-\Th_1)\,
\ol{\chi}_R\,\big]\,\PAV\,\varphi}\,
\grg\, \big\|\,\PAV\,\varphi\,\big\|^2\,-\,c_R\, \|\vp\|^2. 
\end{equation}
\end{lemma}

\proof
To begin with
we introduce a scaled partition of unity. 
Namely, we pick some $\tilde{\mu}\in C_0^\infty(\RR^3,[0,1])$
such that $\tilde{\mu}\equiv1$ on $\ball{2}{0}$ and observe that
$\theta:=\tilde{\mu}^2+(1-\tilde{\mu})^2$ is strictly positive.
We further
set, for $R\grg1$ and $x\in\RR^3$,
$\mu_1(x)\equiv\mu_{R,1}(x):=\tilde{\mu}(x/R)/\theta^{1/2}(x/R)$,
and 
$\mu_2(x)\equiv\mu_{R,2}(x)
:=(1-\tilde{\mu}(x/R))/\theta^{1/2}(x/R)$, so that $\mu_1^2+\mu_2^2=1$.
Since $\mu_1\,\nabla\mu_1+\mu_2\,\nabla\mu_2=\nabla(\mu_1^2+\mu_2^2)/2=0$
it follows that, for $\vp\in\COI$,
\begin{eqnarray}
\lefteqn{  
\SPn{\varphi}{\PAV\big[\DAVC+(1-\Th_1)
\ol{\chi}_R\big]\PAV\,\varphi}\label{eq:ana10}
}
\\
&=&\sum_{j=1,2}\SPb{\varphi}{\PAV\,\big[\,\mu_j \,\DAVC\, \mu_j\,
+(1-\Th_1)\,\nonumber
\mu_j^2\,\ol{\chi}_R\,\big]\,\PAV\,\varphi}
\,=:\,\sum_{j=1,2}Y_j
\,.
\end{eqnarray}
To treat the summand with $j=1$ we use that, by construction,
$\mu_1\ol{\chi}_R=\mu_1$, for every $R\grg1$. 
Taking also Corollary~\ref{cor-ruth} and \eqref{def-EA}
into account
we find, for all $R\grg1$ and $\vp\in \COI$,
\begin{eqnarray}
\label{eq:anna13}
Y_1
&\grg&(1-1/R)\,\SPb{\mu_1\,\varphi}{\PAV\,\big[\DAVC-\Th_1\big]
\,\PAV\,\mu_1\,\vp}
\\
& &\;
\,+\,\|\mu_1\,\PAV\,\varphi\|^2\,-\,\nonumber
\bigO(1/R)\,\|\,\varphi\,\|^2
\\
& \grg& \|\mu_1\,\PAV\,\varphi\|^2\,-\,\nonumber
\bigO(1/R)\,\|\,\varphi\,\|^2\,.
\end{eqnarray}
We next turn to
the summand with $j=2$ in \eqref{eq:ana10}
where $\mu_2^2\,\ol{\chi}_R=0$. 
Applying successively Corollaries~\ref{cor-ruth}
and~\ref{cor-PA-PAV}, Proposition~\ref{prop-lisa},
and Lemma~\ref{le-PA-PAV} we deduce that, for all
$\vp\in\COI$ and every $\ve>0$, 
\begin{eqnarray}
\SPb{\varphi}{\PAV \,\mu_2\, \DAVC\, \mu_2\,\PAV\,\varphi}
&\grg&(1-\ve)\,\SPb{\varphi}{\mu_2\,\PA\, \DA\, \PA\,\mu_2\,\varphi}
\,-\,o_\ve(1)\,\|\vp\|^2\nonumber
\\
&\grg&(1-\ve)\,\big\|\,\PA\, \mu_2\, \varphi\,\big\|^2\,
-\,o_\ve(1)\,\|\vp\|^2\nonumber
\\\nonumber
&\grg&(1-\ve)^2\,\big\|\,\mu_2\, \PA\, \varphi\,\big\|^2\,
-\,o_\ve(1)\,\|\vp\|^2
\\
&\grg&(1-\ve)^3\,\big\|\,\mu_2\, \PAV\, \varphi\,\big\|^2\,
-\,o_\ve(1)\,\|\vp\|^2\,,\label{eq:ana99}
\end{eqnarray}
as $R\to\infty$.
We conclude by combining \eqref{eq:ana10}-\eqref{eq:ana99}
and using $\mu_1^2+\mu_2^2=1$.
\qed

\bigskip

\noindent
{\it Proof of Lemma \ref{identity}: }  
We have to study the contribution to
$T=e^F\wt{\HH}_Ne^{-F}-\wt{\HH}_N$
coming from each term
in \eqref{ana100a}-\eqref{ana100c}.
The terms in \eqref{ana100a} commute with $e^F$
and hence give no contribution. 
In order to estimate the contribution coming from the left
term in \eqref{ana100b} we first observe that 
Corollary~\ref{cor-eFPAVe-F}
implies the following identity on $\COI$,
\begin{eqnarray}
e^F\,\PAV\, \DAVC\, \PAV\, e^{-F}&=&\PAVF\,
\big( \DAVC+i\alpha\cdot\nabla F\big)\,\PAVF\nonumber
\\
&=&\PAVF\,\DAVC\,\PAVF\,+\,\bigO(a)\,.\label{solvej1}
\end{eqnarray}
The term in \eqref{ana100b}  involving 
the cut-off function $\ol{\chi}_R$ yields a contribution
of order $\bigO(a)$, too, due to
Corollary~\ref{cor-ruth}.
To account for the projection on the right in \eqref{ana100b}
we write $Q_N^\bot=\id_{\HR_N}-P_{N-1}\otimes\PAV$
and use Proposition~\ref{prop-lisa} to obtain
$
\big\|Q_N^\bot-e^F\,Q_N^\bot\, e^{-F}\big\|
=
\bigO(a)
$.
Finally, we apply Corollary~\ref{cor-ruth} to \eqref{solvej1}
and Corollary~\ref{cor-ruth-W} to all terms in \eqref{ana100c}
-- this is the only place in this section where we use the assumption
that $B$ is bounded --
and arrive at
$$
\big|\SPn{\vp}{T\,\vp}\big|\,\klg\,
a\,\SPB{\vp}{Q_N\Big\{\DAVC^{(N)}+\sum_{i=1}^{N-1}W_{iN}\Big\}\,Q_N\,\vp}
\,+\,\bigO(a)\,\|\vp\|^2\,.
$$
Since $\HH_{N-1}^\cA\grg\Th_{N-1}$ this completes
the proof of Lemma~\ref{identity}.
\qed

\bigskip

\noindent
{\it Proof of Lemma~\ref{conservation-formdomain}: } 
We drop the $\pm$-signs in
\eqref{mona7} since the they
do not play any role in this proof. It is clear that
we only have to comment on those terms
in \eqref{ana100a}-\eqref{ana100c} that involve unbounded operators.  
Since $\HH_{N-1}\otimes
\id$ commutes with $e^F$ 
and since $\HH_{N-1}\grg\Th_{N-1}$
we first find, for $\vp\in\core$,
\begin{equation}
  \label{mona12}
\begin{split}
\SPb{e^F\varphi}{&\big(\HH_{N-1}\otimes \id\,-\,\Th_{N-1}
\,P_{N-1}^\bot\otimes\id\big)\,e^F\,\varphi}
\\
&\klg\,
\|\,e^{2F}\,\|\,\SPb{\varphi}{\big(\HH_{N-1}\otimes \id
\,-\,\Th_{N-1}\,P_{N-1}^\bot\otimes\id\big)\,\varphi}\,.
\end{split}
\end{equation}
By virtue of Proposition~\ref{prop-W1/2}
we can estimate
$|\SPn{\varphi}{e^F\,Q_N W_{iN}\,Q_N e^F\varphi}|$, for $\vp\in\core$, as
\begin{equation}
  \label{mona13}
\begin{split}
\big\|\,W_{iN}^{1/2}\,Q_N\, e^F\,\varphi\,\big\|^2&\,\klg\,
2\,\big\| \,e^F\,W_{iN}^{1/2}\,Q_N\,\varphi\,\big\|^2+
2\,\big\| \,W_{iN}^{1/2}\,[e^F,Q_N]\,e^{-F}\,e^F\,\varphi\,\big\|^2
\\
&\,\klg\,2\,\|\,e^{F}\,\|^2\,\big\|\,W_{iN}^{1/2}\,Q_N\,\varphi\,\big\|^2
+\bigO(a^2)\,\|\,e^{F}\,\|^2\,\|\vp\|^2\,.
\end{split}
\end{equation}
(If $B$ is unbounded, then the $\bigO$-symbol in \eqref{mona13} depends
on the supremum of $|B|$ on $\supp(\nabla F)$.)
It remains to prove that there are constants $c_3,c_4\in(0,\infty)$ such that
\begin{equation}
  \label{vero1}
\begin{split}
\SPb{\varphi}{e^F\,\PAV\, \DAVC \,\PAV\,&
e^F\varphi}
\\
\klg\,c_3\,\|e^F\|^2\,&
\SPb{\varphi}{\PAV\, \DAVC \,\PAV\, \varphi}\,+\,c_4\,\|e^F\|^2\,
\|\vp\|^2,
\end{split}
\end{equation}
for $\vp\in\COI$. Moreover, since $\VM$ and $\VE$ are bounded
it suffices to prove this estimate
with $\DAVC$ replaced by $\wt{D}:=\DAV-e_0$, which is
positive on the range of $\PAV$. 
We abbreviate $\PPpm:=\PAVpm$ in the rest of this proof
and 
we seek for bounds on both terms on the right side of
\begin{equation}\label{vero2}
\big\|\,\big(\wt{D}\, \PP\,\big)^{1/2}\, e^F\,\varphi\,\big\|^2
\,\klg\,
2\sum_{\sharp=\pm}
\big\|\,\big(\wt{D}\, \PP\,\big)^{1/2}\, e^F\,\PPsharp\,\varphi\,\big\|^2
\,.
\end{equation}
Here the 
norm with $\sharp=-$
equals $\|\big(\wt{D} \PP\big)^{1/2}[ e^F,\Lambda^-]\,\varphi\|$ 
and is not greater than some
$\bigO(a)\,\|e^F\|\,\|\vp\|$
due to Proposition~\ref{prop-lisa}.
We next define
$$
\hat{D}\,:=\,\PP\,\big(\DAV-e_0\,\big)\,\PP\,+\,
\id\,=\,\PP\,\wt{D}\,\PP\,+\,\id\,\grg\, \id\,.
$$
In fact, because of
$\big\|(\wt{D}\PP)^{1/2}\,\hat{D}^{-1/2}\big\|\klg1$
and
 \begin{equation*}
  \label{mona14}
\begin{split}
\big\|\, \hat{D}^{1/2}\, e^F\,\PP\,\varphi\,\big\|^2&\,\klg\,
\big\|\, e^F\,\hat{D}^{1/2}\,\PP\,\varphi\,\big\|^2\,+\,
\big\|\, [\hat{D}^{1/2}, e^F]\,\PP\,\varphi\,\big\|^2
\\
&\,\klg\,\|\, e^{F}\,\|^2\,\big\|\,\PP\,\hat{D}^{1/2}\,\varphi\,\big\|^2
\,+\,
\big\|\,\hat{D}^{1/2}\, [\hat{D}^{-1/2}, e^F]\,\PP\,\hat{D}^{1/2}\,\varphi
\,\big\|^2
\end{split}
 \end{equation*}
we shall see that \eqref{vero1} holds true
as soon as we have shown that 
\begin{equation}\label{albert} 
\big\|\,\hat{D}^{1/2}\,[\hat{D}^{-1/2}, e^F]\,\PP\,\big\|\,=\,
\bigO(a)\,\|e^F\|\,.
\end{equation}
To check whether \eqref{albert} is correct
we first note that, on $\COI$,
\begin{eqnarray}\nonumber
[\hat{D}, e^F]&=&\PP\,[\wt{D}, e^F]\,+\,
[\PP,e^F]\,\wt{D}
\\
&=&\label{mona15}
-\PP\,i\alpha\cdot\nabla F\,e^F
\,+\,\big([\VE\,,\,e^F]\,e^{-F}\big)\,e^F
\,+\,[\PP,e^F]\,\wt{D}
\,.
\end{eqnarray}
We apply the 
norm-convergent integral representation
\begin{equation}\label{int-sqrt}
T^{-1/2}\,=\,\frac{1}{\pi}\int_0^\infty\frac{1}{T+t}\,\frac{dt}{\sqrt{t}}\,,
\end{equation}
which holds for any strictly positive
operator, $T$, on some Hilbert space.
For $\phi,\psi \in \COI$, it implies
\begin{equation}
  \label{mona17}
\begin{split}
\SPb{\hat{D}^{1/2}\,\phi}{ [\hat{D}^{-1/2}, e^F]\,\PP \,\psi}
\,=\,\frac{1}{\pi}
\int_0^\infty \SPB{\hat{D}^{1/2}\,\phi}{
\frac{-1}{\hat{D}+t}\,[\hat{D}, e^F]\,
\frac{\PP}{\hat{D}+t}
\,\psi}\frac{dt}{\sqrt{t}}\,.
\end{split}
\end{equation}
We estimate
the contribution of the first term 
on the right side of \eqref{mona15} to \eqref{mona17}
as 
\begin{equation}
  \label{mona20}
\Big|\SPB{\hat{D}^{1/2}\,\phi}{\frac{\PP}{\hat{D}+t}
\,i\alpha\cdot\nabla F\,e^F\,\frac{\PP}{\hat{D}+t}\,\psi}\Big|
\klg\,\frac{\bigO(a)\,\|e^F\|}{(1+t)^{3/2}}\,\|\phi\|\,\|\psi\|\,,
\qquad t\grg0\, .
\end{equation}
In view of
\eqref{bd-VE2} the second term in \eqref{mona15} can be dealt with
similarly.
To account for
the second term  in \eqref{mona15} we apply
Proposition \ref{prop-lisa} and obtain, for $t\grg0$, 
\begin{equation}\label{mona21}
\Big|\SPB{\phi}{\frac{1}{\hat{D}+t}\,\big\{\,\hat{D}^{1/2}\,[\PP, e^F]\,
e^{-F}\,\big\}\,e^F
\,\frac{\wt{D}}{\hat{D}+t}\,\PP\,\psi}\Big|\klg\,
\frac{\bigO(a)\,\|e^F\|}{1+t}\,
\|\phi\|\,\|\psi\|\,.
\end{equation}
Equations \eqref{mona17}-\eqref{mona21}  
show that \eqref{albert} holds true, 
which completes the proof of Lemma~\ref{conservation-formdomain}.
\qed


\section{The lower bound on $\inf\specess(\HH_N)$}
\label{ssec-lower-bound}

In order to prove the `hard part' of the HVZ theorem,
Theorem~\ref{main-thm-HVZ}(ii),
we employ an idea we learned from \cite{Gr}: One may
use a localization estimate for spectral projections to prove their
compactness.
Adapted to our non-local model the 
argument looks as follows:

\begin{theorem}\label{thm-EI-compact}
Let the assumptions of Theorem~\ref{main-thm-HVZ}(ii)
be fulfilled and
let $I\subset \RR$ be an interval $\sup I<1+\Th_{N-1}$.
Then the spectral projection $E_I(\HH_N^\cA)$
is a compact operator on $\cA_N\HR_N^+$.
In particular,
$$
\specess(\HH_N^\cA)\,\subset\,[1+\Th_{N-1}\,,\,\infty)\,.
$$
\end{theorem}

\proof
Let $g\in C(\RR,(0,\infty))$ satisfy $g(r)\to\infty$, $r\to\infty$,
and $g(|X|)\,E_I(\HH_N^\cA)\in\LO(\cA_N\HR_N^+)$ and set $h:=1/g$.
We 
let $\vr_R$ denote a smoothed characteristic function of the closed ball
in $\RR^{3}$ with radius $R>0$ and center $0$ and set
$\chi_R(X):=\vr_R(x_1)\cdots\vr_R(x_N)$, for
$X=(x_1,\dots,x_N)\in(\RR^3)^N$. First, we argue that
it suffices to show that $E_I(\HH_N^\cA)\,\chi_R\,(-\Delta+1)^{1/8}$
is a (densely defined)
bounded operator from $\HR_N$ to $\cA_N\,\HR^+_N$. 
In fact, let us assume that this is the case.
Since
$(-\Delta+1)^{-1/8}\,h(|X|)$ is compact and
$g(|X|)\,E_I(\HH_N^\cA)$ is bounded,
it then follows that
\begin{eqnarray*}
\lefteqn{
E_I(\HH_N^\cA)\,\big[\chi_R\,h(|X|)\big]\,
g(|X|)\,E_I(\HH_N^\cA)
}
\\
&=&
E_I(\HH_N^\cA)\,\chi_R\,(-\Delta+1)^{1/8}
\,\big[(-\Delta+1)^{-1/8}\,h(|X|)\big]\,
g(|X|)\,E_I(\HH_N^\cA)
\end{eqnarray*}
is compact.
Since $\chi_R\,h(|X|)$ converges to $h(|X|)$
in the operator norm, as $R$ tends to infinity, 
it further follows that
$E_I(\HH_N^\cA)=E_I(\HH_N^\cA)\,h(|X|)\,
g(|X|)\,E_I(\HH_N^\cA)$ is compact, too.

To verify that $E_I(\HH_N^\cA)\,\chi_R\,(-\Delta+1)^{1/8}$
is bounded we set
$
S:=1+\sum_{j=1}^N|\DAV^{(j)}|
$
and write, for some sufficiently large $c>0$,
\begin{eqnarray}
\lefteqn{\nonumber
E_I(\HH_N^\cA)\,\chi_R\,(-\Delta+1)^{1/8}\;=\;
E_I(\HH_N^\cA)\,(\HH_N+c)^{1/2}\times
}
\\
& &\label{tina1}
\times\big\{(\HH_N+c)^{-1/2}\,\PAVN\,S^{1/2}\big\}
\,\big\{S^{-1/2}
\,\chi_R\,(-\Delta+1)^{1/8}\big\}\,.
\end{eqnarray}
Here the left curly bracket in \eqref{tina1} is a bounded
operator from $\HR_N$ to $\HR^+_N$ since
$
\PAVN\,S\,\PAVN\,
\klg\,\HH_N+c
$, provided $c$ is large enough,
due to the positivity of the interaction potentials
and the boundedness of $\VM$ and $\VE$.
To see that the right curly bracket in \eqref{tina1} is a bounded
operator in $\HR_N$ we first notice that
it is a restriction of $S^{-1/2}T^*$, where $T:=(-\Delta+1)^{1/8}\,\chi_R$
is closed. It thus remains to show that
$TS^{-1/2}=T^{**}S^{-1/2}=(S^{-1/2}T^*)^*$ belongs to $\LO(\HR_N)$.
To this end we
recall that
$
(-\Delta^{(i)}+1)^{1/4}\,\vr_R^{(i)}\,\big(|\DAV^{(i)}|+1\big)^{-1}
$
is bounded 
on $L^2(\RR^3_i,\CC^4)$
since $\dom(\DAV)\subset H^{1/2}_\loc(\RR^3,\CC^4)$.
It follows that
$$
(-\Delta^{(i)}+1)^{1/4}\,\chi_R\,
S^{-1}
=
(-\Delta^{(i)}+1)^{1/4}\,\chi_R\,\big(|\DAV^{(i)}|+1\big)^{-1}
\,\big(|\DAV^{(i)}|+1\big)\,
S^{-1}
$$
is bounded, for $i=1,\ldots,N$, and, hence,
$
\chi_R\,(-\Delta+1)^{1/4}\,\chi_R\,S^{-1}\in\LO(\HR_N)
$.
Since $\chi_R\,(-\Delta+1)^{1/4}\,\chi_R$ is a restriction
of $T^*T$ we see that $T^*TS^{-1}\in\LO(\HR_N)$, which
implies $|T|S^{-1/2}\in\LO(\HR_N)$ and, hence, $TS^{-1/2}\in\LO(\HR_N)$.
\qed


\section{Weyl sequences}\label{sec-Weyl}

In this section we prove the `easy part' of our HVZ theorem,
namely Part~(i) of Theorem~\ref{main-thm-HVZ} asserting that
$$
\specess(\HH_N^\cA)\,\supset\,[\,\Th_{N-1}+1\,,\,\infty)\,.
$$
To this end we 
fix some spectral parameter
$\lambda\grg1$ throughout the whole section and
$\{\psi_n\}_{n\in\NN}$ will always denote a
corresponding Weyl sequence
as in Hypothesis~\ref{hyp-Weyl}(i).

In this and the following section we shall repeatedly employ
the following sequence of cut-off functions:
We pick some $\chi\in C^\infty(\RR,[0,1])$ such that
$\chi\equiv0$ on $(-\infty,1-\ve/4]$
and $\chi\equiv1$ on $[1,\infty)$.
Here $\ve\in(0,1)$ is a fixed parameter whose value becomes
important only in Section~\ref{sec-eigenvalues}.
We set $\chi_n:=\chi(|x|/R_n)$, for $x\in\RR^3$
and $n\in\NN$, where $R_n$ is given by Hypothesis~\ref{hyp-Weyl}(i).
Then it holds $\chi_n\,\psi_{n}=\psi_{n}$
and $\|\nabla\chi_n\|_\infty=R_n^{-1}\,\|\nabla\chi\|_\infty\to0$,
as $n\to\infty$.

To begin with we draw two simple conclusions from
our hypotheses:

\begin{lemma}\label{le-ida}
Assume that $A\in L^\infty_\loc(\RR^3,\RR^3)$ 
and $V$
fulfill Hypotheses~\ref{hyp-Weyl}(i)
and \ref{hyp-V-komplett}, respectively.
Then 
\begin{equation}\label{Weyl-DAV}
\lim_{n\to\infty}
\big\|\,\big(\DAV-\lambda\big)\,\psi_n\,\big\|\,=\,0\,,\quad
\lim_{n\to\infty}\big\|\,(\DAVC-\lambda)\PAV\,\psi_n\,\big\|\,=\,0\,.
\end{equation}
\end{lemma}

\proof
The first identity is clear from the hypotheses. To treat the second
we  
employ the cut-off functions defined in the paragraph preceeding
the statement of this lemma and abbreviate
$\VME:=\VM+\VE$. By means of Proposition~\ref{prop-lisa} 
and $\|\VME\,\chi_n\|\to0$ we then obtain
\begin{equation*}
\big\|\VME\,\PAV\,\psi_n\big\|\,\klg\,\big\|\VME\,\chi_n\,\PAV\,\psi_n
\big\|
\,+\,\big\|\VME\,\big[\PAV\,,\,\chi_n\big]\,\psi_n\big\|\,\longrightarrow\,0,
\end{equation*}
as $n$ tends to infinity.
Therefore, the second identity
follows from the first.
\qed

\begin{lemma}\label{le-PApsinto1} 
Assume that $A\in L^\infty_\loc(\RR^3,\RR^3)$
and $V$ fulfill Hypotheses~\ref{hyp-Weyl}(i)
and~\ref{hyp-V-komplett}, respectively.
Let $\ve>0$ and set $I_\ve:=(\lambda-\ve,\lambda+\ve)$.
Then we have, as $n$ tends to infinity,
\begin{equation} \label{heinrich}
\big\|\,E_{I_\ve}(\DAV)\,\psi_n\,\big\|\,\to\,1\,,
\quad\textrm{in particular,}\quad
\big\|\,\PAV\,\psi_n\,\big\|\,\to\,1
\,.
\end{equation}
\end{lemma}

\proof
Clearly, $\big\|\,E_{I_\ve}(\DAV)\,\psi_n\,\big\|\klg1$
since $\psi_n$ is normalized. Suppose that there is some $\delta>0$
such that 
$\liminf\big\|\,E_{I_\ve}(\DAV)\,\psi_n\,\big\|^2\klg 1-\delta$.
Then we have
$\lim_{\ell\to\infty}\big\|\,E_{I_\ve}(\DAV)\,\psi_{n_\ell}\,\big\|^2
\klg 1-\delta$, for an appropriate subsequence, and
\begin{equation*}
\begin{split}
\lim_{\ell\to\infty}
\big\|\,(\DAV-&\lambda)\,\psi_{n_\ell}\,\big\|^2
\,\grg\,\ve^2\liminf_{\ell\to\infty}
\int_{\RR}(1-\id_{I_\ve}(s))\, 
d\big\|\,E_s(\DA)\,\psi_{n_\ell}\,\big\|^2
\\
&=\,\ve^2\,-\,\ve^2\,\lim_{\ell\to\infty}
\big\|\,E_I(\DAV)\,\psi_{n_\ell}\,\big\|^2\,\grg\,\ve^2\,\delta
\,>\,0\,.
\end{split}
\end{equation*}
This is a contradiction to \eqref{Weyl-DAV}.
\qed

\bigskip

\bigskip

In the following we show that $\Th_{N-1}+\lambda\in\specess(\HH_N)$
by means of a suitable Weyl sequence. Instead of applying 
Weyl's criterion directly to $\HH_N$ we shall, however, use
a slightly strengthend
version of it in Lemma~\ref{le-Weyl-HN} (see, e.g., \cite{CFKS})
which allows to work with quadratic forms.
This is important since, for instance, it seems that one cannot expect
Proposition~\ref{prop-W1/2} to hold with $W^{1/2}$
replaced by $W$. (At least not for large nuclear charges $Z\grg118$.)
To construct the Weyl sequence 
we pick, for every $n\in\NN$, 
some
\begin{equation}\label{def-Phin}
\Phi_n=\cA_{N-1}\Phi_n
\in\PAVNme\,\coreNme\;\;\textrm{such that}\;
\left\{
\begin{array}{l}
\SPb{\Phi_n}{\HH_{N-1}^\cA\,\Phi_n}\,<\,\Th_{N-1}+\frac{1}{n}\,,\\
\|\Phi_n\|=1\,.
\end{array}
\right.
\end{equation}
This is possible since $\HH_{N-1}$ is defined as a Friedrichs extension
starting from $\PAVNme\coreNme$.
We further set
\begin{equation}
\Upsilon_n(x)\,:=\,
\int_{\RR^{3(N-2)}}\big|\Phi_n(x,X')\big|^2\,dX'\,.
\end{equation}
Next, we pick $0<a<\min\{m,\triangle_0\}$, $r\in(0,1-\ve/4)$,
and $r'\in(0,1)$
such that 
\begin{equation}\label{def-srr'}
(1-r)\,a\,>\,(1+r)\,\tau\,,\qquad
s\,:=\,r+r'-1\,>\,0\,.
\end{equation}
Here $\tau$ appears in \eqref{var-bd-B}.
We further pick 
some cut-off function,
$\vt\in C^\infty(\RR,[0,1])$, such that $\vt\equiv0$ on $(-\infty,s/2]$
and $\vt\equiv1$ on $[s,\infty)$.
By Lemma~\ref{cor-H1/2} we know that 
$|\D{0}^{(1)}|^{1/2}\,\Phi_n\in\HR_{N-1}$, where the superscript
$(1)$ again indicates that the operator acts on the first variable.
Therefore, we find a subsequence, $\{R_{k_n}\}_{n\in\NN}$,
of $\{R_k\}_{k\in\NN}$ such that, for every $n\in\NN$,
\begin{equation}\label{int-Upsilon}
\int\limits_{\RR^{3(N-1)}}
\big|\, 
|\D{0}^{(1)}|^{1/2}\,\vt(x_1/R_{k_n})\,\Phi_n(X)\,\big|^2\,dX
\,<\,\frac{1}{n}\,,
\end{equation}
As a candidate for 
a Weyl sequence we then try $\{\cA_N\Psi_n\}_{n\in\NN}$,
where 
\begin{equation}\label{def-Psin}
\Psi_n\,:=\,\Phi_n\otimes \PAV\psi_{k_n}\in\PAVN\core\,,\qquad n\in\NN\,.
\end{equation}
To simplify the notation we again write $n$ instead of $k_n$
in the following.
Finally, we 
pick some $c>1$ and set
$$
f(t)\,:=\,(t+c)^{-1/2}\,(t-\Th_{N-1}-\lambda)\,,\qquad t>-c\,.
$$

\begin{lemma}\label{le-Weyl-HN}
Let the assumptions of Theorem~\ref{main-thm-HVZ}(i)
be fulfilled. If, in the situation described above,
$c>1$ is sufficiently large, then
$\cA_N\,\Psi_n\in\dom(f(\HH_N))$, for every $n\in\NN$, and
\begin{equation}\label{Weyl-crit-CFKS}
\underset{n\to\infty}{{\rm w-lim}}\,\cA_N\,\Psi_n=0\,,\quad
\liminf_{n\to\infty}\|\cA_N\Psi_n\|>0\,,\quad
\lim_{n\to\infty}\big\|\,f(\HH_N)
\,\cA_N\Psi_n\,\big\|\,=\,0\,.
\end{equation}
In particular, $\Th_{N-1}+\lambda\in\specess(\HH_N)$.
\end{lemma}

\proof
First, suppose that \eqref{Weyl-crit-CFKS} holds true.
If $c>1$ is chosen sufficiently large, then $f$
is strictly monotonically increasing on $\spec(\HH_N)$.
If $I$ is some small open interval around $\Th_{N-1}+\lambda$
we thus get $E_I(\HH_N)=E_{f(I)}(f(\HH_N))$.
By \eqref{Weyl-crit-CFKS} and the Weyl criterion 
applied to $f(\HH_N)$
it follows that $\infty=\dim\Ran(E_{f(I)}(f(\HH_N)))=\dim\Ran(E_I(\HH_N))$.

To verify \eqref{Weyl-crit-CFKS} we first notice that
$\Psi_n\rightharpoonup0$, as $n\to\infty$,
because of \eqref{Weyl-DA}.
Exactly as in \cite[\textsection4]{MoVu}
we can also check that $\liminf\|\cA_N\Psi_n\|>0$.
So it suffices to show that
$\|f(\HH_N)\,\Psi_n\|\to0$,
as $\HH_N$ commutes with $\cA_N$. 
Since $\psi_n$ and $\Phi_n$ are normalized 
and $\Psi_n=\Phi_n\otimes\PAV\psi_n\in\PAVN\core$
we obtain
\begin{eqnarray}
\lefteqn{
\big\|\,f(\HH_N)
\,\Psi_n\,\big\|\nonumber
}
\\
&\klg&\label{ida0}
\big\|(\HH_N+c)^{-\frac{1}{2}}\,(\HH_{N-1}-\Th_{N-1})^{\frac{1}{2}}\otimes\id_{\HR^+}\big\|
\,
\big\|(\HH_{N-1}-\Th_{N-1})^{\frac{1}{2}}\,\Phi_n\big\|
\\
& &
\,+\,\label{ida1}
\big\|\,(\DAVC-\lambda)\,\PAV\,\psi_n\,\big\|
\\
&+&
\!\!\sum_{i=1}^{N-1}\!
\big\|(\HH_N+c)^{-\frac{1}{2}}\,\PAVN\,W_{iN}^{\frac{1}{2}}\big\|
\,\big\|\label{ida2}
W_{iN}^{\frac{1}{2}}(\Phi_n\otimes\PAV\,\psi_n)
\big\|.
\end{eqnarray}
We first show that the operator norm in \eqref{ida2} is actually finite.
In fact,
\begin{eqnarray*}
\big\|\,(\HH_N+c)^{-1/2}\,\PAVN\,W_{iN}^{1/2}\,\big\|
&=&\big\|\,W_{iN}^{1/2}\,\PAVN\,(\HH_N+c)^{-1/2}\,\big\|\,\klg\,1\,,
\end{eqnarray*}
since $W\grg0$, $\PAV\DAVC\PAV\grg-C'\PAV$, 
$\HH_{N-1}\grg\Th_{N-1}\,\PAVNme$, and, hence,
$$
\|\,W_{iN}^{1/2}\,\PAVN\,\phi\,\|^2
\,=\,
\SPb{\phi}{\PAVN\,W_{iN}\PAVN\,\phi}
\,\klg\,
\SPb{\phi}{(\HH_N+c)\,\phi}\,,
$$
for $\phi\in \PAVN\core$.
Using similar estimates and \eqref{def-Phin} 
it is straightforward to check
that the term in \eqref{ida0} converges to zero
provided $c>1$ is sufficiently large.
The norm in \eqref{ida1} tends to zero by Lemma~\ref{le-ida}.
The claim now follows from Lemma~\ref{le-WW} below
which
implies that the remaining norm in \eqref{ida2}
tends to zero, too.
\qed

\bigskip

The first inequality of the following lemma is used in the
proof of Lemma~\ref{le-Weyl-HN} and the second one
in Section~\ref{sec-eigenvalues}.

\begin{lemma}\label{le-WW}
There are $\kappa,C\in(0,\infty)$ such that, for all 
$n\in\NN$,
$$
\int_{\RR^6}W(x,y)\,\Upsilon_n(y)\,\big|\PAV\psi_n(x)\big|^2\,d(x,y)
\,\klg\,
\sup_{{|x-y|\grg\atop (1-r')R_n}}W(x,y)
+
C\,e^{-\kappa R_n}
\,+\frac{1}{n}\,.
$$
If $B$ is bounded, then there is some $C'\in(0,\infty)$ such that,
for all $n\in\NN$,
\begin{eqnarray*} 
\lefteqn{
\int_{\RR^6}W(x,y)\,\Upsilon_n(y)\,\big|\PAV\,\psi_n(x)\big|^2\,d(x,y)
}
\\
&\klg&
\sup_{{|x-y|\grg\atop (1-\ve)R_n}}W(x,y)
\,\big\|\,\PAV\,\psi_n\,\big\|^2
\,+\,
C'\,(1+\|B\,\|_\infty)\,e^{-a\ve R_n/2}
\\
& &
\,+\,C'\int\limits_{\{|y|\grg\ve R_n/2\}}\!\!\Upsilon_n(y)\,dy
\,(1+\|B\,\|_\infty)\,\big\|\,(\DAV+i)\,\psi_n\,\big\|^2
\,.
\end{eqnarray*}
\end{lemma}

\proof
For $n\in\NN$, we pick a weight function, 
$F_n\in C^\infty(\RR^3,[0,\infty))$,
with
$F_n\equiv0$ on $\RR^3\setminus\ball{R_n}{0}$,
$F_n\grg(1-r)\,a\,R_n-a'$ on $\ball{rR_n}{0}$ and
$\|\nabla F_n\|_\infty\klg a$.
Here $a$ and $r$ are the parameters from \eqref{def-srr'}
and $a'>0$ is some fixed, $n$-independent constant.
Since $\psi_n=\chi_n\,\psi_n$ and
$\id_{\ball{rR_n}{0}}\chi_n=0$ we obtain
\begin{eqnarray}
\lefteqn{\nonumber
\int\limits_{\{|x-y|< (1-r')R_n\}}\!\!\!\! \id_{\ball{rR_n}{0}}(x)\,
W(x,y)\,\Upsilon_n(y)\,\big|\PAV\,\psi_n(x)\big|^2\,
d(x,y)
}
\\
&\klg &\nonumber
\big\|\,\id_{\ball{rR_n}{0}}\,e^{-F_n}\,\big\|_\infty\,
\sup_{|y|\klg(r+1-r')R_n}
\big\|\,
W_y^{1/2}\,e^{F_n}\,\big[\,\PAV\,,\,\chi_n\,e^{-F_n}\,\big]
\,\psi_n\,\big\|^2
\,\|\Upsilon_n\|_1
\\
&\klg&\label{clarissa0}
C'\,e^{-(1-r)aR_n}\,\sup_{|y|\klg(1+r)R_n}(1+|B(y)|)
\,\klg\,C''\,e^{-([1-r]a-[1+r]\tau)R_n }
\,.
\end{eqnarray}
In the last two steps we make use of Proposition~\ref{prop-W1/2}
and \eqref{var-bd-B}.
Next, if $|x-y|\klg (1-r')R_n$ and $\id_{\ball{rR_n}{0}}(x)=0$,
then $|y|\grg (r+r'-1)R_n=sR_n$, and 
by the choice of $\vt$ (see the paragraph below \eqref{def-srr'})
it follows that
\begin{eqnarray}
\lefteqn{\nonumber
\int\limits_{\{|x-y|\klg (1-r')R_n\}}\!\!\!\! 
\big(1-\id_{\ball{rR_n}{0}}(x)\big)\,
W(x,y)\,\Upsilon_n(y)\,\big|\PAV\,\psi_n(x)\big|^2\,
d(x,y)
}
\\
& &\qquad\quad\klg\,
\sup_{|x|\grg rR_n}\int_{\RR^3}
W(x,y)\,\vt(y/R_n)\,\Upsilon_n(y)\,dy
\:\big\|\,
\PAV\,\psi_n\,\big\|^2.\label{clarissa}
\end{eqnarray}
On account of \eqref{def-srr'}, Kato's inequality,
and \eqref{int-Upsilon}the first asserted estimate
follows from
\eqref{clarissa0} and \eqref{clarissa}.
The second one is derived similarly by means of 
Lemma~\ref{le-josephine} and the replacements
$r\mapsto1-\ve/2$, $r'\mapsto\ve$.
Note that $\id_{\ball{(1+\ve/2)R_n}{0}}\,\chi_n=0$,
which is used to derive the analogue of \eqref{clarissa0}.
\qed


\section{Existence of eigenvalues}\label{sec-eigenvalues}

In this section we prove 
Theorem~\ref{main-thm-ev} which asserts
that $\HH_N^\cA$ possesses infinitely many
eigenvalues below $\inf\specess(\HH_N^\cA)=1+\Th_{N-1}$. 
We proceed along the lines
of \cite[\textsection6]{MoVu} with a few changes. 
In particular, we replace the
arguments of \cite{MoVu} that employ explicite
position or momentum space representations of $\PO$ 
by more abstract ones. Throughout this section
we always assume without further notice 
that the assumptions of Theorem~\ref{main-thm-ev}, i.e.,
Hypothesis~\ref{hyp-ex},
are fulfilled.

\bigskip

\noindent
{\it Proof of Theorem~\ref{main-thm-ev}: }
We proceed by induction on $N$ and start with the induction step.
So, we pick $N\in\NN$, $N\grg2$, and assume that $\HH_{N-1}^\cA$
possesses infinitely many eigenvalues below $\Th_{N-2}+1$.
In particular, we can pick a 
normalized ground state of $\HH_{N-1}^\cA$, 
which we denote by $\Phi$.
Moreover, we denote the transposition
operator which flips the $i^{\mathrm{th}}$
and $N^{\mathrm{th}}$ electron variable by $\pi_{iN}$,
$1\klg i<N$, and set $\pi_{NN}:=\id$. 
The vectors $\psi_1,\psi_2,\ldots$ are the elements of the sequence
appearing in Hypothesis~\ref{hyp-ex}.

Now, let $d\in\NN$.
By Lemma~\ref{le-lin-indep} below
we know that, for all sufficiently large $m_0\in\NN$,
the set $\{\cA_N\big(\Phi\otimes\PAV\,\psi_n\big)\}_{n=m_0}^{m_0+d}$,
where
$$
\cA_N\big(\Phi\otimes\PAV\,\psi_n\big)\,=\,
\frac{1}{N}
\sum_{i=1}^N(-1)^{N-i}\,
\pi_{iN}(\Phi\otimes\PAV\psi_n)\,,\qquad n\in\NN\,,
$$
is linearly independent.
Our goal then is to show that the expectation of
$$
\wh{\HH}_N\,:=\,\HH_N\,-\,\Th_{N-1}\,-\,1
$$
with respect to any
linear combination of the vectors 
$\{\cA_N\big(\Phi\otimes\PAV\,\psi_n\big)\}_{n=m_0}^{m_0+d}$
is strictly negative
provided $m_0\in\NN$ is large enough.
Since $d$ is arbitrary
the assertion of Theorem~\ref{main-thm-ev} then follows from
the minimax principle.
For $c_{m_0},\ldots,c_{m_0+d}\in\CC$, and 
\begin{equation}\label{def-Psi}
\Psi\,:=\,
\sum_{n=m_0}^{m_0+d} c_n\sum_{i=1}^N
\frac{(-1)^{N-i}}{N^{1/2}}\,\pi_{iN}(\Phi\otimes\PAV\psi_n)\,,
\end{equation}
we obtain as in
\cite{MoVu} by means of the anti-symmetry of $\Phi$,
\begin{align}
\SPn{\Psi}{\wh{\HH}_N\,\Psi}\,
&\klg\,\label{sergey1}
\sum_{n=m_0}^{m_0+d}|c_n|^2\,
\SPb{\Phi\otimes\PAV\psi_n}{\wh{\HH}_N\,(\Phi\otimes\PAV\psi_n)}
\\
+\,(N-1)&\sum_{n,m=m_0}^{m_0+d}|c_n||c_m|\,
\big|\SPb{\pi_{1N}(\Phi\otimes\PAV\psi_n)}{\wh{\HH}_N\,(\Phi\otimes\PAV
\psi_m)}\big|\label{sergey2}
\\
+&\label{sergey3}
\sum_{{n,m=m_0\atop m\not= n}}^{m_0+d}|c_n||c_m|\,
\big|\SPb{\Phi\otimes\PAV\psi_n}{\wh{\HH}_N\,(\Phi\otimes\PAV\psi_m)}\big|
\,.
\end{align}
Combining the eigenvalue equation $(\HH_{N-1}-\Th_{N-1})\,\Phi=0$
with
Lemmata~\ref{le-sergey2}--\ref{le-sergey4}, 
Hypothesis~\ref{hyp-ex}, and \eqref{heinrich},
we find some $\delta_0>0$ such that the scalar product 
in \eqref{sergey1} is bounded from above
by $-\delta_0\,R_{m_0}^{-1}+o(R_{m_0}^{-1})$, 
as
$m_0$ gets large.
Here the numbers $R_1,R_2,\ldots$ are those appearing
in Hypothesis~\ref{hyp-ex}.
Lemmata~\ref{le-sergey5} and~\ref{le-sergey6} imply that
the scalar products in
\eqref{sergey2} and \eqref{sergey3} are of order $\bigO(R_{m_0}^{-K})$,
as $m_0\to\infty$,
for every $K\in\NN$. 
By the Cauchy-Schwarz inequality we find some $\delta_0'>0$
such that 
$$
\SPn{\Psi}{\wh{\HH}_N\,\Psi}\,\klg\,-\delta_0'\sum_{n=m_0}^{m_0+d}|c_n|^2\,,
$$
for all $c_{m_0},\ldots,c_{m_0+d}\in\CC$, if
$m_0$ is sufficiently large (depending on $d$).
This concludes the induction step.

Finally, the case $N=1$ is treated in the same way
as the induction step $N\to N+1$  
(setting $\Th_{0}:=0$ and ignoring 
$\Phi$, $W$, and the term \eqref{sergey2}).
\qed

\bigskip

To show that the contribution coming from the 
(one-particle) kinetic energy of $\psi_n$
decreases faster than its negative potential energy
we make use of the requirement that the $\psi_n$
have vanishing lower spinor components,
$\psi_n=(\psi_{n,1},\psi_{n,2},0,0)^\top$, $n\in\NN$.
This has also been used in \cite{MoVu} together with explicite
formulas for $\Pp{0}$.
We replace these arguments by the following observation:

\begin{lemma}\label{le-sergey2}
There is some $C\in(0,\infty)$ such that
$$
0\,\klg\,\SPb{\PA\,\psi_n}{(\DA-1)\,
\PA\,\psi_n}\,\klg\,C\,R_n^{-2}\,,\qquad n\in\NN\,.
$$
\end{lemma}

\proof
Since the last two components of $\psi_n$ are zero we
have $(\beta-\id)\,\psi_n=0$.
If we denote the projection onto the first two spinor components,
$L^2(\RR^3,\CC^4)\ni(\vp_1,\vp_2,\vp_3,\vp_4)^\top
\mapsto(\vp_1,\vp_2,0,0)^\top$,
by $p$ then we also have 
$p\,\alpha_i\,\psi_n=0=p\,\alpha_i\,\partial_i\psi_n$,
$i=1,2,3$, and, therefore,
$p\,(\DA-1)\,\psi_n=0$.
Moreover,
$[\,p\,,\DA^2]=0$ and, hence, $[\,p\,,|\DA|^{-1}]=0$.
This implies
\begin{eqnarray*}
\lefteqn{
\big|\SPb{\PA\,\psi_n}{\big(\DA-1\big)\,
\PA\,\psi_n}\big|
}
\\
&=&\Big|\frac{1}{2}\,
\SPb{\psi_n}{p\,\big(\DA-1\big)\,
\psi_n}
\,+\,
\frac{1}{2}\,\SPb{\sgn(\DA)\psi_n}{\big(\DA-1\big)\,
\,\psi_n}\Big|
\\
&= &\Big|\frac{1}{2}\,\SPb{\big(\DA-1\big)\,\psi_n}{|\DA|^{-1}\,
\big(\DA-1\big)
\,\psi_n}
\\
& &\qquad\qquad\qquad\,+\;
\frac{1}{2}\,\SPb{\psi_n}{|\DA|^{-1}\,p\,
\big(\DA-1\big)
\,\psi_n}\Big|
\\
&\klg&
\frac{1}{2}\:\big\|\,(\DA-1\big)
\,\psi_n\,\big\|^2\;=\;\bigO(1/R_n^2)
\,.
\end{eqnarray*}
In the last step we apply Hypothesis~\ref{hyp-ex}.
\qed

\bigskip

In the following we split $\VC$ into a singular and regular part,
$\VC=\VCt+\VCr$, where $\VCt$ is defined in \eqref{def-wtVC}.
By Hypothesis~\ref{hyp-Vogel} $\VCr$ is bounded.

\begin{lemma}\label{le-sergey1}
As $n$ tends to infinity,
\begin{eqnarray*}
\lefteqn{
\SPb{\PAV\,\psi_n}{(\DAVC-1)
\,\PAV\,\psi_n}
}
\\
&=&
\SPb{\PAV\,\psi_n}{\VCr\,
\PAV\,\psi_n}
\,+\,\SPb{\PA\,\psi_n}{(\DA-1)\,\PA\,\psi_n}
\,+\,
o(R_n^{-1})\,.
\end{eqnarray*}
\end{lemma}

\proof
We let $\chi_n$, $n\in\NN$, denote the cut-off functions
introduced in the paragraph preceeding Lemma~\ref{le-ida}.
Then the assertion follows from
Corollary~\ref{cor-PA-PAV} applied to $\D{A,\VCt}-1$
with $\zeta=\chi_n$, since
by Lemma~\ref{le-sergey2} and Hypothesis~\ref{hyp-V-komplett},
$$
\big(\|\chi_n\,V\|+\|\nabla\chi_n\|\big)\,
\sqrt{\SPb{\PA\,\psi_n}{(\DA-1)\,
\PA\,\psi_n}\,\|\psi_n\|^2}\,=\,o(R_n^{-1})\,.
$$
\qed

\bigskip

In the next lemma
we single out
the leading order negative contribution to \eqref{sergey1}.

\begin{lemma}\label{le-sergey3}
There is some constant $C\in(0,\infty)$
such that, for all sufficiently large $n\in\NN$,
$$
\SPb{\PAV\psi_n}{\VCr\,
\PAV\,\psi_n}\,\klg\,v_\star(\delta,R_n)
\:\|\,\PAV\,\psi_n\,\|^2
\,+\,C\,e^{-R_n/C}\,,
$$
where $v_\star(\delta,R_n)$ is given by \eqref{def-vstar}.
\end{lemma}

\proof
We pick some even function 
$f\in C^\infty(\RR,[0,\infty))$ such that
$f\equiv1$ on $[\delta,\infty)$,
$f\equiv0$ on $[0,\delta/2]$, 
and $|f'|\klg 4/\delta$.
(Recall \eqref{supp-psin-ex}.)
For some $a\in(0,\delta\,\min\{\triangle_0,m\}/4)$, we
define exponential weights, $F_n(x):=a\,R_n\,f(|x|/R_n)$, $n\in\NN$.
Using the notation introduced in \eqref{def-Sdelta}\&\eqref{def-vstar}
we then obtain, for all sufficiently large $n\in\NN$,
\begin{eqnarray*}
\SPb{\PAV\,\psi_n}{\VCr\,
\PAV\,\psi_n}
&\klg&
\SPb{\PAV\,\psi_n}{\id_{S_\delta(R_n)}\,\VCr\,
\PAV\,\psi_n}
\\
& &\,+\,\|\VCr\|\,\big\|\id_{\RR^3\setminus S_\delta(R_n)}\,e^{-F_n}\big\|\,
\big\|\,e^{F_n}\,\PAV\,e^{-F_n}\,\big\|\,.
\end{eqnarray*}
where, by \eqref{def-vstar} and Pythagoras' theorem,
\begin{eqnarray*}
\lefteqn{
\SPb{\PAV\,\psi_n}{\id_{S_\delta(R_n)}\,\VCr\,
\PAV\,\psi_n}
}
\\
&\klg&
v_\star(\delta,R_n)\,\big(\,\|\,\PAV\,\psi_n\,\|^2-
\|\,\id_{\RR^3\setminus S_\delta(R_n)}\,
\PAV\,\psi_n\,\|^2\big)
\\
&\klg&
v_\star(\delta,R_n)\,\big\|\,\PAV\,\psi_n\,\big\|^2
\,+\,|v_\star(\delta,R_n)|\,\big\|\id_{\RR^3\setminus
S_\delta(R_n)}\,e^{-F_n}\big\|^2\,
\big\|\,e^{F_n}\,\PAV\,e^{-F_n}\,\big\|^2
\,.
\end{eqnarray*}
By \eqref{def-Sdelta}, \eqref{supp-psin-ex}, and the choice
of $F_n$ we know that $\|\id_{\RR^3\setminus
S_\delta(R_n)}\,e^{-F_n}\|\klg C\,e^{-a\delta R_n/2}$,
which implies the assertion of the lemma.
\qed

\bigskip

From now on, we always assume 
that the induction hypothesis
made in the proof of Theorem~\ref{main-thm-ev} is fulfilled
and that $\Phi$ is a normalized ground state eigenvector
of $\HH_{N-1}^\cA$. So,
$\HH_{N-1}^\cA\,\Phi=\Th_{N-1}\,\Phi$, $\Th_{N-1}<\Th_{N-2}+1$.
Given $\delta\in(0,\frac{1}{N})$ we pick
some $\ve\in(0,1)$ as in Hypothesis~\ref{hyp-ex}(i). Then the following
assertion is valid: 

\begin{lemma}\label{le-sergey4}
As $n$ tends to infinity, we have, for $1\klg i\klg N-1$,
$$
\SPb{\Phi\otimes\PAV\psi_n}{W_{iN}\,\Phi\otimes\PAV\psi_n}
\klg\sup_{{|x-y|\grg\atop (1-\ve)R_n}} W(x,y)
\,\|\,\PAV\,\psi_n\,\|^2+
\bigO(R_n^{-\infty})\,.
$$
\end{lemma}

\proof
This follows from Lemma~\ref{le-WW}
with $\Upsilon_n(y)=\int_{\RR^{3(N-2)}}|\Phi(y,X')|^2dX'$ and
the exponential decay
of $\Phi$, 
which is ensured by Theorem~\ref{main-thm-exp} and the induction
hypothesis.
\qed

\bigskip

Now, we turn to the discussion of the terms in \eqref{sergey2}.

\begin{lemma}\label{le-sergey5}
As $n$ and $m$ tend to infinity, 
$$
\sigma_{nm}\,:=\,
\big|\SPb{\pi_{1N}(\Phi\otimes\PAV\psi_n)}{\wh{\HH}_N\,(\Phi\otimes\PAV
\psi_m)}\big|
\,=\,\bigO\big(R_{\min\{n,m\}}^{-\infty}\big)\,.
$$
\end{lemma}

\proof
We pick $\chi\in C^\infty_0(\RR^3,[0,1])$
such that $\chi\equiv1$ on $\ball{1/4}{0}$
and $\chi\equiv0$ outside $\ball{1/2}{0}$
and set $\chi_n:=\chi(\,\cdot\,/R_n)$
and $\ol{\chi}_n:=1-\chi_n$, for $n\in\NN$.
As in \cite{MoVu} we find
\begin{eqnarray*}
\sigma_{nm}&\klg&
\Big|\SPB{\big\{\chi_n\,\PAV\psi_n\big\}\otimes
\Phi}{\Phi\otimes
\big\{(
\DAVC-1)\,\PAV\psi_m\big\}}\Big|
\\
& &+\,
\Big|\SPB{\big\{\PAV\psi_n\big\}\otimes
\Phi}{\big\{\ol{\chi}_n^{(1)}\,\Phi\big\}\otimes
\big\{(
\DAVC-1)\,\PAV\psi_m\big\}}\Big|
\\
& &+\,\sum_{i=1}^{N-1}
\Big|\SPB{\big\{\chi_n\,\PAV\,\psi_n\big\}\otimes
\Phi
}{W_{iN}\,\Phi\otimes(\PAV\,\psi_m)}\Big|
\\
& &+\,
\sum_{i=1}^{N-1}
\Big|\SPB{\big\{\PAV\psi_n\big\}\otimes
\Phi
}{W_{iN}\,(\ol{\chi}_n^{(1)}\,\Phi)\otimes(\PAV\,\psi_m)}\Big|
\\
&=:&Y_1+Y_2+\sum_{i=1}^{N-1}Y_{3i}\,+\sum_{i=1}^{N-1}Y_{4i}\,.
\end{eqnarray*}
For the first two summands we find
\begin{eqnarray*}
Y_1+Y_2&\klg&\big\|\,(
\DAVC-1)\,\PAV\,\psi_m\,\big\|\,\Big(\big\|\,\chi_n\,\PAV\psi_n\,\big\|
\,+\,\big\|\,\ol{\chi}_n^{(1)}\,\Phi\,\big\|\Big)\,,
\end{eqnarray*}
where the right side is of order 
$\bigO\big(R_{\min\{n,m\}}^{-\infty}\big)$
due to the exponential localization of $\Phi$ and the
support properties of $\psi_n$ and $\chi_n$.
Moreover, we observe that, for $i=2,\ldots,N-1$,
\begin{eqnarray*}
Y_{3i}&\klg&
\big\|\,\chi_n\,\PAV\psi_n\,\big\|
\,\big\|\,W_{iN}^{1/2}\,\Phi\,\big\|\,
\|\Phi\|\,\sup_{y\in\RR^3}\big\|\,W_{y}^{1/2}\,\PAV\,\psi_m\,\big\|\,,
\\
Y_{4i}&\klg&
\big\|\,\PAV\psi_n\,\big\|
\,\big\|\,W_{iN}^{1/2}\,\Phi\,\big\|\,
\big\|\,\ol{\chi}_n^{(1)}\,\Phi\,\big\|
\,\sup_{y\in\RR^3}\big\|\,W_{y}^{1/2}\,\PAV\,\psi_m\,\big\|\,.
\end{eqnarray*}
Here the norms $\|W_{iN}^{1/2}\,\Phi\|$, $i=2,\ldots,N-1$, 
are actually finite since $\Phi\in\Ker(\HH_{N-1}-\Th_{N-1})$
implies
$$
\big\|\,W_{iN}^{1/2}\,\Phi\,\big\|^2
\,=\,
\SPb{\Phi}{\PAVNme\,W_{iN}\,\PAVNme\,\Phi}
\,\klg\,(\Th_{N-1}+C)\|\Phi\|^2\,,
$$
for some constant $C\in(0,\infty)$.
Finally,
\begin{eqnarray*}
Y_{31}&\klg&
\sup_{y\in\RR^3}\big\|\,W_{y}^{1/2}\,\chi_n\,\PAV\,\psi_n\,\big\|\,
\|\Phi\|^2\,
\sup_{y\in\RR^3}\big\|\,W_{y}^{1/2}\,\PAV\,\psi_m\,\big\|\,,
\\
Y_{41}&\klg&\sup_{y\in\RR^3}
\big\|\,W_{y}^{1/2}\,\PAV\,\psi_n\,\big\|\,\|\Phi\|\,
\big\|\,\ol{\chi}_n^{(1)}\,\Phi\,\big\|\,
\sup_{y\in\RR^3}\big\|\,W_{y}^{1/2}\,\PAV\,\psi_m\,\big\|\,.
\end{eqnarray*}
We pick $f\in C^\infty(\RR,[0,\infty))$
such that $f\equiv0$ on $[1,\infty)$, $f\equiv1$
on $(-\infty,1/2]$,
and $-3\klg f'\klg0$, and set $F_n(x)=a\, R_n\,f(|x|/R_n)$,
$x\in\RR^3$, $n\in\NN$, where $a\in(0,\min\{\triangle_0,m\}/3)$.
Since $\chi_n\,\psi_n=0$, we find
\begin{eqnarray*}
\lefteqn{
\sup_{y\in\RR^3}\big\|\,W_{y}^{1/2}\,\chi_n\,\PAV\,\psi_n\,\big\|
}
\\
&\klg&
\sup_{|x|\klg R_n/2}e^{-F_n}\,
\sup_{y\in\RR^3}\big\|\,W_y^{1/2}\,\big[\PAV\,,\,\chi_n\,e^{F_n}\big]
\,e^{-F_n}\,\big\|\,\|\psi_n\|\,.
\end{eqnarray*}
This estimate, the exponential decay of $\Phi$, 
and Lemma~\ref{le-josephine}
imply that the terms $Y_{3i}$ and $Y_{4i}$, $1\klg i\klg N-1$,
vanish of order $\bigO\big(R_{\min\{n,m\}}^{-\infty}\big)$ also.
\qed

\bigskip

Finally, we discuss the terms in \eqref{sergey3}.

\begin{lemma}\label{le-sergey6}
As $n$ tends to infinity, it holds, for all $m>n$,
$$
\big|\SPb{\Phi\otimes\PAV\psi_n}{\wh{\HH}_N\,(\Phi\otimes\PAV\psi_m)}\big|
\,=\,\bigO(R_n^{-\infty})\,.
$$
\end{lemma}

\proof
We pick a family of smooth weight functions, $\{F_{k\ell}\}_{k,\ell\in\NN}$,
such that $F_{k\ell}\equiv0$ on $\supp(\psi_k)$,
$F_{k\ell}$ is constant outside some ball containing
$\supp(\psi_k)$ and $\supp(\psi_\ell)$, 
$\|\nabla F_{k\ell}\|_\infty\klg a<1$, and
$$
g_{k\ell}\,:=\,\|e^{-F_{k\ell}-F_{\ell k}}\|_\infty
\,\klg\,C\,e^{-a'\,\min\{R_k,R_\ell\}}\,,
\qquad k,\ell\in\NN\,,
$$
where $a,a'\in(0,1)$ and $C\in(0,\infty)$ do not depend
on $k,\ell\in\NN$.
In view of \eqref{supp-psin-ex}
it is easy to see that such a family exists.
Then we observe that
\begin{eqnarray*}
\lefteqn{
\big|\SPb{\Phi\otimes\PAV\psi_n}{\wh{\HH}_N\,(\Phi\otimes\PAV\psi_m)}\big|
}
\\
&\klg&
\big|\SPb{\PAV\,\psi_n}{\big(\DAV-1\big)\,\psi_m}\big|
\,+\,\big|\SPb{\PAV\,\psi_n}{(\VM+\VE)\,\PAV\,\psi_m}\big|
\\
& &\quad
\,+\,
\sum_{1\klg i<N}\big|\SPb{W_{iN}^{1/2}\,\Phi\otimes\PAV\,\psi_n}{
W_{iN}^{1/2}\,\Phi\otimes\PAV\,\psi_m}\big|
\\
&\klg&
g_{nm}\,
\big\|\,e^{F_{nm}}\,\PAV \,e^{-F_{nm}}\,\big\|\,\|\psi_n\|
\,\Big\{
\,\big\|\,\big(\DAV-1\big)\,\psi_m\,\big\|
\\
& &\quad\quad+
\,\big\|\,\big(i\alpha\cdot\nabla F_{mn}\,+\,\VM\,\,+\,[e^{F_{mn}}\,,\,\VE]
\,e^{-F_{mn}}\,\big)\,\psi_m\,\big\|\,\Big\}
\\
& &+\,g_{nm}
\big\|\,e^{F_{mn}}\,(\VM+\VE)\,e^{-F_{mn}}\,\big\|\,
\sup_{{k\not=\ell}}
\big\|\,e^{F_{k\ell}}\,\PAV\,e^{-F_{k\ell}}\,\psi_k\,\big\|^2
\\
& &+\,g_{nm}\,
(N-1)\,\sup_{{k\not=\ell}}\sup_{y\in\RR^3}
\big\|\,W_y^{1/2}\,e^{F_{k\ell}}\,\PAV\,e^{-F_{k\ell}}\,\psi_k\,\big\|^2
\,.
\end{eqnarray*}
By virtue of Proposition~\ref{prop-W1/2}
and Lemma~\ref{le-josephine} we know that all terms behind the factors
$g_{nm}$ appearing here are uniformly bounded which shows that the
assertion holds true.
\qed

\bigskip

Applying the above arguments in an easier situation 
we obtain the following lemma.

\begin{lemma}\label{le-lin-indep}
For every $d\in\NN$, there is some $n_0\in\NN$ such that the set
of vectors
$\{\cA_N(\Phi\otimes\PAV\,\psi_n)\}_{n=m_0}^{m_0+d}$ is linearly
independent, for all $m_0\in\NN$, $m_0\grg n_0$.
\end{lemma}

\proof
We pick $\Psi$ as in \eqref{def-Psi} and estimate $\|\Psi\|^2$
from below by an obvious analogue of \eqref{sergey1}-\eqref{sergey3}
with $\wh{\HH}_N$ replaced by the identity.
Now, by virtue of Lemma~\ref{le-PApsinto1}
there is some $m_1\in\NN$ such that 
$\|\Phi\otimes\PAV\psi_n\|=\|\PAV\psi_n\|\grg1/2$,
for all $n\grg m_1$.
The proof of Lemma~\ref{le-sergey5} shows that
$$
\big|\SPb{\pi_{1N}(\Phi\otimes\PAV\,\psi_n)}{\Phi\otimes\PAV\,\psi_m}\big|
\,=\,
\bigO(R^{-\infty}_{\min\{n,m\}})\,.
$$
Furthermore, by employing the exponential weights
from the proof of Lemma~\ref{le-sergey6}
we see that
$$
\big|\SPb{\Phi\otimes\PAV\,\psi_n}{\Phi\otimes\PAV\,\psi_m}\big|
\,=\,
|\SPn{\psi_n}{\PAV\,\psi_m}|\,\klg\,C\,e^{-\min\{R_n,R_m\}/C}\,.
$$
Altogether we find some $C'\in(0,\infty)$
such that, for $d\in\NN$ and all sufficiently large
$m_0\in\NN$,
$$
\|\Psi\|^2\,\grg\,
\frac{1}{2}\sum_{n=m_0}^{m_0+d}|c_n|^2
\,-\,\frac{C'\,(N-1)}{R_{m_0}}\sum_{n,m=m_0}^{m_0+d}|c_n||c_m|
\,-\,\frac{C'}{R_{m_0}}\sum_{{n,m=m_0\atop n\not=m}}^{m_0+d}|c_n||c_m|\,.
$$
Hence, the Cauchy-Schwarz inequality implies that,
for sufficiently large $m_0$, $\Psi$ in \eqref{def-Psi} is zero
if and only if $c_{m_0}=\ldots=c_{m_0+d}=0$.
\qed


\begin{appendix}

\section{Proofs of Proposition~\ref{prop-W1/2} and Lemma~\ref{le-josephine}}
\label{app-proofs}

\bigskip

First, we recall a usefull resolvent identity.
We remind the reader
that $\VCt=S\,|\VCt|$ denotes the polar
decomposition of the potential defined in \eqref{def-wtVC}
and set
\begin{equation}\label{def-M(z)}
M(z)\,:=\,|\VCt|^{1/2}\,
\R{0}{z}\,|\VCt|^{1/2}\,,\qquad z\in\vr\big(\D{0}\big)\,.
\end{equation}

\begin{lemma}[\cite{Ka1,Ne1,Ne2}]\label{le-Ne-for}
Assume that $\VC$ fulfills Hypothesis~\ref{hyp-Vogel} 
and let $\VCt$ be given by \eqref{def-wtVC}.
Then there exist $\eta_0>0$ and $\gamma_0\in(\gamma,1)$ such that,
for every $\eta\in\RR\setminus(-\eta_0,\eta_0)$, we have
$\|M(i\eta)\|\klg\gamma_0$
and 
\begin{equation}\label{for-Nenciu}
\R{0,\VCt}{i\eta}\,=\,\R{0}{i\eta}\,-\,
\ol{\R{0}{i\eta}\,|\VCt|^{1/2}}\,
\big(1+S\,\ol{M(i\eta)}\big)^{-1}\,S\,|\VCt|^{1/2}\,
\R{0}{i\eta}\,.
\end{equation}
\end{lemma}

\noindent{\it Notes:}
The inequality 
$\|\,|\cdot|^{-1/2}\,\R{0}{i\eta}\,|\cdot|^{-1/2}\,\|\klg1$
has been conjectured in \cite{Ne1} and proved in \cite{Ka1}.
By means of this inequality and the arguments of
\cite[Pages~2\&3 (with $k_s\equiv1$)]{Ne2} we
find some $\gamma_0\in(\gamma,1)$ such that
$\|M(i\eta)\|\klg\gamma_0$ provided $|\eta|$ is large enough.
The resolvent formula \eqref{for-Nenciu} then follows from
\cite[Lemma~2.2 \& Theorem~2.2]{Ne1}.
\qed

\bigskip

\noindent
{\it Proof of Proposition~\ref{prop-W1/2}:}
We pick some $\zeta\in C_0^\infty(\RR^3,[0,1])$ such that
$\zeta=1$ in a neighbourhood of $\cY$
and
$\supp(\zeta)\Subset\mathring{\cN}$. We
set $\ol{\zeta}:=1-\zeta$. 
Moreover, we pick some $\vr\in C_0^\infty(\RR^3,[0,1])$
with $\vr\equiv1$ on $\ball{1/2}{0}$ and $\supp(\vr)\subset\ball{1}{0}$
and set $\vr_y(x):=\vr(x-y)$, $x\in\RR^3$.
On account of Proposition~\ref{prop-lisa}
it suffices to consider
\begin{eqnarray}
\lefteqn{\nonumber
\SPb{\vr_y\,W_y^{1/2}\,\phi}{
\big[\,\PAV\,,\,e^F\,\chi\,\big]\,e^{-F}\,\psi}
}
\\
&=&\label{anna1}
\int_\Gamma\SPB{\vr_y\,W_y^{1/2}\,\phi}{
\zeta\,T(z)\,\psi
}\,\frac{dz}{2\pi}
\,+\,\int_\Gamma\SPB{\vr_y\,W_y^{1/2}\,\phi}{
\ol{\zeta}\,T(z)\,\psi
}\,\frac{dz}{2\pi}
\,,
\end{eqnarray}
for $\phi\in H^{1/2}(\RR^3,\CC^4)$ and $\psi\in \HR$,
where, by \eqref{lisa6} and \eqref{lisa5},
\begin{eqnarray}\label{anna1a}
T(z)&:=&\RAV{z}\,\wt{T}\,e^F\,\RAV{z}\,e^{-F}\,,\qquad z\in\Gamma\,,
\\
\wt{T}&:=&
i\alpha\cdot\label{anna1b}
(\chi\nabla F+\nabla\chi)\,+\,\big[\,e^F\,\chi\,,\,\VE\,\big]\,e^{-F}
\,=\,
\bigO\big(\|\nabla\chi\|_\infty+a\big)\,.
\end{eqnarray}
To study the first integral in \eqref{anna1} we write,
using \eqref{jim1adjoint},
\begin{eqnarray*}
\zeta\,\RAV{z}&=&\zeta\,\R{0,\VCt}{z}
\,+\,\nonumber
\R{0,\VCt}{z}\,i\alpha\cdot\nabla\zeta\,
\big(\R{0,\VCt}{z}-\RAV{z}\big)
\\
& &\;\;-\,\label{anna2}
\R{0,\VCt}{z}\,\zeta\,\big\{\,V-\VCt
\,+\,\alpha\cdot A\,\big\}\,\RAV{z}
\,,
\end{eqnarray*}
where $\VCt=S|\VCt|$ is defined in \eqref{def-wtVC}.
Since $\dom\big(\D{0,\VCt}\big)
\subset H^{1/2}(\RR^3,\CC^4)$ due to Lemma~\ref{le-BdMetc}(e)
we find $C,C'\in(0,\infty)$ such that, for all
$y\in\RR^3$ and all $z\in\Gamma$,
\begin{equation}\label{anna3}
\big\|\,W_y^{1/2}\,\R{0,\VCt}{z}\,\big\|
\,\klg\,C\,\big\|\,|\D{0}|^{1/2}\,\R{0,\VCt}{z}\,\big\|
\,\klg\,C'\,.
\end{equation}
By definition of $T(z)$ and $\wt{T}$
and by \eqref{harry} we thus have
\begin{eqnarray}
\lefteqn{
\int_\Gamma\Big|\SPB{\vr_y\,W_y^{1/2}\,\phi}{
\zeta\,\big\{\,T(z)-\R{0,\VCt}{z}\,
\wt{T}\,e^F\,\RAV{z}\,e^{-F}\,\big\}\,\psi
}\Big|\,\frac{|dz|}{2\pi}\nonumber
}
\\\label{anna4}
& &\qquad\qquad\qquad\qquad
\klg\,C_{a,\zeta}\,\big(a+\|\nabla\chi\|_\infty\big)\,
\int_{\RR}\frac{d\eta}{1+\eta^2}\:\|\phi\|\,\|\psi\|\,.
\end{eqnarray}
To treat the remaining part of the first integral in \eqref{anna1}
we employ the first resolvent formula and \eqref{for-Nenciu}
to obtain, for $\eta\in\RR$ with $|\eta|\grg\eta_0$
and $z=e_0+i\eta$
($\eta_0$ is the parameter appearing in Lemma~\ref{le-Ne-for}),
\begin{eqnarray}
\lefteqn{
\R{0,\VCt}{z}\,=\,\R{0}{i\eta}\,+\,e_0\,\label{anna1999}
\R{0,\VCt}{z}\,\R{0,\VCt}{i\eta}
}
\\
&- &\ol{\R{0}{i\eta}\,|\VCt|^{1/2}}
\,\big(1+S\,\ol{M(i\eta)}\big)^{-1}
\,S\,|\VCt|^{1/2}\,|\D{0}|^{-1/2}\,|\D{0}|^{1/2}\,
\R{0}{i\eta}\,.\nonumber
\end{eqnarray}
Here the operator $(1+S\,\ol{M(i\eta)})^{-1}$ is uniformly bounded
for $|\eta|\grg\eta_0$. Moreover,
\begin{equation}\label{anna2000}
\big\|\,W_y^{1/2}\,\ol{\R{0}{i\eta}\,|\VCt|^{1/2}}\,\big\|
\,\klg\,C\,,
\end{equation}
uniformly in $y\in\RR^3$ and $\eta\in\RR$, which is a simple
consequence of Kato's inequality.
In view of \eqref{anna4}, \eqref{anna1999}, and \eqref{ralf3}
it is therefore clear that it suffices to discuss the contribution
coming from the bare resolvent $\R{0}{i\eta}$ in \eqref{anna1999}.
On account of \eqref{harry}, \eqref{lisa8},
and \eqref{anna1b} we find by means of the Cauchy-Schwarz inequality,
\begin{eqnarray*}
\lefteqn{
\int\limits_{|\eta|\grg\eta_0}\Big|\SPB{
\R{0}{-i\eta}\,W_y^{1/2}\,\vr_y\,\zeta\,\phi}{
\wt{T}\,e^F\,\RAV{e_0+i\eta}\,e^{-F}\,\psi
}\Big|\,\frac{d\eta}{2\pi}
}
\\
&\klg &
\|\wt{T}\|\,
\big\|\,\big(|\D{0}|^{-1/2}\,W_y^{1/2}\big)\,\vr_y\,\zeta\,
\,\phi\,\big\|\,\|\psi\|
\,=\,\bigO\big(a+\|\nabla\chi\|_\infty\big)\,\|\phi\|\,\|\psi\|\,.
\end{eqnarray*}
Since the remaining part of the integral over $\{|\eta|<\eta_0\}$
does not pose any further problem,
we see altogether that the first integral in \eqref{anna1}
is absolutely convergent and of order
$\bigO\big(\|\nabla\chi\|_\infty+a\big)$.

Next, we treat the second integral on the right side of \eqref{anna1}.
Since we eventually have to change the gauge locally,
we pick a (smooth)
partition of unity, $\{J_\nu\}_{\nu\in\ZZ^3}$, on $\RR^3$
such that 
$\supp(J_\nu)\subset\ball{1}{\nu}$, for every $\nu\in\ZZ^3$.
We can certainly assume that $\sum_{\nu\in\ZZ^3}|\nabla J_\nu|\klg C$,
for some $C\in(0,\infty)$.
We set
$$
\sG(\chi)\,:=\,\big\{\,\nu\in\ZZ^3\,:\:J_\nu\,\chi
\not=0\,\big\}\,,\qquad \mu\,:=\,\sum_{\nu\in\sG(\chi)}J_\nu\,,\qquad
\ol{\mu}\,:=\,1-\mu\,,
$$
so that $\mu\,\chi=\chi$, $\ol{\mu}\,\chi=0$,
and re-write the operator defined in \eqref{anna1b} as
\begin{eqnarray*}
\wt{T}&=&i\alpha\cdot
(\chi\nabla F+\nabla\chi)\,+\,\chi\,\big[\,e^F\,,\,\VE\,\big]\,e^{-F}
\,+\,\big[\,\chi\,,\,\VE\,\big]
\\
&=&
\mu\,\Big\{i\alpha\cdot(\chi\nabla F+\nabla\chi)\,+\,
\chi\,\big[\,e^F\,,\,\VE\,\big]\,e^{-F}\,+\,\big[\,\chi\,,\,\VE\,\big]\Big\}
\,-\,\big\{\ol{\mu}\,\VE\,\chi\big\}\,\mu
\\
&=:&\mu\,U_1\,-\,U_2\,\mu\,=\,
\sum_{\nu\in\sG(\chi)}(J_\nu\,U_1-U_2\,J_\nu)
\,.
\end{eqnarray*}
For every $\nu\in\ZZ^3$, there is some gauge potential,
$g_\nu\in C^2(\RR^3,\RR)$, such that
$A-\nabla g_\nu=A_\nu$, where $A_\nu$ is defined by
\begin{equation}\label{gauge-Ay}
A_\nu(x)\,:=\,\int_0^1B\big(\nu+t(x-\nu)\big)\wedge t(x-\nu)\,dt\,,\qquad
x\in\RR^3\,.
\end{equation}
By virtue of \eqref{jim1adjoint}
we obtain with $\wt{A}_\nu:=\nabla g_\nu$,
\begin{eqnarray}
\lefteqn{
\vr_y\,\ol{\zeta}\,\RAV{z}\,\wt{T}\,=\,\sum_{\nu\in\sG(\chi)}
\vr_y\,\ol{\zeta}\,\RAV{z}\,(J_\nu\,U_1-U_2\,J_\nu)\nonumber
}
\\
&=&\nonumber
\sum_{\nu\in\sG(\chi)}
\vr_y\,\ol{\zeta}\,\R{\wt{A}_\nu}{z}\,(J_\nu\,U_1-U_2\,J_\nu)
\\
& &+\nonumber
\sum_{\nu\in\sG(\chi)}
\R{\wt{A}_\nu}{z}\,i\alpha\cdot\nabla
(\vr_y\,\ol{\zeta})\,\big(\R{\wt{A}_\nu}{z}-\RAV{z}\big)\,
(J_\nu\,U_1-U_2\,J_\nu)
\\
& &-\label{mathilda}
\sum_{\nu\in\sG(\chi)}
\R{\wt{A}_\nu}{z}\,\vr_y\,\ol{\zeta}\,(V+\alpha\cdot A_\nu)\,\RAV{z}
\,(J_\nu\,U_1-U_2\,J_\nu)
\\
&=:&
S_1\,+\,S_2\,-\,S_3
\,.\nonumber
\end{eqnarray}
The proof of Proposition~\ref{prop-W1/2} is finished as soon
as we have shown the following lemma.

\begin{lemma}\label{le-Si}
In the situation above there exists a 
$\chi$- and $F$-independent constant, 
$C_{a_0}\in(0,\infty)$, such that,
for $j=1,2,3$,
\begin{eqnarray*}
I_j&:=&\int_\Gamma\Big|
\SPB{W_y^{1/2}\,\phi}{S_j\,e^F\,\RAV{z}\,e^{-F}\,\psi}
\Big|\,|dz|
\\
&\klg& C_{a_0}\,
\Big(1+\min\big\{|B(y)|\,,\,\|B\,\|_{\infty,\chi}\big\}\Big)
\big(\|\nabla\chi\|_\infty+a\big)\,
\|\phi\|\,\|\psi\|\,.
\end{eqnarray*}
\end{lemma}
 
\noindent{\it Proof of Lemma~\ref{le-Si}: }
In our estimations below, that involve non-local operators,
we exploit the fact that the interference between
spatially seperated regions decays exponentially.
Therefore, we start by introducing appropriate
exponential weight functions:
We pick some $\tilde{a}\in(\tau,\min\{\triangle_0,m\})$
and a convex, even function $\tilde{f}\in C^\infty(\RR,[0,\infty))$
such that $\tilde{f}\equiv0$ on $[-2,2]$, 
$\tilde{f}(t)=\tilde{a}|t|-3\tilde{a}$, for $|t|\grg 4$, and
$0\klg \tilde{f}'\klg1$ on $(0,\infty)$.
We define
$f_\nu(x):=\tilde{f}(|x-\nu|)$, $x\in\RR^3$,
so that $f_\nu\equiv0$ on $\ball{2}{\nu}$
and $f_\nu\grg \tilde{a}\,\dist(\,\cdot\,,\ball{2}{\nu})-\tilde{a}$
with equality outside
$\ball{4}{\nu}$.
Moreover, $|\nabla f_\nu|\klg \tilde{a}$.
We further pick some 
non-decreasing $\theta\in C^\infty(\RR,\RR)$
such that $\theta(t)=t$, for $t\klg1$,
$\theta\equiv2$ on $[3,\infty)$ and $\theta'\klg1$.
We set $\theta_{\nu,y}(t):=(|\nu-y|+1)\,\theta\big(t/(|\nu-y|+1)\big)$,
$t\in\RR$, and $f_{\nu,y}:=\theta_{\nu,y}\circ f_\nu$,
so that $f_{\nu,y}$ is bounded,
$f_{\nu,y}=f_\nu\grg \tilde{a}\,\dist(\,\cdot\,,\ball{2}{\nu})-\tilde{a}$ 
on $\ball{1}{y}\supset\supp(\vr_y)$, and $|\nabla f_{\nu,y}|\klg \tilde{a}$.
By construction $e^{f_{\nu,y}}\,J_\nu=J_\nu$. Setting
$$
\wt{\psi}_{\nu,y}(z)\,:=\,
\big(J_\nu\,U_1
-e^{f_{\nu,y}}\,U_2\,e^{-f_{\nu,y}}\,J_\nu\big)\,e^F\,\RAV{z}\,e^{-F}\,\psi
$$
we thus have
$$
(J_\nu\,U_1-U_2\,J_\nu)\,e^F\,\RAV{z}\,e^{-F}\,\psi\,=\,
e^{-f_{\nu,y}}\,\wt{\psi}_{\nu,y}(z)
\,.
$$
Observing that $\ol{\mu}\,\chi=0$ implies
$$
e^{f_{\nu,y}}\,U_2\,e^{-f_{\nu,y}}\,
\,=\,\ol{\mu}\,\big[\,e^{f_{\nu,y}}\,\VE\,e^{-f_{\nu,y}}\,,\,\chi\,\big]
$$
and employing \eqref{bd-VE1} and \eqref{harry}
we further find some constant $C\in(0,\infty)$ such that,
for all $z=e_0+i\eta\in\Gamma$, $\nu\in\ZZ^3$, and $y\in\RR^3$,
\begin{equation}\label{est-psit}
\|\wt{\psi}_{\nu,y}(z)\|\,\klg\,C\,
\frac{a+\|\nabla\chi\|_\infty}{\sqrt{1+\eta^2}}\:\|\psi\|\,.
\end{equation}
Taking these remarks into account we obtain
\begin{eqnarray*}
I_1&\klg&
\sum_{\nu\in\sG(\chi)}\int_\Gamma\Big|
\SPB{e^{-f_{\nu,y}}\,\R{\wt{A}_\nu}{\ol{z}}\,e^{f_{\nu,y}}\,
W_y^{1/2}\,(\vr_y\,e^{-f_{\nu,y}})\,\ol{\zeta}\,\phi}{
\wt{\psi}_{\nu,y}(z)}
\Big|\,|dz|\,.
\end{eqnarray*}
Now, since $\wt{A}_\nu=\nabla g_\nu$ is a gradient we
have $\rot\,\wt{A}_\nu=0$, whence \eqref{Clifford},
\eqref{eq-hyp-W}, and Hardy's inequality imply
$$
\big\|W_y\,\vp\,\big\|
\klg C\,\big\|\nabla(e^{ig_\nu}\,\vp)\,\big\|= C\,
\big\|(-i\nabla+\wt{A}_\nu)\,\vp\,\big\|\klg C\,
\big\|\,|\D{\wt{A}_\nu}|\,\vp\,\big\|\,,
$$
for $\vp\in \COI$,
with some $\nu$- and $y$-independent constant $C\in(0,\infty)$.
Standard arguments now imply that
$|\D{\wt{A}_\nu}|^{-1/2}W_y^{1/2}$ is a bounded operator
whose norm is uniformly bounded in $\nu$ and $y$.
Setting 
$$
\phi_\nu\,:=\,|\D{\wt{A}_\nu}|^{-1/2}W_y^{1/2}
\,(\vr_y\,e^{-f_{\nu,y}})\,\ol{\zeta}\,\phi
$$ 
and applying \eqref{Cauchy2} we thus find
\begin{eqnarray*}
I_1&\klg&\sum_{\nu\in\sG(\chi)}
\Big(\int_\Gamma\big\|\,e^{-f_{\nu,y}}\,\R{\wt{A}_\nu}{\bar{z}}
\,e^{f_{\nu,y}}
\,|\D{\wt{A}_\nu}|^{1/2}\,\phi_\nu\,\big\|^2\,|dz|\Big)^{1/2}
\\
& &\qquad\qquad\cdot
\Big(\int_\Gamma\big\|\,\wt{\psi}_{\nu,y}(z)\,\big\|^2\,|dz|\Big)^{1/2}
\\
&\klg&C\sum_{\nu\in\sG(\chi)}\|\phi_\nu\|\,
\big(a+\|\nabla\chi\|_\infty\big)
\Big(\int_\RR\frac{d\eta}{1+\eta^2}\Big)^{1/2}\,\|\psi\|
\\
&\klg&C'\sum_{\nu\in\ZZ^3}\sup\big\{
e^{-f_{\nu,y}(x)}:\,x\in\ball{1}{y}\big\}
\,\|\phi\|\,
\big(a+\|\nabla\chi\|_\infty\big)\,\|\psi\|
\\
&\klg&
C''\,\big(a+\|\nabla\chi\|_\infty\big)\,\|\phi\|\,\|\psi\|\,.
\end{eqnarray*}
For $I_2$,
we obtain the estimate by means of \eqref{ralf3}, \eqref{harry},
and \eqref{est-psit},
\begin{eqnarray*}
I_2&\klg&
\sum_{\nu\in\sG(\chi)}
\int_\Gamma|dz|\,\big\|\,\,|\D{\wt{A}_\nu}|^{-1/2}\,W_y^{1/2}\,\big\|
\|\phi\|\,
\big\|\,|\D{\wt{A}_\nu}|^{1/2}\,\R{\wt{A}_\nu}{z}\,\big\|
\\
& &\;\cdot\,\big\|\,
|\nabla
(\vr_y\,\ol{\zeta})|\,e^{-f_{\nu,y}}\,\big\|\,
\big\|\,e^{f_{\nu,y}}\,
\big(\RAV{z}-\R{\wt{A}_\nu}{z}\big)\,e^{-f_{\nu,y}}\,
\,\big\|\,\big\|\,\wt{\psi}_{\nu,y}(z)\,\big\|
\\
&\klg&C\,\big(\|\nabla\chi\|_\infty+a\big)
\sum_{\nu\in\ZZ^3}\sup\big\{e^{-f_{\nu,y}(x)}:\,x\in\ball{1}{y}\big\}
\,\|\phi\|\,\|\psi\|
\\
&\klg&
C'\,\big(\|\nabla\chi\|_\infty+a\big)\,\|\phi\|\,\|\psi\|\,.
\end{eqnarray*}
To derive a bound on $I_3$ we employ the special properties
of the gauge transformed vector potentials $A_\nu$.
Namely, we make use of the bound
\begin{equation}\label{eva22}
\vr_y(x)\,e^{-f_{\nu,y}(x)}\,|A_\nu(x)|\,\klg\,
\vr_y(x)\,\frac{|x-\nu|}{2}\,
\Big(b_1\,e^{-(\tilde{a}-\tau)|x-\nu|+3\tilde{a}}\,+|B(\nu)|
\,e^{-\tilde{a}|x-\nu|+3\tilde{a}}\Big)\,,
\end{equation}
for all $\nu\in\ZZ^3$ and $x\in\RR^3$, which follows from
\eqref{var-bd-B}.
Since also
$|B(\nu)|\klg|B(y)|+|B(\nu)-B(y)|$ 
and $|x-y|\klg1$, if $\vr_y(x)\not=0$,
we
infer again from \eqref{var-bd-B} that
\begin{eqnarray}
\lefteqn{\label{josephine}
\sum_{\nu\in\sG(\chi)}\big\|\,\vr_y\,e^{-f_{\nu,y}}\,|A_\nu|\,\big\|_\infty
}
\\
&\klg&C\nonumber
\sum_{\nu\in\sG(\chi)}e^{-\hat{a}\,|y-\nu|}\,
\Big(1+\min\big\{|B(y)|\,,\,\sup_{\nu\in\sG(\chi)}|B(\nu)|\big\}\Big)
\,,
\end{eqnarray}
for some sufficiently small $\hat{a}>0$.
Using these observations and 
the uniform boundedness of 
$\ol{\zeta}\,e^{f_{\nu,y}}\,V\,e^{-f_{\nu,y}}$,
which is implied by Hypothesis~\ref{hyp-V-komplett}
and the choice of $\zeta$,
we find some $\chi$-, $F$-, and $y$-independent constant
$C'\in(0,\infty)$ such that
\begin{eqnarray*}
I_3&\klg&\sum_{\nu\in\sG(\chi)}
\int_\Gamma\big\|\,\R{\wt{A}}{\bar{z}}\,W_y^{1/2}\,\big\|\,\|\phi\|
\,
\Big\{\,\big\|\,\vr_y\,e^{-f_{\nu,y}}\,\big\|_\infty\,
\big\|\ol{\zeta}\,e^{f_{\nu,y}}\,V\,e^{-f_{\nu,y}}\,\big\|
\\
& &\;\qquad+\,
\big\|\,\vr_y\,e^{-f_{\nu,y}}\,|A_\nu|\,\big\|_\infty\Big\}
\,
\big\|\,e^{f_{\nu,y}}\,\RAV{z}\,e^{-f_{\nu,y}}\,\big\|\,
\big\|\,\wt{\psi}_{\nu,y}(z)\,\big\|\,|dz|
\\
&\klg&
C'\,\Big(1+\min\big\{|B(y)|\,,\,\|B\,\|_{\infty,\chi}\big\}\Big)
\,\big(\|\nabla\chi\|_\infty+a\big)\,\|\phi\|\,\|\psi\|\,.
\end{eqnarray*}
This completes the proof of Lemma~\ref{le-Si} and, at the same time, 
the proof
of Proposition~\ref{prop-W1/2}.
(The last assertion of Proposition~\ref{prop-W1/2} follows by
inspecting the arguments above.)
\qed

\bigskip

\noindent
{\it Proof of Lemma~\ref{le-josephine}:}
We use the notation introduced in the proofs 
of Proposition~\ref{prop-W1/2} and Lemma~\ref{le-Si} 
in the following.
We already know from Corollary~\ref{cor-H1/2} that 
the vector
$\PAV\,\psi$ belongs to $\dom(W_y^{1/2})$, but we do not have 
any control on the norm on the left in \eqref{petra} yet.
It is certainly sufficient to derive the asserted bound
with $W_y^{1/2}$ replaced by $\vr_y\,W_y^{1/2}$.
As in the proof of Corollary~\ref{cor-H1/2} we first
pick some $\zeta\in C_0^\infty(\RR^3,[0,1])$
(independent of $\psi$) such that $\zeta\equiv1$
on some large open ball containing $\cY$ and set
$\ol{\zeta}=1-\zeta$. By the closed graph theorem
$|\D{0}|^{1/2}\zeta\RAV{i}$ is bounded whence
$$
\big\|\,\vr_y\,W_y^{1/2}\,\zeta\,\PAV\,\psi\,\big\|
\,\klg\,C\,
\big\|\,|\D{0}|^{1/2}\zeta\RAV{i}\,\big\|\,
\big\|\,\big(\DAV-i\big)\,\psi\,\big\|\,.
$$
We denote the characteristic function of the support
of $\psi$ by $\chi$.
To treat the remaining piece of the norm we 
set 
$$
\wt{\psi}_\nu\,:=\,\PAV\,\big(\D{A,V}-i\big)\,J_\nu\,\psi\,,
$$
and
write
analogously to \eqref{mathilda},
\begin{eqnarray*}
\lefteqn{
\big|\SPb{W_y^{1/2}\,\phi}{\vr_y\,\ol{\zeta}\,\PAV\,\psi}\big|
\,\klg\,\sum_{\nu\in\sG(\chi)}
\big|\SPb{W_y^{1/2}\,\phi}{\vr_y\,\ol{\zeta}\,\RAV{i}\,
\wt{\psi}_\nu}\big|
}
\\
&\klg&
\sum_{\nu\in\sG(\chi)}
\big|\SPb{W_y^{1/2}\,\vr_y\,\ol{\zeta}\,\phi}{\R{\wt{A}_\nu}{i}\,
\wt{\psi}_\nu}\big|
\\
& &+
\sum_{\nu\in\sG(\chi)}
\big|\SPb{W_y^{1/2}\,\phi}{\R{\wt{A}_\nu}{i}\alpha\cdot\nabla(
\vr_y\,\zeta)\,\big(\RAV{i}-\R{\wt{A}_\nu}{i}\big)\,\wt{\psi}_\nu
}\big|
\\
& &+
\sum_{\nu\in\sG(\chi)}
\big|\SPb{W_y^{1/2}\,\phi}{\R{\wt{A}_\nu}{i}\,
\vr_y\,\ol{\zeta}\,(V+\alpha\cdot A_\nu)\,\RAV{i}\,\wt{\psi}_\nu
}\big|
\\
&=:&Q_1\,+\,Q_2\,+\,Q_3\,,
\end{eqnarray*}
where $\phi\in H^{1/2}(\RR^3,\CC^4)$.
Again we use exponential weights 
constructed in the beginning of the proof of Lemma~\ref{le-Si}
and abbreviate
$$
\wh{\psi}_{\nu,y}\,
:=\,
\big(e^{f_{\nu,y}}\,\PAV\,e^{-f_{\nu,y}}\big)\,
e^{f_{\nu,y}}\big(\D{A,V}-i\big)e^{-f_{\nu,y}}\,J_\nu\,\psi\,,
$$
so that by Corollary~\ref{cor-eFPAVe-F}
$$
\|\wh{\psi}_{\nu,y}\|\klg C
\big\|\,\big(\D{A,V}-i\big)\,\psi\,\big\|\,+\,
\bigO\big(\|\nabla J_\nu\|_\infty+\tilde{a}\big)\,\|\psi\|\,
\klg\,C'\,\big\|\,\big(\D{A,V}-i\big)\,\psi\,\big\|,
$$
where $C,C'\in(0,\infty)$ neither depend on $\nu$ nor $y$.
Using also \eqref{RA-conjugated} we thus obtain
\begin{eqnarray*}
Q_1&\klg&\sum_{\nu\in\sG(\chi)}
\big\|\,|\D{\wt{A}_\nu}|^{-1/2}\,W_y^{1/2}\,\big\|
\,\big\|\,\vr_y\,e^{-f_{\nu,y}}\,\big\|\,\|\phi\|
\\
& &\;\cdot
\,\big\|\,|\D{\wt{A}_\nu}|^{1/2}\,
e^{f_{\nu,y}}\,\R{\wt{A}_\nu}{i}\,e^{-f_{\nu,y}}\,\big\|
\,\|\wh{\psi}_{\nu,y}\|
\\
&\klg&
C''\,\|\phi\|\,\big\|\,\big(\D{A,V}-i\big)\,\psi\,\big\|\,.
\end{eqnarray*}
Using \eqref{josephine} we further find
\begin{eqnarray*}
Q_3&\klg&\sum_{\nu\in\sG(\chi)}
\big\|\,|\D{\wt{A}_\nu}|^{-1/2}\,W_y^{1/2}\,\big\|\,\|\phi\|\,
\big\|\,|\D{\wt{A}_\nu}|^{1/2}\,\R{\wt{A}_\nu}{i}\,\big\|
\\
& &\;\cdot\,
\Big\{\,\big\|\,\vr_y\,e^{-f_{\nu,y}}\,\big\|_\infty\,
\big\|\ol{\zeta}\,e^{f_{\nu,y}}\,V\,e^{-f_{\nu,y}}\,\big\|+
\big\|\,\vr_y\,e^{-f_{\nu,y}}\,|A_\nu|\,\big\|_\infty\Big\}
\,\|\wh{\psi}_{\nu,y}\|
\\
&\klg&C'''\,
\Big(1+\min\big\{
|B(y)|\,,\,\|B\,\|_{\infty,\psi}\big\}\Big)\,
\|\phi\|\,\big\|\,\big(\D{A,V}-i\big)\,\psi\,\big\|\,.
\end{eqnarray*}
The remaining term, $Q_2$, can be dealt with similarly.
\qed

\end{appendix}

\bigskip

\noindent
{\bf Acknowledgement:} It is a pleasure to thank Hubert Kalf,
Sergey Morozov, and Heinz Siedentop for useful remarks and
helpful discussions. Moreover, we thank Sergey Morozov for
making parts of his manuscripts \cite{Mo} available to us prior
to publication.



\begin{thebibliography}{99}



\bibitem{BFS1}
V.~Bach, J.~Fr\"ohlich, and I.~Sigal.
\newblock Quantum electrodynamics of confined non-relativistic
particles.
\newblock {\em Adv. Math.}, \textbf{137}: 299--395, 1998.


\bibitem{BaMa}
V.~Bach and O.~Matte.
\newblock Exponential decay of eigenfunctions of the
Bethe-Salpeter operator.
\newblock {\em Lett. Math. Phys.}, \textbf{55}: 53--62, 2001.

 
\bibitem{BaEv1}
A.~A.~Balinsky and W.~D.~Evans.
\newblock
On the virial theorem for the relativistic operator of 
Brown and Ravenhall, and the absence of embedded eigenvalues.  
\newblock
{\em Lett. Math. Phys.}, \textbf{44}: 233--248, 1998. 


\bibitem{BaEv2}
A.~A.~Balinsky and W.~D.~Evans.
\newblock Stability of one-electron molecules
in the Brown-Ravenhall model.
\newblock {\em Comm. Math. Phys.}, \textbf{202}: 481--500, 1999.


\bibitem{BaEv4}
A.~A.~Balinsky and W.~D.~Evans.
\newblock 
On the spectral properties of the Brown-Ravenhall operator. 
\newblock 
{\em J. Comput. Appl. Math.}, \textbf{148}: 239--255, 2002.


\bibitem{BeGe}
A.~Berthier and V.~Georgescu.
\newblock
On the point spectrum of Dirac operators.
\newblock
{\em J. Funct. Anal.}, \textbf{71}: 309--338, 1987.


\bibitem{BeSa}
H.~A.~Bethe and E.~E.~Salpeter.
\newblock Quantum mechanics of one- and two-electron atoms.
\newblock In: {Handbuch der Physik, XXXV}. (Pages 88--436.)
\newblock Edited by S.~Fl\"ugge.
\newblock Springer, Berlin, 1957.


\bibitem{BdMP}
A.~M.~Boutet de Monvel and R.~Purice.
\newblock 
A distinguished self-adjoint extension for the Dirac operator
with strong local singularities and arbitrary behaviour at infinity.
\newblock 
{\em Rep. Math. Phys.}, \textbf{34}: 351--360, 1994.


 
\bibitem{BrRa}
G.~E.~Brown and D.~G.~Ravenhall.
\newblock On the interaction of two electrons.
\newblock {\em Proc. Roy. Soc. London A}, \textbf{208}:
552--559, 1951.


\bibitem{CaSi}
R.~Cassanas and H.~Siedentop
\newblock The ground-state energy of heavy atoms according 
to Brown and Ravenhall: 
absence of relativistic effects in leading order.  
\newblock{\em J. Phys. A}, \textbf{39}:   10405--10414, 2006.




\bibitem{CFKS}
H.~L.~Cycon, R.~G.~Froese, W.~Kirsch, and B.~Simon.
\newblock{\em Schr\"odinger operators.}
\newblock Texts and Monographs in Physics,
Springer, Berlin-Heidelberg, 1987.


\bibitem{DiSj}
M.~Dimassi and J.~Sj\"{o}strand.
\newblock {\em Spectral asymptotics in the semi-classical limit.}
\newblock London Math. Soc. Lecture Note Series, \textbf{268}.
\newblock Cambridge University Press, Cambridge, 1999.
 

\bibitem{DEL}
J.~Dolbeault, M.~J.~Esteban, and M.~Loss.
\newblock
Relativistic hydrogenic atoms in strong magnetic fields.
\newblock
{\em Preprint}, 2006.


\bibitem{EPS}
W.~D.~Evans, P.~Perry and H.~Siedentop.
\newblock The spectrum of relativistic one-electron atoms
according to Bethe and Salpeter.
\newblock {\em Comm. Math. Phys.}, \textbf{178}: 733--746, 1996. 


\bibitem{GeMa}
V.~Georgescu and M.~M\v{a}ntoiu.
\newblock On the spectral theory of singular Dirac type Hamiltonians.
\newblock {\em J. Operator theory}, \textbf{46}: 289--321, 2001.


\bibitem{Gr}
M.~Griesemer.
\newblock Exponential decay and ionization thresholds in
non-relativistic quantum electrodynamics.
\newblock {\em J. Funct. Anal.}, \textbf{210}: 321--340, 2004.


\bibitem{GrSi}
M.~Griesemer and H.~Siedentop.
\newblock 
A minimax principle for the eigenvalues in spectral gaps.  
\newblock 
{\em J. London Math. Soc. (2)}, \textbf{60}: 490--500, 1999.


\bibitem{GrTix}
M.~Griesemer and C.~Tix.
\newblock 
Instability of a pseudo-relativistic model of 
matter with self-generated magnetic field.  
\newblock {\em J. Math. Phys.}, \textbf{40}: 1780--1791, 1999.


\bibitem{HLSS}
C.~Hainzl, M.~Lewin, E.~S\'er\'e, and J.~P.~Solovej.
\newblock A minimization method for relativistic electrons in a
mean-field approximation of quantum electrodynamics.
\newblock {\em Preprint, arXiv:0706.1486v1}, 18 Pages, 2007.


\bibitem{HNW}
B.~Helffer, J.~Nourrigat, and X.~P.~Wang.
\newblock
Sur le spectre de l'\'{e}quation de Dirac (dans $\RR^3$ ou $\RR^2$)
avec champ magnetic.
\newblock
{\em Ann. Sci. \'{E}cole Normale Superieur, $4^e$ S\'erie}, \textbf{22}:
515--533, 1989.


\bibitem{HoeSi}
G.~Hoever and H.~Siedentop.
\newblock Stability of the Brown-Ravenhall operator.
\newblock {\em Math. Phys. Electr. J.}, \textbf{5}: Paper 6, 11 pages, 1999.


\bibitem{J-A1}
D.~H.~Jakuba{\ss}a-Amundsen.
\newblock The HVZ theorem for a pseudo-relativistic operator.
\newblock {\em Ann. Henri~Poincar\'e}, \textbf{8}: 337--360, 2007.


\bibitem{J-A2}
D.~H.~Jakuba{\ss}a-Amundsen.
\newblock The HVZ theorem for the Brown-Ravenhall
operator with constant magnetic field.
\newblock {\em Preprint, arXiv:0709.1034}, 27 Pages, 2007.


\bibitem{Kato}
T.~Kato.
\newblock {\em Perturbation theory for linear operators.}
\newblock 
Classics in mathematics, Springer, Berlin-Heidelberg, 1995.


\bibitem{Ka1}
T.~Kato.
\newblock Holomorphic families of Dirac operators.
\newblock {\em Math. Z.}, \textbf{183}: 399--406, 1983.


\bibitem{LL}
E.~H.~Lieb and M.~Loss.
\newblock
Stability of a model of relativistic quantum electrodynamics.
\newblock
{\em Comm. Math. Phys.}, \textbf{228}: 561--588, 2002.


\bibitem{LSS}
E.~H.~Lieb, H.~Siedentop, and J.~P.~Solovej.
\newblock Stability and instability of relativistic electrons in 
classical electromagnetic fields. 
\newblock 
{\em J. Statist. Phys.}, \textbf{89}: 37--59, 1997.


\bibitem{Mo1}
S.~Morozov.
\newblock
Essential spectrum of multiparticle Brown-Ravenhall operators in 
external field.
\newblock
{\em Preprint, arXiv:0802.0453,} 29 Pages, 2008.

\bibitem{Mo}
S.~Morozov.
\newblock
PhD thesis, Universit\"at M\"unchen. {\em In preparation}.


\bibitem{MoVu}
S.~Morozov and S.~Vugalter.
\newblock Stability of atoms in the Brown-Ravenhall model.
\newblock {\em Ann. Henri Poincar\'e}, \textbf{7}: 661--687, 2006.


\bibitem{Ne1}
G.~Nenciu.
\newblock Self-adjointness and invariance of the essential spectrum
for Dirac operators defined as quadratic forms.
\newblock {\em Comm. Math. Phys.}, \textbf{48}: 235--247, 1976.


\bibitem{Ne2}
G.~Nenciu.
\newblock Distinguished self-adjoint extension for the Dirac operator
with potentia dominated by multicenter Coulomb potentials.
\newblock {\em Helvetica Phys. Acta}, \textbf{50}: 1--3, 1977.


\bibitem{RTdA}
R.~Richard and R.~Tiedra de Aldecoa.
\newblock On the spectrum of magnetic Dirac operators with Coulomb-type
perturbations.
\newblock {\em Preprint}, arXiv: math-ph/0611072v1, 2006.


\bibitem{Su1}
J.~Sucher.
\newblock Foundations of the relativistic theory of many-electron
atoms.
\newblock {\em Phys. Rev. A}, \textbf{22}: 348--362, 1980.


\bibitem{Su2}
J.~Sucher.
\newblock 
Relativistic many-electron Hamiltonians.
\newblock {\em Physica Scripta}, \textbf{36}: 271--281, 1987.



\bibitem{Th}
B.~Thaller.
\newblock {\em The Dirac equation.}
\newblock Texts and Monographs in Physics, Springer,
Berlin-Heidelberg, 1992.


\bibitem{Tix1}
C.~Tix.
\newblock Strict positivity of a relativistic
Hamiltonian due to Brown and Ravenhall.
\newblock {\em Bull. London Math. Soc.}, \textbf{30}: 283--290.


\bibitem{Tix2}
C.~Tix.
\newblock Self-adjointness and spectral properties of a
pseudo-relativistic Hamiltonian due to Brown and Ravenhall.
\newblock {\em Preprint,} ${mp}_-{arc}$ 97-441, 
20 pages, 1997.

 
\bibitem{Tix3}
C.~Tix.
\newblock Lower bound for the ground state energy of the
no-pair Hamiltonian.
\newblock {\em Phys. Lett. B}, \textbf{405}: 293--296, 1997.


\bibitem{Xia}
J.~Xia.
\newblock On the contribution of the Coulomb singularity of
arbitrary charge to the Dirac Hamiltonian.
\newblock {\em Trans. Amer. Math. Soc.}, \textbf{351}: 1989--2023.


\end{thebibliography}
\end{document}